\documentclass[a4paper]{scrartcl}
\usepackage{epsfig}
\usepackage{amssymb}
\usepackage{graphicx}
\usepackage{amsmath}
\usepackage{array}

\usepackage[center]{subfigure}
\usepackage{slashed}
\usepackage{bbm}
\usepackage{mathrsfs}

\newcommand{\bra}[1]{\left\langle #1 \right|}
\newcommand{\ket}[1]{\left| #1 \right\rangle}

\newcommand{\unitop}{\mathbbm{1}}
\newcommand{\Lc}{{\Lambda_c}}
\newcommand{\Lcst}{{\Lambda_c^{\!\ast}}}

\newcommand{\be}{\begin{equation}}
\newcommand{\ee}{\end{equation}}
\newcommand{\ba}{\begin{eqnarray}}
\newcommand{\ea}{\end{eqnarray}}
\begin{document}

%%%%%%%%%%%%%%%%
%%%%% Titlepage:
%%%%%%%%%%%%%%%%
 \begin{titlepage}\begin{flushright}
 SI-HEP-2011-05\\
\end{flushright}
 \vfill
 \begin{center}
 {\Large\bf
 Form Factors and Strong Couplings
of Heavy Baryons from QCD Light-Cone Sum Rules\\[.5cm]}
 {\large\bf
 A.~Khodjamirian, Ch.~Klein, Th.~Mannel, Y.-M.~Wang}\\[0.5cm]
 {\it  Theoretische Elementarteilchenphysik, Department Physik,\\
 Universit\"at Siegen, D-57068 Siegen, Germany }
 \end{center}
 \vfill
 \begin{abstract}
 We derive QCD light-cone sum rules for the hadronic matrix elements
of the heavy baryon transitions to nucleon.  In the correlation
functions  the $\Lambda_c,\Sigma_c$  and $\Lambda_b$ -baryons are
interpolated by three-quark currents and  the nucleon distribution
amplitudes are used. To eliminate the contributions of negative
parity heavy baryons, we combine the sum rules obtained from
different kinematical structures. The results are then less
sensitive to the choice of the interpolating current. We predict
the  $\Lambda_{b}\to p$  form  factor and calculate the widths of
the $\Lambda_{b}\to p\ell\nu_l$ and $\Lambda_{b}\to p \pi$ decays.
Furthermore, we consider double dispersion relations for the same
correlation functions and derive the light-cone sum rules for the
$\Lambda_cND^{(*)}$ and $\Sigma_cND^{(*)}$ strong couplings. Their
predicted values  can be used in the models of charm production in
$p\bar{p}$ collisions.
 \end{abstract}
 \vfill
 \end{titlepage}

\section{Introduction}

Our understanding  of heavy flavour physics
is incomplete without a deeper insight in the processes
with  heavy-flavoured  baryons.
The (electro)weak decays of $\Lambda_{b}$,
$\Lambda_c$  and other $b$- or $c$-baryons,
such as $\Lambda_b\to p \ell\nu_\ell$, $\Lambda_c\to \Lambda
\ell\nu_\ell$, $\Lambda_b \to \Lambda \ell \bar{\ell} $ and
$\Lambda_b \to \Lambda \gamma$,
can provide valuable information on the
underlying quark-flavour operators, in particular, on their spin structure.
Some exclusive decay channels
of heavy baryons are already being investigated at Tevatron  and LHC
(see e.g., \cite{exp}).
For a comprehensive analysis of these processes
the heavy-to-light baryon form factors have to be calculated
in QCD. These form factors  are also used
in the factorization estimates for  nonleptonic
decays e.g.,  for $\Lambda_b\to p \pi$.

Another topical problem, which stems from a
different physical context, concerns  the charmed baryon
and meson strong couplings to the nucleon, for example, the
$\Lambda_cD N $ or $\Lambda_cD^* N $ couplings.
These strong interaction parameters can be used as normalization
inputs in the  models of charm production in the proton-antiproton
collisions, such as the future experiment PANDA \cite{PANDA}.

The lattice QCD studies do not yet access
the heavy-to-light baryon form factors, whereas strong couplings
in general remain a problem for the lattice simulations.
The situation is more advanced for the non-lattice techniques.
Among them, the method of QCD light-cone sum rules \cite{lcsr} (LCSR),
well developed to calculate the heavy-to-light meson form factors \cite{Bpi},
is flexible enough to predict also the baryonic matrix elements.
One possibility to derive LCSR for the baryon form factors is to
consider, in full analogy with the meson case, the vacuum $\to$
nucleon correlation function and express the result of the
operator-product expansion (OPE) near the light-cone in terms of
the nucleon distribution amplitudes (DA's). The latter have been
worked out in \cite{BFMS,BLMS}. This approach was applied to the
nucleon  electromagnetic form factors \cite{LWS,BLW06,NuclDAs09},
where the second nucleon in the correlation function was
interpolated with the three-quark current e.g., with the Ioffe
current \cite{Ioffe}. In order to access heavy baryons, e.g.,
$\Lambda_{c}$ or $\Lambda_{b}$, one has to use a three-quark
current with one heavy $c$ or $b$ quark. Matching the QCD
calculation result for the correlation function to the hadronic
dispersion relation in the variable of the heavy-baryon momentum
squared, one obtains the LCSR for a heavy-to-light  form factor.
Furthermore, a sum rule  for the (heavy baryon)-(heavy meson)-nucleon
strong coupling can be obtained from the double dispersion relation for
the same correlation function. Originally, this method was used to
calculate the $D^*D\pi$ coupling in \cite{BBKR}.

The aim of this paper is to calculate the heavy-to-light baryon
form factors and strong couplings from the LCSR with the nucleon
DA's. In the literature, one can find  several applications of
LCSR \cite{Huang:2004vf,Wang:2008sm,Wang:2009hra,Azizi:2009wn,Aliev:2010yx} or other QCD sum
rule techniques \cite{Dai:1996xv,Huang:1998rq,Navarra:1998vi,Marques de Carvalho:1999ia} to the heavy-baryon
form factors and  strong couplings. There is however an important
problem in the sum rules for baryons which is absent in the case
of mesons. Whatever three-quark current one uses to interpolate a
given baryon, not only the ground state with the positive parity
($J^P=1/2^+$), but also a heavier baryon resonance with the
negative parity ($J^P=1/2^-$) couples to that current. As a
result, e.g., in the hadronic dispersion relation for the
isospin-zero charmed baryons, the ground state $\Lambda_c$ is
accompanied by a $\simeq 300$  MeV heavier negative-parity
resonance $\Lambda_c (2595)$ \cite{PDG}, which we hereafter denote
as $\Lambda_c^*$. In the $b$-baryon spectrum, the $\Lambda_b^*$
resonance with $J^P=1/2^-$ is expected to have a similar mass
difference with respect to $\Lambda_b$. Note that
$\Lambda_{c(b)}^*$ is a $P$-wave state in terms of quark model,
and not a radial excitation of $\Lambda_{c(b)}$. Also the mass
difference of heavy-light baryons with $J^P=1/2^-$ and $J^P=1/2^+$
is smaller, than in the case of light-quark baryons. Consequently,
the influence of negative-parity states on the sum rules for
heavy-light baryon matrix elements is expected to be more
significant than in the case of the nucleon form factors. Hence, a
usual quark-hadron duality ansatz for the hadronic spectral
density in QCD sum rules -- one lowest $\Lambda_{c(b)}$ resonance
plus continuum approximated with OPE -- is not accurate. Moreover,
we expect that this simplified ansatz is one of the main reasons
why  there is a substantial dependence of the sum rule results on
the choice of the interpolating heavy-baryon current in the
correlation function.

Several approaches were suggested in the literature to isolate the
negative-parity baryons  in QCD sum rules. In \cite{Bagan:1993ii},
the heavy quark limit was employed, making use of the fact that
the contributions of positive (negative) parity baryons in the
two-point sum rules are proportional to $[1+\not\!v]$ $([1-\not \!
v])$, where $v$ is the 4-velocity vector of the heavy baryon.
Hence, one can introduce a parity projection matrix for the
correlation function and construct two separate sum rules for the
positive and negative parity baryons. However, such a procedure
only works for infinitely heavy baryons, hence, at finite masses
it cannot guarantee a clean separation of the negative parity
states, especially in the case of charmed baryons. Another
possibility suggested in \cite{Jido:1996ia} is to introduce the
step function $\theta(x_0)$ in the correlation function and
separate the contributions from baryons with different parities.
One advantage of this ansatz is that it also works for  light
baryons.

In what follows, we adopt a new approach, including the
contributions of negative parity baryons explicitly in the
hadronic dispersion relations. The idea is very simple: we
use a linear combination of the sum rules obtained from different
kinematical structures of the same correlation function, so
that the negative-parity baryon terms are cancelled out
and only the ground-state baryon contribution is retained. The
advantage of this procedure is that it does not rely on the heavy
quark limit. Also, as we shall see from the numerical results,
the form factor determination from LCSR becomes largely insensitive
(within the uncertainties of our calculation)
to the choice of the interpolating baryonic current.

The main phenomenological results obtained in this paper include
the form factors of $\Lambda_b\to p$ transition and the strong
couplings  $\Lambda_cND^{(*)}$ and $\Sigma_cND^{(*)}$. The plan of
the paper is as follows. In Sect.~2 we introduce the correlation
functions and discuss the choice of the quark currents. In Sect.~3
we derive the hadronic dispersion relations for these correlation
functions. In Sect.~4 the LCSR for the heavy-baryon $\to$ nucleon
form factors are obtained, calculating the correlation functions
in terms of the nucleon DA's and matching them to the dispersion
relations. Since the form factors enter LCSR together with the
decay constants of heavy baryons, in Sect.~5 we describe the
two-point QCD sum rules used for these constants. In Sect.~6, the
LCSR for the strong couplings are derived. The details of the
numerical analysis of the form factors and strong couplings are
collected in Sect.~7. Sect.~8 contains our predictions for the
exclusive semileptonic $\Lambda_b \to p \ell \nu_\ell$ and
nonleptonic $\Lambda_b\to p\pi$ decays based on the form factors
obtained from LCSR. Sect.~9 is reserved for the concluding
discussion. The paper contains several appendices where the bulky
expressions of the nucleon DA's (App.~A), LCSR for the form
factors (Apps.~B, C), the two-point sum rules for the decay
constants (App.~D) and the double spectral densities used in LCSR
for the strong couplings (App.~E) are collected.

\section{Correlation function and interpolating
currents}
As a first  step to derive  the LCSR,
we introduce  the following vacuum-to-nucleon correlation function:
\begin{equation}
\Pi_a(P,q)=i\int d^4z\ e^{iq\cdot
z}\bra{0}T\left\{\eta(0),j_a(z)\right\}\ket{N(P)} \,.
\label{eq:corr}
\end{equation}
In the above, the
current $\eta$  interpolating  a heavy-light baryon and
the current $ j_a$ of the heavy-light transition
($a$ indicates a certain Lorentz structure) enter the $T$-product,
sandwiched between the nucleon on-shell state $|N\rangle$ with the
four-momentum $P$ ($P^2=m_N^2$) and the vacuum. The heavy-quark mass
$m_Q$ is finite, and the calculation is
applicable to both charmed ($m_Q=m_c$) and beauty ($m_Q=m_b$)
baryons. Moreover, a
generalization to the case of strange baryons is possible in the
same framework ($m_Q\to m_s$), provided the external momentum
transfer $q$ is deep spacelike. For the heavy quarks this condition is
fulfilled if $q^2 \ll m^2_{Q}$. Note that the $N$-state is taken
as an initial one in (\ref{eq:corr}) for simplicity, in order to
directly use the definitions of the nucleon DA's from  \cite{BFMS}.
In the  phenomenological applications of our interest $N$ is a proton.

For definiteness, in what follows we consider the correlation function
(\ref{eq:corr}) with the $c$-quark, selecting the flavour configuration
$udc$ for the baryon interpolating current and,
correspondingly, $\bar{c} u$ for the transition  current.
With this choice, we first derive LCSR for the $\Lambda_c\to p$
and $\Sigma_c\to p$ form factors. Switching
from $c$ to $b$ quark in the $\Lambda_c\to p$  sum rules
(and accordingly adjusting the relevant scales in the correlation function),
we obtain the LCSR  for $\Lambda_b\to p$ form factors, our first
phenomenological goal. Furthermore, the  flavour configuration chosen in (\ref{eq:corr}) leads, via double dispersion relations, to LCSR for the
strong couplings of charmed baryons $\Lambda_c,\Sigma_c$ with the nucleon
and $D^{(*)}$-mesons.

In what follows, we consider the heavy-light transition currents
with pseudoscalar, vector  and axial-vector quantum numbers:
\be j_a = \bar{c}\,\Gamma_a u\,,~ ~\mbox{with} ~
\Gamma_a=m_c i\gamma_5,\gamma_\mu, \gamma_\mu\gamma_5\,,
\label{eq:jb}
\ee
respectively. For the sake of renormalization invariance, the
quark mass is inserted in the pseudoscalar current.

For the heavy-light baryon interpolating current we have
the following general structure:
\be
\eta = \epsilon^{ijk} \left(u_i\,C\,\Gamma_b\,d_j\right)
\widetilde{\Gamma}_b\,c_k\,,
\label{eq:eta}
\ee
%%%%%%%
where the first fermion  field $u_i$
should be hereafter understood as $u_i^{{\rm T}}$, $C$ is the charge
conjugation matrix, and the sum goes over the colour indices
$i,j,k$. There are multiple choices for the Dirac structures
$\Gamma_b$ and $\widetilde{\Gamma}_b$ in the above current. The
discussion of the optimal choice of the baryon interpolating
current goes back to the early papers
\cite{Ioffe,Shuryak:1981fza,Chung:1981cc}. In the
$\Lambda_c$-baryon, the isospin of the light diquark $[ud]$ is
zero, excluding the structures $\gamma_{\mu}$ and $\sigma_{\mu
\nu}$ for $\Gamma_b$ in the $\eta$ current. Still there is a
freedom to choose in (\ref{eq:eta}) the following combinations:
$\Gamma_b=\gamma_5$, $\widetilde{\Gamma}_b=\unitop$, leading to
\be \eta= \eta^{({\cal P})}_{\Lambda_c} =
\left(u\,C\,\gamma_5\,d\right)c, ~~ \label{eq:etaP} \ee or
$\Gamma_b=\gamma_5\gamma_\lambda, ~~
\widetilde{\Gamma}_b=\gamma^\lambda$, for which \be
\eta=\eta^{({\cal A})}_{\Lambda_c} =
\left(u\,C\,\gamma_5\gamma_\lambda\,d\right)\gamma^\lambda\,c \,.
\\
\label{eq:etaA}
\ee
In addition, a simpler current
\begin{eqnarray}
\eta^{({\cal S})}_{\Lambda_c} =
 \left(u\,C\, d\right) \gamma_5\, c\,,
\label{eq:eta3}
\end{eqnarray}
is also possible, as well as any linear combination of all  three
above currents.

The heavy-quark limit \cite{Shuryak:1981fza} provides another
guiding principle for choosing an optimal heavy baryon current, at
least at the qualitative level.  In particular, since the light
diquark system in the current (\ref{eq:eta3}) is in the $P$-wave,
this current is not expected \cite{Bagan:1993ii} to have a
considerable overlap with the ground-state $\Lambda_c$. Hence, in
what follows  we will only leave (\ref{eq:etaP}) and
(\ref{eq:etaA}) under consideration, which we denote as the {\em
pseudoscalar}  and {\em axial-vector\,} currents, respectively.
The correlation functions (\ref{eq:corr}) with the pseudoscalar
(axial-vector) interpolating  current and with the transition
currents listed in (\ref{eq:jb}) are denoted as
$\Pi^{(\cal{P})}_{5}$, $\Pi^{(\cal{P})}_{\mu}$,
$\Pi^{(\cal{P})}_{\mu 5}$ ($\Pi^{(\cal{A})}_{5}$,
$\Pi^{(\cal{A})}_{\mu}$, $\Pi^{(\cal{A})}_{\mu 5}$), respectively.

Turning to the $\Sigma_c$ baryon where
the light diquark
$[ud]$ has isospin one, we again adopt two  different currents:
the {\em Ioffe} current \cite{Ioffe} with $\Gamma_b=\gamma_\lambda$,
and $\widetilde{\Gamma}_b=\gamma^\lambda \gamma_5$ :
\be \eta^{({\cal I})}_{\Sigma_c}=
\left(u\,C\,\gamma_\lambda\,d\right) \gamma^\lambda \gamma_5
\,c\,,
\label{eq:etasigma1}
\ee
%%%%%%%%%%%%%%%%%%%%%%%%%%%%%%
and the {\em tensor} current with $\Gamma_b=\sigma_{\mu \nu} $ and
$\widetilde{\Gamma}_b= \sigma^{\mu \nu} \gamma_5$ : \be
\eta^{({\cal T})}_{\Sigma_c} =
 \left(u\,\sigma_{\mu \nu} \,d\right) \sigma^{\mu \nu} \gamma_5 c
 \,.
 \label{eq:etasigma2} \ee

The Ioffe current is used in
the LCSR for the nucleon form factors \cite{BLW06}. One
advantage of this current is that the power corrections terms are
small. On the other hand, the tensor current provides a reduced continuum
contributions, at least in the sum rules with light baryons
\cite{Chung:1981cc}.

\section{Accessing the form factors  with hadronic dispersion
relations }
We begin with the hadronic transitions involving $\Lambda_c$.
Following the usual procedure of the QCD sum rule derivation, we insert
in the correlation function (\ref{eq:corr})
a total set of charmed-baryon states between the interpolating current
$\eta=\eta_{\Lambda_c}^{(i)}$ (where $i={\cal P}$ or ${\cal A})$
and the transition current $j_a$. In the resulting hadronic dispersion relation
the contributions of the lowest state $\Lambda_c$
and its negative-parity partner $\Lambda_c^*$ enter.
The residue of the  $\Lambda_c$-pole  contains  the
product of two hadronic matrix
elements. The first one is the coupling of $\Lambda_c$ with the
interpolating current $\eta^{(i)}_{\Lambda_c}$ (the decay constant),  defined as
\be
 \bra{0}\eta^{(i)}_{\Lambda_c}\ket{\Lc(P-q)} = m_{\Lc}\lambda_{\Lc}^{(i)}\,\,
u_{\Lc}(P-q)\,,
\label{eq:lambdai}
\ee
where $u_{\Lambda_c}(P-q)$ is the $\Lambda_c$-bispinor
with the four-momentum $(P-q)$.
The second hadronic matrix element
$\bra{\Lambda_c(P-q)} j_a\ket{N(P)}$ is defined in terms of $N\to \Lambda_c$
form factors which are equal to the $\Lambda_c\to N$  form factors
of our interest, up to an inessential general phase.

For the pseudoscalar transition current there is only one form factor, which
we define as:
\be
 \bra{\Lambda_c(P-q)}m_c\bar{c} \,i\gamma_5\ u \ket{N(P)} =
(m_{\Lambda_c}+m_N) G(q^2) \bar{u}_{\Lambda_c} (P-q)
 i\gamma_5\,u_N(P)\,,
\label{eq:ff1}
\ee
where the mass-dependent factor on r.h.s. is introduced to keep the form factor
$G(q^2)$ dimensionless. The hadronic matrix elements with  the vector and axial-vector
transition currents  contain three form factors each:
\be
\bra{\Lambda_c(P-q)}\bar{c} \,\gamma_\mu\, u\ket{N(P)}
=
\bar{u}_{\Lambda_c} (P-q) \bigg\{f_1(q^2)\,\gamma_\mu
+ i\frac{f_2(q^2)}{m_{\Lc}}\,\sigma_{\mu\nu}q^\nu
+\frac{f_3(q^2)}{m_{\Lc}}\,q_\mu\bigg\}u_N(P) \,,
\\
\label{eq:ff2}\\
\ee
\be
\bra{\Lambda_c(P-q)}\bar{c} \,\gamma_\mu \gamma_5 \, u \ket{N(P)}
 =\bar{u}_{\Lambda_c} (P-q) \bigg\{g_1(q^2)\,\gamma_\mu + i\frac{g_2(q^2)}{m_{\Lc}}
 \,\sigma_{\mu\nu}q^\nu
+\frac{g_3(q^2)}{m_{\Lc}}\,q_\mu\bigg\} \gamma_5 u_N(P) \,.
\\
\label{eq:ff3}
\ee

Taking the divergence of the axial-vector current
one obtains the following relation:
\begin{eqnarray}
G(q^2)  =  g_1(q^2) - {q^2 \over m_{\Lambda_c} (m_{\Lambda_c} +m_{N}) }
g_3(q^2)\,. \label{pseudoscalar FF relation}
\end{eqnarray}
 The $\Lcst$-pole term in the dispersion relation contains the decay constant
\be
 \bra{0}\eta^{(i)}_{\Lambda_c}\ket{\Lcst(P-q)} =
m_{\Lcst}\lambda_{\Lcst}^{(i)}\,\gamma_5 u_{\Lcst}(P-q)\,,
\label{eq:lambdaist}
\ee
which is multiplied with the form factor of the
$N\to \Lcst $ transition. For the
pseudoscalar transition current we define this form factor as
\be
 \bra{\Lcst(P-q)}m_c\bar{c}\, i\gamma_5\ u \ket{N(P)}
= (m_{\Lcst}-m_N) \widetilde{G}(q^2)i\bar{u}_{\Lcst} (P-q)
u_N(P)\,.
\label{eq:ff1star}
\ee
For the vector and axial-vector
currents the definitions of the corresponding form factors
$\tilde{f}_{1,2,3}(q^2)$ and $\tilde{g}_{1,2,3}(q^2)$  are obtained
from (\ref{eq:ff2}) and (\ref{eq:ff3}),
respectively, adding $\gamma_5$ after
the $\Lambda_c$ bispinor and replacing $\Lc\to \Lcst$.

Taking into account
the equation of motion $(\slashed{P}-m_N)u_N(P)=0$, we decompose
the correlation function (\ref{eq:corr}) in independent invariant amplitudes.
In the case of the pseudoscalar transition current $j_5$
there are two amplitudes:
\be
\Pi^{(i)}_5(P,q) = \left[\Pi_1^{(i)}((P-q)^2,q^2)+
\slashed{q}\,\Pi_2^{(i)} ((P-q)^2,q^2)
\,\right]i\gamma_5 u_N(P)\,,
\label{eq:corrps}
\ee
for both $i={\cal P}$ or ${\cal A}$. In the case of the vector current $j_\mu$ there are six invariant amplitudes:
\ba
\Pi^{(i)}_\mu(P,q) =
\left(\widetilde{\Pi}_1^{(i)}\,P_\mu\,+\widetilde{\Pi}_2^{(i)}\,P_\mu\slashed{q}\,+
\widetilde{\Pi}^{(i)}_3\,\gamma_\mu
\,+\widetilde{\Pi}_4^{(i)}\,\gamma_\mu\slashed{q}\,+
\widetilde{\Pi}_5^{(i)}\,q_\mu\,+\widetilde{\Pi}_6^{(i)}\,q_\mu\slashed{q}\,\right)u_N(P)\,,
 \label{eq:corrvect}
\ea
where the dependence of $\widetilde{\Pi}^{(i)}_{1-6}$ on $(P-q)^2$ and $q^2$ is not
shown for brevity; a similar decomposition for the correlation
function $\Pi^{(i)}_{\mu5} $ with the axial-vector current $j_{\mu 5}$ reads:
 \ba
\Pi^{(i)}_{\mu 5}(P,q) =
\left(\bar{\Pi}_1^{(i)}\,P_\mu\,+\bar{\Pi}_2^{(i)}\,P_\mu\slashed{q}\,+
\bar{\Pi}^{(i)}_3\,\gamma_\mu
\,+\bar{\Pi}_4^{(i)}\,\gamma_\mu\slashed{q}\,+
\bar{\Pi}_5^{(i)}\,q_\mu\,+\bar{\Pi}_6^{(i)}\,q_\mu\slashed{q}\,\right)
\gamma_5 u_N(P)\,.
 \label{eq:corraxi}
\ea

Employing the above definitions of decay constants and form
factors, and summing over the helicities of the $\Lc$ and
$\Lcst$, we obtain the hadronic dispersion relations for
each invariant amplitude in (\ref{eq:corrps}), (\ref{eq:corrvect}) and
(\ref{eq:corraxi}). In  the case of the pseudoscalar current one
has:
\begin{eqnarray}
\Pi^{(i)}_1((P-q)^2,q^2)=  \frac{ m_{\Lc}(m^2_\Lc-m^2_N)\lambda_{\Lc}^{(i)}
G(q^2)}{m_{\Lc}^2-(P-q)^2}\nonumber \\
+\frac{m_{\Lcst}(m_{\Lcst}^2-m_N^2)\lambda_{\Lcst}^{(i)}
\tilde{G}(q^2)}{m_{\Lcst}^2-(P-q)^2}\
+ \int\limits_{s_0^h}^{\infty} ds \, { \rho^{(i)}_1(s, q^2) \over
s -(P-q)^2 }\, , \label{eq:dispP1}
\end{eqnarray}
and
\begin{eqnarray}
\Pi^{(i)}_2((P-q)^2,q^2)= -\frac{ m_{\Lc}(m_\Lc+m_N)\lambda_{\Lc}^{(i)}
G(q^2)}{m_{\Lc}^2-(P-q)^2}\nonumber \\
+\frac{ m_{\Lcst}(m_{\Lcst}-m_N)\lambda_{\Lcst}^{(i)}
\tilde{G}(q^2)}{m_{\Lcst}^2-(P-q)^2}\,
+ \!\!\!\int\limits_{s_0^h}^{\infty} ds \, { \rho^{(i)}_2(s, q^2)
\over s -(P-q)^2 }\, , \label{eq:dispP2}
\end{eqnarray}
where the hadronic spectral densities of all excited and
continuum  states with the quantum numbers of $\Lc$ and
$\Lcst$ are denoted as $\rho_{1,2}^{(i)}$, and $s_0^h$ is
the corresponding threshold. Possible subtractions are neglected,
having in mind the subsequent Borel transformation.

In the case of the vector transition current the dispersion relations
for the six independent invariant amplitudes
have the same structure as (\ref{eq:dispP1}) and (\ref{eq:dispP2}).
Instead of writing
them down one by one,  we present one combined expression
for the correlation function, written in terms of
the hadronic contributions:
\begin{eqnarray}
\Pi^{(i)}_\mu(P,q) = \frac{\lambda_{\Lc}^{(i)}
m_{\Lc}}{m_{\Lc}^2-(P-q)^2} \bigg [ 2 f_1(q^2)\,P_\mu
-2\frac{f_2(q^2)}{m_{\Lc}} \,P_\mu \not \! q
\nonumber\\
+(m_{\Lc}-m_N)
\bigg ( f_1(q^2)- \frac{  m_{\Lc}+m_N }{ m_{\Lc} } f_2(q^2) \bigg )\gamma_{\mu}
%\nonumber\\
 + \bigg ( f_1(q^2)- \frac{m_{\Lc}+m_N }{ m_{\Lc} } f_2(q^2) \bigg )  \gamma_{\mu}  \not \! q
\nonumber\\
+  \bigg ( -2 f_1(q^2) + \frac{m_{\Lc}+m_N}{m_{\Lc} } (f_2(q^2)+f_3(q^2)) \bigg )q_{\mu}
%\nonumber \\
  +\frac{1}{m_{\Lc} } (f_2(q^2)-f_3(q^2))q_{\mu} \not \! q
\bigg ]  u_N(P)  \nonumber
\end{eqnarray}
%%%%%%%%%%%%%%
\begin{eqnarray}
+ \frac{\lambda_{\Lcst}^{(i)} m_{\Lcst}}{m_{\Lcst}^2-(P-q)^2}
\bigg [- 2\tilde{f}_1(q^2)\,P_\mu
+2\frac{\tilde{f}_2(q^2)}{m_{\Lcst}} \,P_\mu \not \! q
\nonumber \\
+ (m_{\Lcst}+m_N) \bigg ( \tilde{f}_1(q^2)+
\frac{m_{\Lcst}-m_N}{m_{\Lcst} } \tilde{f}_2(q^2)
\bigg )\gamma_{\mu}
%\nonumber \\
- \bigg ( \tilde{f}_1(q^2)+\frac{m_{\Lcst}-m_N }{ m_{\Lc}}
\tilde{f}_2(q^2) \bigg )  \gamma_{\mu}  \not \! q
\nonumber\\
+  \bigg ( 2 \tilde{f}_1(q^2) + \frac{  m_{\Lcst}-m_N}{
m_{\Lcst} } (\tilde{f}_2(q^2)+\tilde{f}_3(q^2)) \bigg
)q_{\mu}
%\nonumber \\
  -\frac{1}{ m_{\Lcst} }
(\tilde{f}_2(q^2)-\tilde{f}_3(q^2))q_{\mu} \not \! q
\bigg ]  u_N(P)    \nonumber \\
 + \int\limits_{s_0^h}^{\infty}\frac{ds}{s-(P-q)^2}
\bigg(\bar{\rho}_{1}^{(i)}(s, q^2)\,P_\mu\,
+\tilde{\rho}_2^{(i)}\,P_\mu\slashed{q}\nonumber \\
+\tilde{\rho}_3^{(i)}\,\gamma_\mu\,
+\tilde{\rho}_4^{(i)}\,\gamma_\mu\slashed{q}\,
+\tilde{\rho}_5^{(i)}\,q_\mu\,
+\tilde{\rho}_6^{(i)}\,q_\mu\slashed{q}\,\bigg)u_N(P)\,.
\label{eq:hadrvec}
\end{eqnarray}
Collecting the coefficients at each bispinor structure in the
above, we equate their sum to the amplitude $
\widetilde{\Pi}^{(i)}_{1,...,6}$ which multiplies the same
structure in the decomposition (\ref{eq:corrvect}). The analogous
hadronic decomposition for the correlation function
$\Pi^{(i)}_{\mu 5}(P,q)$ with the axial-vector current can be
obtained from (\ref{eq:hadrvec}) by replacing $f_i \to g_i$,
$\tilde{f}_i \to \tilde{g}_i$, changing the sign of $m_N$ and
adding $\gamma_5$ before the nucleon spinor.

The hadronic dispersion relations for the
correlation function with  the $\Sigma_c$ interpolating currents
are obtained from the relations for the $\Lc$ presented above
by simple replacements $\Lambda_c \to \Sigma_c $ and $\Lambda_c^* \to
\Sigma_c^* $. We identify the $\Sigma_c^*(1/2^-)$ state with
the resonance $\Sigma_c(2800)$ \cite{PDG} whose mass
is close to the expected one:
\be m_{\Sigma_c^*}\simeq
m_{\Sigma_c}+(m_{\Lambda_c^*}-m_{\Lambda_c})\simeq 2764 \,\, {\rm
MeV}. \label{eq:Sstarmass}
\ee
The dispersion relations obtained above will be used in the following
section to derive the LCSR.

\section{Light-Cone Sum Rules for the form factors}

\begin{figure}[t]
\centering
\includegraphics[scale=0.5]{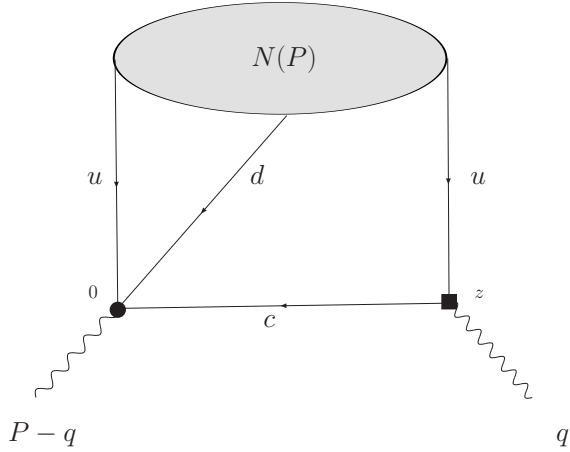}
\vspace{-0.3cm} \caption{ \em Diagrammatic representation of
the correlation function (\ref{eq:corr}). The wavy lines (oval) represent
the external currents (the nucleon DA).}
\label{fig:diag}
\end{figure}

We now turn to the computation of  the correlation function
(\ref{eq:corr}) for the $\Lambda_c \to p$ transition, employing two
different interpolating currents for $\Lambda_c$ and, in each
case, the three transition currents listed in (\ref{eq:jb}).
Throughout this calculation we neglect the light-quark masses
everywhere; the only two mass parameters in the correlation
function are the $c$-quark mass $m_c$ and the nucleon mass $m_N$,
the latter entering the nucleon DA's. The external 4-momenta $P-q$
and $q$ are taken spacelike, $(P-q)^2,~q^2\ll m_c^2$,  to justify
the expansion of the product of the two currents  in
(\ref{eq:corr}) near the light-cone ($z^2\sim 0$). The OPE result
is obtained as a sum over nucleon DA's of growing twist,
convoluted with the hard-scattering amplitudes  formed by the
virtual $c$-quark propagator, as shown in the diagram of
Fig.~\ref{fig:diag}. We include all three-particle nucleon DA's
from twist 3 to twist 6. The contributions of soft gluons emitted
from the $c$-quark and absorbed by the nucleon, demand the
knowledge of the four-particle (three-quark-gluon) nucleon DA's.
Their analysis has just started \cite{Braun:2008ia,Braun:2011aw}.
In fact, the soft-gluon  contributions to OPE are expected to be
suppressed by extra powers of the virtual $c$-quark propagator.
Another future improvement of LCSR is possible, if one calculates
the $O(\alpha_s)$ corrections to the correlation function
corresponding to the hard gluon exchanges between the quark lines
in the diagram of Fig.~\ref{fig:diag}.
%%FIG 1%%%%%%%%%%%%%%%%%%%%%%
%%%%%%%%%%%%%%%%%%%%%%%%%

The nucleon DA's at $z^2\to 0$ are defined according to \cite{BFMS}:
\begin{eqnarray}
 \bra{0} \epsilon^{ijk} u_\alpha^i(0) u_\beta^j(z
) d_\gamma^k(0) \ket{N(P)} = \mathcal{V}_1 \left(\!\not\!{P}C
\right)_{\alpha \beta} \left(\gamma_5 u_N \right)_\gamma  +
\mathcal{A}_1 \left(\!\not\!{P} \gamma_5 C \right)_{\alpha \beta}
\left(u_N \right)_\gamma   &&  \nonumber \\
+ \mathcal{T}_1\left(P^\nu i \sigma_{\mu\nu} C\right)_{\alpha
\beta} \left(\gamma^\mu\gamma_5 u_N \right)_\gamma
+ ...
%\{\mbox{twist $\geq$ 4 terms}\} \oplus O(z^2)\,,
&&
\label{nucleon:DA}
\end{eqnarray}
where $\alpha,\beta,\gamma$ are Dirac indices. The terms shown above
receive their contributions from the lowest twist-3 DA's.
The complete, rather bulky decomposition is presented in App.~A.
In (\ref{nucleon:DA}), the calligraphic notations
$\mathcal{F}=\{\mathcal{V}_1, \mathcal{A}_1,\mathcal{T}_1\}$
denote the integrals over the  twist-3 nucleon DA's:
\begin{eqnarray}
\mathcal{F}=\int d x_1 d x_2 d x_3
 \delta(1-x_1-x_2-x_3)e^{-i x_2 P \cdot z }
F(x_i,\mu)\,,
\label{eq:F}
\end{eqnarray}
 denoted by the same
noncalligraphic  letters  $F=\{V_1,A_1,T_1\}$,
 where ~$x_i=\{x_1,x_2,x_3\}$, ($0 \leq x_i \leq 1$) are  the longitudinal momentum fractions
of the quarks in the nucleon and $\mu$ is the normalization scale.
The twist-3  DA's:
\begin{eqnarray}
V_1(x_i, \mu) &=& 120 x_1 x_2 x_3 [  \phi_3^0(\mu)  +
\phi_3^+(\mu) (1- 3 x_3) ] \,, \nonumber \\
A_1(x_i, \mu) &=& 120 x_1 x_2 x_3 (x_2-x_1)\phi_3^-(\mu) \,, \nonumber \\
T_1(x_i, \mu) &=& 120 x_1 x_2 x_3 [  \phi_3^0(\mu)  - {1 \over
2}(\phi_3^+(\mu)-\phi_3^-(\mu)) (1- 3 x_3) ] \,. \label{eq:tw3DA}
\end{eqnarray}
are derived in \cite{BFMS} where one can find the details. The
expressions for the nucleon DA's of twist 4, 5, 6  as well as the
relations for their normalization and shape parameters, such as
$\phi_3^{0,\pm}$ in (\ref{eq:tw3DA}),  are presented in App.~A.

Substituting the decomposition  (\ref{nucleon:DA}) in the
correlation functions $\Pi_{5,\,\mu,\,\mu5}^{(i)}$, we isolate
the invariant amplitudes. The integration over
the variable $x_1$ is performed easily, while the virtual $c$-quark momentum in the
chosen configuration  is equal to $(x_2P-q)$ and contains no
$x_1$-dependence. The result is represented as a sum of integrals
over the remaining variable $x_2\equiv x$.

For the correlation function with the
pseudoscalar transition current, the invariant amplitudes defined
in (\ref{eq:corrps}) can be transformed to the following form:
\begin{eqnarray}
\Pi^{(i)}_{j}((P-q)^2,q^2) = \frac{m_c}{4}\sum_{n=1,2,3} \int\limits_0^1 dx
\frac{\omega^{(i)}_{jn} (x,(P-q)^2,q^2) }{ D^n}\,
\label{eq:QCDcorr}
\end{eqnarray}
with the denominator
\begin{eqnarray}
 D= m_c^2-(xP-q)^2= m_c^2-x(P-q)^2-\bar{x}
q^2+x\bar{x}\,m_N^2\,,
\label{eq:den}
\end{eqnarray}
and $\bar{x}=1-x$. The functions $\omega^{(i)}_{jn}$ are
distinguished by their indices: $i={\cal P}, {\cal A}$ (baryon
current), $j=1,2$ (the number of the invariant amplitude) and
$n=1,2,3$ (the power of the denominator). They depend linearly on
$(P-q)^2$, $q^2$ and polynomially or logarithmically on $x$. Note
that in (\ref{eq:QCDcorr}) we do not show the terms which vanish
after Borel transformation in $(P-q)^2$. Since the invariant
amplitudes in the form (\ref{eq:QCDcorr}) will now be used in  the
LCSR for the form factors, we also replace $(P-q)^2$  in  the
numerators $\omega^{(i)}_{jn} (x,(P-q)^2,q^2)$  by $ s(x)-D/x$,
where \be s(x)= (m_c^2-\bar{x}q^2+x\bar{x}m_N^2)/x\,.
\label{eq:sx} \ee The transformed functions
$\omega^{(i)}_{jn}(x,q^2) $ are presented in App.~B. For the
correlation functions with the vector and axial-vector transition
currents, expressions  similar to (\ref{eq:QCDcorr}) (without  the
factor $m_c$)  are obtained for the invariant amplitudes
$\widetilde{\Pi}^{(i)}_{j}$ and $\bar{\Pi}^{(i)}_{j}$
$(j=1,...,6)$. The corresponding numerator functions
$\widetilde{\omega}^{(i)}_{jn}$ and $\bar{\omega}^{(i)}_{jn}$,
respectively, are also given in App.~B. Furthermore, App.~C
contains the numerator functions $\omega^{(i)}_{jn}$,
$\widetilde{\omega}^{(i)}_{jn}$ and $\bar{\omega}^{(i)}_{jn}$
($i={\cal I}, {\cal T}$) for the correlation functions with the
$\Sigma_c$ interpolating currents.

After computing the OPE expressions for all
invariant amplitudes in the integral
 form (\ref{eq:QCDcorr}), we use
the hadronic dispersion relations for these amplitudes presented in the
previous section. At this point, we notice that each form factor  enters more than
one dispersion relation. E.g.,
in the case of pseudoscalar current, there are two linearly
independent relations (\ref{eq:dispP1}) and (\ref{eq:dispP2}) for
$\Pi^{(i)}_{1}$ and $\Pi^{(i)}_{2}$, respectively, both containing
the $\Lc$- and $\Lambda_c^{\ast}$-pole terms. Combining them,
we eliminate the $\Lcst$ contributions, obtaining a linear combination of dispersion relations containing only the hadronic matrix elements
for  the ground-state $\Lc$-baryon:
 \ba
 \frac{ m_{\Lc} (m_{\Lc}+m_N)  (m_{\Lc}+
m_{\Lambda_c^{\ast}}) \lambda_{\Lc}^{(i)} G(q^2)}{m_{\Lc}^2-(P-q)^2}+
\int\limits_{s_0^h}^{\infty} ds \, \frac{ \rho_1^{(i)}(s,
q^2)-(m_{\Lambda_c^{\ast}}+m_N)\rho_2^{(i)}(s, q^2)}{s -(P-q)^2 }
&& \nonumber \\
 = \bigg [ \Pi_1^{(i)}((P-q)^2, q^2)- (m_{\Lambda_c^{\ast}}+m_N
)\Pi_2^{(i)}((P-q)^2, q^2) \, \bigg ] \,. && \label{eq:sr12} \ea
The contributions of the hadronic states above the threshold
$s_0^h$  are approximated using quark-hadron duality:
\begin{eqnarray}
\int\limits_{s_0^h}^{\infty} {ds \over s-(P-q)^2} \bigg
[\rho_1^{(i)}(s,q^2)-(m_{\Lambda_c^{\ast}}+m_N)\rho_2^{(i)}(s,
q^2) \bigg]
\nonumber\\
= {1 \over \pi}\int\limits_{s_0}^{\infty} {ds \over s-(P-q)^2}
\bigg [{\rm Im}_{s} \Pi_1^{(i)}(s, q^2)
-(m_{\Lambda_c^{\ast}}+m_N){\rm Im}_{s} \Pi_2^{(i)}(s, q^2) \bigg
]\,, \label{eq:dual}
\end{eqnarray}
where $s_0$ is the effective threshold parameter.
The spectral densities ${\rm Im}_{s}\Pi^{(i)}_{1,2}$
are calculated from the  OPE result (\ref{eq:QCDcorr}).
To this end, the integrals in (\ref{eq:QCDcorr}) are transformed to the dispersion  form in $(P-q)^2$,
transforming the integration variable $x$ to $s(x)$ defined
in (\ref{eq:sx}), so that
\begin{equation}
x(s) =\frac{1}{2m_N^2}\bigg[ m_N^2+q^2-s + \sqrt{(s-q^2-m_N^2)^2+ 4 m_N^2 (m_c^2-q^2)}\bigg ]\,.
\end{equation}

The final step in obtaining LCSR is the
Borel transformation $(P-q)^2\to M^2$, introducing
the Borel parameter $M^2$ in the charmed baryon channel. The resulting
sum rule for the form factor reads:
\ba
 \,G(q^2)={ e^{m_{\Lc}^2/M^2}\over m_{\Lc} (m_{\Lc}+m_N)  (m_{\Lc}+
m_{\Lcst})\lambda_{\Lc}^{(i)}}\,\frac1{\pi}\int\limits _{m_c^2}^{s_0} ds e^{-s/M^2}\bigg [ {\rm
Im}_{s}\Pi_1^{(i)}(s, q^2)
\nonumber \\
-(m_{\Lambda_c^{\ast}}+m_N) {\rm
Im}_{s}\Pi_2^{(i)}(s, q^2) \bigg ]\,. \label{eq:srLc} \ea

In the case of the vector transition current, the same
procedure of eliminating the $\Lcst$-contributions yields the
following LCSR for the two most important form factors:
\begin{eqnarray}
 f_1(q^2)  =\frac{e^{m_{\Lc}^2/M^2}}{2 m_{\Lc} (m_{\Lc}+m_{\Lambda_c^{\ast}})\lambda_{\Lc}^{(i)}}\frac1{\pi}\int\limits_{m_c^2}^{s_0}
ds~ e^{-s/M^2}
    \bigg [
(m_{\Lc}+ m_N )\Big({\rm Im}_{s}\tilde{\Pi}^{(i)}_1(s, q^2)
\nonumber \\
- ( m_{\Lambda_c^{\ast}} -m_N) {\rm Im}_{s}\tilde{\Pi}^{(i)}_2(s,
q^2)\Big) + 2 {\rm Im}_{s}\tilde{\Pi}^{(i)}_3(s, q^2)+ 2
(m_{\Lambda_c^{\ast}}-m_{\Lc}) {\rm Im}_{s}\tilde{\Pi}^{(i)}_4(s,
q^2)\bigg] \,, \label{eq:srf1}
\end{eqnarray}
%%%%%%%%%%%%%%%%%%%%%
\begin{eqnarray}
f_2(q^2)=
\frac{e^{m_{\Lc}^2/M^2}}{2(m_{\Lc}+m_{\Lambda_c^{\ast}})\lambda_{\Lc}^{(i)} }\frac1{\pi}\int\limits_{m_c^2}^{s_0} ds~ e^{-s/M^2}\, \Bigg [ {\rm Im}_{s}\tilde{\Pi}^{(i)}_1(s, q^2) \nonumber\\
-(m_{\Lambda_c^{\ast}}- m_N )  {\rm Im}_{s}\tilde{\Pi}^{(i)}_2(s,
q^2)  - 2
 {\rm Im}_{s}\tilde{\Pi}^{(i)}_4(s, q^2)  \Bigg]\, .
\label{eq:srf2}
\end{eqnarray}
The LCSR for the axial-vector form factors $g_1(q^2)$ and
$g_2(q^2)$ can be obtained from the above sum rules  for $f_1(q^2)$ and
$f_2(q^2)$, respectively, by replacing
$ {\rm Im}_{s}\tilde{\Pi}^{(i)}_j \to  {\rm Im}_{s}\bar{\Pi}^{(i)}_j$ and changing the sign of $m_N$.
%In what follows we assume
%the same threshold parameter $s_0$
%for all sum rules in the  $\Lambda_c$ channel.
Furthermore, the LCSR obtained above are easily transformed to the
case of $\Lambda_b\to p $ transition by replacing the $c$-quark by $b$-quark in the correlation function.
Finally, to obtain LCSR  for  the $\Sigma_c\to p$ form factors
we repeat the whole procedure for the correlation functions
with the $\Sigma_c$-interpolation currents.

In practice, all three procedures: transformation
to the dispersion integral form, subtraction of continuum
and Borel transformation are unified in the following substitution rules for the integrals in (\ref{eq:QCDcorr}),
similar to the ones used in \cite{BLW06}:
\begin{eqnarray}
\int dx \frac{\omega(x) }{D} &\to&
\int\limits_{x_0}^1 \frac{dx}{x}\omega(x)
\exp\left(-\frac{s(x)}{M^2}\right)\,,
\nonumber
\\
\int dx \frac{\omega(x) }{D^2} &\to&
\frac{1}{M^2} \int\limits_{x_0}^1 \frac{dx}{x^2} \omega(x)
\exp\left(-\frac{s(x)}{M^2}\right)
+  \frac{\omega(x_0)\,e^{-s_0 /M^2} }{m_c^2+x_0^2 m_N^2-q^2} \, , \nonumber \\
%%%%%%%%%%
\int dx \frac{\omega(x) }{D^3} &\to& \frac{1}{2
M^4} \int\limits_{x_0}^1 \frac{dx}{x^3}\omega(x)\exp\left(-\frac{s(x)}{M^2}\right)
%\nonumber \\  &&
+
 \frac{1}{2 M^2}  \frac{\omega(x_0)\,e^{-s_0 /M^2} }{x_0\left(m_c^2+x_0^2 m_N^2-q^2\right)}
 \nonumber \\  &&-   \frac12  \frac{x_0^2\, e^{-s_0 /M^2}}{(m_c^2+x_0^2 m_N^2-q^2)}
\frac{d}{dx}\left(
\frac{\omega(x)}{x\left(m_c^2+x^2 m_N^2-q^2\right)}
\right)\Bigg|_{x=x_0}\,.
\label{eq:subst}
\end{eqnarray}
with $x_0=x(s_0)$ and
for any numerator function $\omega(x)$ in (\ref{eq:QCDcorr}).
The surface terms appearing on r.h.s. of the above relations
originate from the transformations of the
integrals with the power of denominator $n=2,3$ to the ``canonical'' dispersion
form with the first power of $s-(P-q)^2$ in denominator.

\section{Decay constants of charmed baryons}

To obtain decay constants of the $\Lambda_c$  baryon, the
following two-point correlation function of the interpolating
current and its Dirac-conjugate is considered:
\begin{eqnarray}
F^{(i)}(q) &=& i\int d^4z\ e^{iq\cdot
z}\bra{0}T\left\{\eta_{\Lc}^{(i)}(z),\overline{\eta}_{\Lc}^{(i)}(0)\right\}\ket{0} \nonumber \\
&=&  F^{(i)}_1(q^2)  \not \! q +  F^{(i)}_2(q^2)  \,,
\label{eq:2ptcorr}
\end{eqnarray}
where $i={\cal P},{\cal A}$, At $q^2\ll m_c^2$ this correlation
function containing two invariant amplitudes is  calculated from
local OPE in terms of the perturbative and vacuum-condensate
contributions up to dimension 6. The OPE results will be taken
from the literature. Note that it is consistent to use two-point
sum rules without the $\alpha_s$  corrections, since  the latter
are also not taken into account  in LCSR. The hadronic dispersion
relation for (\ref{eq:2ptcorr}) reads:
\begin{eqnarray}
F^{(i)}(q) &=& {|\lambda_{\Lc}^{(i)}|^2 m_{\Lc}^2 \over  m_{\Lc}^2
-q^2} (\not \! q + m_{\Lc} ) + {|\lambda_{\Lcst}^{(i)}|^2
m_{\Lcst}^2 \over m_{\Lcst}^2 -q^2} (\not \! q - m_{\Lcst} )
\nonumber \\
&& + \int\limits_{\overline{s}_0^h}^{\infty} {ds \over s-q^2}
\left [\rho^{(i)}_1(s) \not \! q + \rho^{(i)}_2(s)  \right ]\,.
\label{eq:2ptdisp}
\end{eqnarray}
Here again the $\Lcst$-contribution ``contaminates'' the sum
rules. To tackle this problem we use the linear combination of the
dispersion relations for the invariant amplitudes $F_1^{(i)}$ and
$F_2^{(i)}$ and eliminate the $\Lcst$-state contribution,
obtaining  the following QCD sum rule:
\begin{eqnarray}
|\lambda_{\Lc}^{(i)}|^2
 = \frac{e^{m_{\Lc}^2 /M_{2pt}^2}}{m_{\Lc}^2  (m_{\Lc}+m_{\Lcst})} {1 \over \pi} \int\limits
_{m_c^2}^{s_0^{2pt}} ds~e^{-s/M_{2pt}^2}   \left [ m_{\Lcst}{\rm
Im} F_1^{(i)}(s) + {\rm Im} F_2^{(i)}(s)\right ] \,,
\end{eqnarray}
where $M^2_{2pt}$ and $s_0^{2pt}$ are the  Borel and threshold
parameters.

For the decay constant $\lambda_{\Lc}^{({\cal P})}$ induced  by the pseudoscalar interpolation current, the
OPE results for the spectral densities of the  invariant amplitudes
$F_{1,2}^{({\cal P})}$  are taken from
\cite{Bagan:1992tp} and presented in App.~\ref{two-point sum rules}.
To access the decay constant of $\Lambda_c$ induced by the axial-vector current,
we  employ  the results from \cite{Bagan:1993ii}
where a linear combination of two different interpolating currents
$J_{\Lambda_c}^{(1)} +bJ_{\Lambda_c}^{(2)}$  were used. At $b=-1/5$
we recover a superposition  of the currents of our choice:
\begin{eqnarray}
J_{\Lambda}^{(1)}- {1 \over 5} J_{\Lambda}^{(2)}= { \sqrt{6} \over
10} \left (\eta_{\Lc}^{({\cal A})} + 4 \eta_{\Lc}^{({\cal P})}\right)\,.
\end{eqnarray}
The correlation function of this combined current with its conjugate
has a decomposition similar to (\ref{eq:2ptcorr}), with two invariant
amplitudes which we denote as $\tilde{F}_{1,2} $.
From that we derive the sum rule for the
linear combination of the decay constants:
\begin{eqnarray}
\left|\lambda_{\Lc}^{({\cal A})} +
4\lambda_{\Lc}^{({\cal P})}\right |^2
% \nonumber \\
= \frac{50\, e^ {m_{\Lc}^2 /M_{2pt}^2} }{3 m_{\Lc}^2 (m_{\Lc}+m_{\Lcst})} {1 \over \pi} \int\limits
_{m_c^2}^{s^{2pt}_0} ds~e^{ -s/M_{2pt}^2}  [ m_{\Lcst} {\rm Im}
\tilde{F}_1 (s)
 + {\rm Im} \tilde{F}_2 (s)] \,. &&
\label{eq:Lc}
\end{eqnarray}
The OPE expressions for the spectral densities ${\rm Im} \tilde{F}_{1,2}
(s)$ combined from the results obtained  in \cite{Bagan:1993ii} are
collected in App.~D. To resolve the
ambiguity of the relative sign between $\lambda_{\Lc}^{({\cal A})}$ and $\lambda_{\Lc}^{({\cal P})}$ in this sum rule, we assume
that they are of the same order of magnitude.
The numerical analysis reveals that this is only
possible if the two decay constants in (\ref{eq:Lc}) have the same sign.
This allows us to obtain  $\lambda_{\Lc}^{({\cal A})}$ using the value of $\lambda_{\Lc}^{({\cal P})}$ calculated above.

Decay constants of the $\Sigma_c$ baryon have been calculated in Ref.
\cite{Bagan:1991sc}  from  the similar two-point QCD sum rules. Using
Fierz transformation, we relate the mixed current $J_{\Sigma_c}$ used in that paper
with a linear combination of the two currents of our choice:
\begin{eqnarray}
J_{\Sigma_c} = (u^T C \gamma_5 c) u + b (u^T C  c) \gamma_5 u =
{1-b \over 4} \eta_{\Sigma_c}^{({\cal I})}   + {1+b \over 8}
\eta_{\Sigma_c}^{({\cal T})}  \,,
\end{eqnarray}\
where $b$ is the mixing parameter.
The following sum rule for the combination of decay constants
is then obtained:
\begin{eqnarray}
 \left|2 (1-b)\lambda_{\Sigma_c}^{({\cal I})} + (1+b)
\lambda_{\Sigma_c}^{({\cal T})} \right|^2
%&&
%\nonumber  \\
\!=\!  \frac{64\,e^{m_{\Sigma_c}^2/M_{2pt}^2}}{ m_{\Sigma_c} ^2
(m_{\Sigma_c}+m_{\Sigma_c^{\ast}}) }{1 \over \pi} \int\limits_{m_c^2}^{s^{2pt}_0} ds
e^{-s/M_{2pt}^2}  \bigg( m_{\Sigma^{\ast}}{\rm Im} \bar{F}_1 (s)   +
{\rm Im} \bar{F}_2 (s)\bigg) \,,
%&&
\nonumber \\
\end{eqnarray}
where the spectral densities $\bar{F}_1 (s)$ and $\bar{F}_2 (s)$
obtained from \cite{Bagan:1991sc} are given in App.~\ref{two-point
sum rules}. Choosing subsequently $b=1$ and $b=-1$ we obtain
separate sum rule for $\lambda_{\Sigma_c}^{({\cal I})}$ and for
$\lambda_{\Sigma_c}^{({\cal T})}$. Finally, the two-point QCD sum
rules for the decay constants of $\Lambda_b$-baryon are obtained
from the $\Lambda_c$-sum rules, replacing $c\to b$.

\section{LCSR for the strong couplings}

The strong coupling constants of $\Lambda_c$-baryon with nucleon and
$D$- or $D^*$-meson  are formally defined as hadronic matrix
elements:
%\footnote{Here, the notion $D^{(\ast)}$ in the final state
%denotes $\bar{D}^{(\ast)}(\bar{c} q)$ meson   for brevity. }:
\begin{eqnarray}
\langle \Lambda_c(P-q) |D(-q)  N(P)\rangle &=& g_{ \Lambda_c
ND} \, \bar{u}_{\Lambda_c}(P-q)\,i\gamma_5\,u_N(P),
\label{CC1} \\
\langle  \Lambda_c(P-q) |D^*(-q)  N(P) \rangle &=&
\bar{u}_{\Lambda_c}(P-q)\left(g^V_{ \Lambda_c
ND^*}\slashed{\epsilon}+i\,\frac{g^T_{ \Lambda_c ND^*}}{m_{
\Lambda_c}+m_N}\sigma_{\mu\nu}\epsilon^\mu q^\nu\right)u_N(P).
\nonumber \hspace{1 cm}
\end{eqnarray}

Different from the $D^*D\pi$ coupling  that can be
measured in kinematically allowed $D^*\to D\pi$ decays, a direct
measurement of the baryonic strong couplings is not possible
because at least one of the hadrons has to be off-shell. E.g., in
the hadronic dispersion relation for the $\Lambda_c\to N$ form
factor, the residue of the $D$- or $D^*$-pole  (for pseudoscalar
or vector transition current) is proportional to the
$\Lambda_cND^{(*)}$  coupling. This pole  is located at
$q^2=m^2_{D^{(*)}}$, beyond the physical regions
$q^2\leq(m_\Lc-m_N)^2$ (for semileptonic decays and scattering)
and $q^2 \geq (m_\Lc+m_N)^2$ (for the baryon pair production).

In the heavy mass limit for the $c$-quark we may obtain
relations between the coupling constants appearing in (\ref{CC1}).
In this case the masses of the  $\Lambda_c$ and $D^{(*)}$ become equal to  $m_c$.
The heavy mesons $D$ and $D^*$ form a spin
symmetry doublet which can be represented by
\begin{equation}
{\cal D} (v) = {\cal N} (1+\slashed{v}) \left(- i \gamma_5 D + \gamma^\mu
D_\mu^* \right)  \quad \mbox{with} \quad v^\mu D_\mu^* = 0
\end{equation}
where $D$ and $D^*$ represent the charmed meson fields and ${\cal N}$ is
a normalization factor. Likewise, the spinor of the
$\Lambda_c$ is equal to the $c$-quark spinor, since the light
degrees of freedom are in a spinless state. Thus in the $m_c\to \infty$ limit,
the two hadronic amplitudes in (\ref{CC1}) can be unified in one:
\begin{equation}
{\cal A} = \bar{u}_{\Lambda_c} (v) {\cal D} (v) {\cal M} u_N (P)\,,
\end{equation}
with a Dirac-structure ${\cal M }$ and the nucleon bispinor $u_N$ accumulating
the light degrees of freedom in the process.
The quantity ${\cal M}$ may be expanded in
the basis Dirac matrices $1$, $\gamma_5$, $\gamma_\mu$,
$\gamma_\mu \gamma_5$ and $\sigma_{\mu \nu}$; due to parity and
Lorentz invariance only the unit matrix remains. Hence we obtain
\begin{equation}
 {\cal A} =  g \bar{u}_{\Lambda_c} \left (- i \gamma_5 D + \gamma^\mu D_\mu^* \right) u_N (P) \, ,
\end{equation}
where $g$ is the strong coupling in the heavy quark limit.
Comparing this with (\ref{CC1}) we obtain the heavy mass relations
\begin{equation}
g_{ \Lambda_c ND} = - g^V_{ \Lambda_c ND^*} \quad \mbox{and} \quad
g^T_{ \Lambda_c ND^*} = 0 \,. \label{HQET coupling 1}
\end{equation}
Following the same procedure, one can also  derive the following
relation for the three strong couplings involving $\Sigma_c$
baryon:
\begin{eqnarray}
g_{\Sigma_c N D} + 3 g^V_{ \Sigma_c ND^*} ={ 3 m_{\Sigma_c} +
m_{N} - 2 P \cdot v \over m_{\Sigma_c} +m_N } \,\, g^T_{ \Sigma_c
ND^*}\,. \label{HQET coupling 2}
\end{eqnarray}
Here, the four-velocity vector is defined as $v=-q/{m_D^{(*)}}$;
hence, up to $O(1/m_c)$ corrections, $ P\cdot v=
(m_{\Sigma_c}^2-m_{D^{(*)}}^2)/(2m_{D^{(*)}})$.

The couplings (\ref{CC1})  play an important role in  various
models of strong interactions formulated in terms of virtual hadron
exchanges, like e.g., in the production of a charmed baryon pair in
the $p\bar{p}$ collision, with a virtual $D^{(*)}$ exchange in $t$-
channel. In the forthcoming publication \cite{pbarp2charm}
we shall consider this process in the PANDA energy region.
It is tempting to formulate the hadronic exchange models in terms of
effective Lagrangians involving propagation and couplings of
hadronic fields rather than quarks and gluons. An effective
Lagrangian involving the hadronic couplings discussed above has the following form:
\begin{eqnarray} \label{effL}
 \mathscr{L}_{\Lambda_c D^{(\ast)} N}=\,\bar{\Lambda}_c\, \bigg[ i  a_{\Lambda_c ND}  \gamma_5 \, D +\bigg(a^V_{\Lambda_c ND^*}\gamma^\mu
 +  \frac{a^T_{\Lambda_c ND^*}}{m_{\Lambda_c}+m_N}\sigma^{\mu \nu} \partial_\nu \bigg) \, D^*_\mu \bigg ]\,N  \, +\,h.\,c.
\label{eq:Leff}
\end{eqnarray}
where we have defined new couplings $a_i$. However, these
couplings are not necessarily the same as $g_i$ appearing in
(\ref{CC1}), since this depends on the kinematic region where
 the effective Lagrangian (\ref{eq:Leff}) is applied.
The latter  assumes point-like baryons,
and hence it can only be used at impact parameters large compared
to the size of the baryons, which means it is restricted to small
momentum transfers $t=q^2$. This in turn means that the exchanged
$D^{(*)}$ meson is far off shell and thus for such an application
the couplings $a_i$ will be different from $g_i$. Nevertheless,
one may use the coupling constants calculated here as an input
for a more elaborated hadronic models based on Regge poles.

The possibility to calculate strong couplings from LCSR is based
on the fact that they enter double dispersion relations for the
same correlation function (\ref{eq:corr}).
E.g., to access  the $\Lambda_cND$ coupling, we employ
 the hadronic double dispersion relation for
the correlation  function $\Pi^{(i)}_5(P,q)$ with the pseudoscalar
transition current (\ref{eq:corrps}), choosing one of the
interpolating currents $i={\cal P},{\cal A}$. The double
dispersion relation is obtained by analytically continuing the
imaginary parts of the invariant amplitudes
$\mbox{Im}_s\Pi^{(i)}_{1,2}(s,q^2)$ in the second variable $q^2$.
The result is given by the following expression containing the
$\{\Lc,D\}$ double pole (the ground-state contribution), and, in
addition, the  $\{\Lcst,D\}$ double pole:
\begin{eqnarray}
\Pi^{(i)}_5(P,q)&= &  \frac{ \lambda_{\Lc}^{(i)} m_D^2 f_D
m_{\Lc} g_{\Lambda_c ND}}{(m_{\Lc}^2-(P-q)^2)(m_D^2-q^2)
}\,\bigg[(m_\Lc-m_N)-\slashed{q}
 \bigg]i\gamma_5 u_N(P) \nonumber \\
&& +  \frac{\lambda_{\Lambda_c^{\ast}}^{(i)}  m_D^2 f_D
m_{\Lambda_c^{\ast}} g_{\Lambda_c^{\ast} ND} } {
(m_{\Lambda_c^{\ast}}^2-(P-q)^2) (m_D^2-q^2)
 }\,\bigg[(m_{\Lambda_c^{\ast}}+m_N) +\slashed{q}
 \bigg]i\gamma_5 u_N(P)\nonumber \\
&&+ \ldots \,,
%&& + \int\limits_{s_0^h}^{+\infty} { ds \over s -(P-q)^2 }
%\int_{s_0^{h'}}^{+\infty} { ds^{\prime} \over s^{\prime} -(P-q)^2
%} \rho^h_5 (s, s^{\prime}) u_N(P) \, ,
\label{eq:ddispDLN}
\end{eqnarray}
where  the ellipses indicate the contributions of excited and
continuum states in both $\Lc$ and $D$  channels which have a
generic form of dispersion integrals over  the hadronic double
spectral density.
 %$(P-q)^2\to s$, $q^2\to s'$.

Similarly, for the $\Lc D^*N$ couplings  we employ the hadronic double dispersion relation
for the correlation function with the vector transition current:
\ba
\Pi^{(i)}_{\mu}(P,q)&=& { m_{\Lambda_c} m_{D^{\ast}}  f_{D^{\ast}}
\lambda_{\Lc}^{(i)} \over ( m_{\Lambda_c}^2 -(P-q)^2 )
(m_{D^{\ast}}^2 -q^2) }  \bigg \{ - 2 g^{V}_{\Lambda_c ND^*}P_\mu +
2 \frac{g^{T}_{ \Lambda_c ND^*}}{m_{
\Lambda_c}+m_N} P_\mu  \not\! q  \nonumber \\
&&  -( g^{V}_{\Lambda_c ND^*}-g^{T}_{\Lambda_c ND^*} ) \bigg[(m_{\Lc}
-m_N) \gamma_{\mu} + \gamma_{\mu} \not \! q \bigg]\nonumber \\
&& + \bigg [ (2 - { m_{ \Lambda_c}^2- m_N^2 \over m_{D^{\ast}}^2
})  g^{V}_{\Lambda_c
ND^*}  - g^{T}_{ \Lambda_c ND^*}  \bigg ] q_{\mu} \nonumber \\
&&  + \bigg [  -{ g^{T}_{ \Lambda_c ND^*} \over m_{ \Lambda_c}
+m_N} +{ m_{ \Lambda_c} -m_N \over  m_{D^{\ast}}^2 }g^{V}_{
\Lambda_c ND^*}    \bigg]  q_{\mu} \not \!  q  \bigg \} u_N(P)
\nonumber
\\
&& + \frac{ m_{\Lcst} m_{D^{\ast}}  f_{D^{\ast}}
\lambda_{\Lcst}^{(i)} }{ ( m_{\Lcst}^2 -(P-q)^2 ) (m_{D^{\ast}}^2
-q^2) }  \bigg \{
2g^{V}_{\Lcst ND^*} P_\mu - 2 \frac{g^{T}_{ \Lambda_c^{\ast}ND^*}}{m_{\Lcst}+m_N} P_\mu  \not \! q  \nonumber \\
&&  -  \bigg [ g^{V}_{\Lcst ND^*}
+\frac{m_{\Lcst}-m_N }{ m_{\Lcst}+m_N}  g^{T}_{\Lcst ND^*}  \bigg ]
\bigg[(m_{\Lcst}+m_N)  \gamma_{\mu}   -\gamma_{\mu}  \not \! q \bigg] \nonumber \\
&& + \bigg [ (-2 + { m_{ \Lcst}^2- m_N^2 \over
m_{D^{\ast}}^2 })  g^{V}_{\Lcst ND^*} -
{m_{\Lcst} -m_N \over  m_{\Lcst} + m_N  }
 g^{T}_{ \Lcst ND^*}  \bigg ] q_{\mu}  \nonumber \\
&& +  \bigg [ { g^{T}_{ \Lcst  ND^*} \over
m_{\Lcst} + m_N  } + { m_{\Lcst} + m_N \over
m_{D^{\ast}}^2 } g^{V}_{ \Lcst ND^*} \bigg ]
q_{\mu} \not \! q \bigg \}u_N(P)
\nonumber \\
&& + \ldots ~~.
%&& + \int_{s_0^h}^{+\infty} { ds \over s -(P-q)^2 }
%\int_{s_0^{h'}}^{+\infty} { ds^{\prime} \over s^{\prime} -(P-q)^2
%} \rho^h_{\mu} (s, s^{\prime})u_N(P) \,.
\label{eq:ddispDstLN}
\ea

Decomposition of the relations (\ref{eq:ddispDLN}) and (\ref{eq:ddispDstLN}) in terms of invariant
amplitudes and elimination of the $\Lcst$ contributions are similar to
the steps done in the derivation of the LCSR for  the form factors.

The OPE results for the invariant
amplitudes obtained in Sect.~4  have now to be considered
in the deep spacelike region for both variables  $(P-q)^2,q^2\ll m_c^2$ .
The new elements needed for the quark-hadron duality approximation of the higher states are the double  spectral densities of the invariant amplitudes.
These amplitudes were already obtained in the form (\ref{eq:QCDcorr}).
It suffices to find  double spectral representations
for the master integrals of the type $\int dx x^K D^{-n}$ with $k\geq 0$ and $n=1, \, 2 \,, 3$.
For $n=1$, we obtain
\begin{eqnarray}
 \int\limits_0^1 d x   { x^k \over  D } =  {1 \over 2 \pi} \sum \limits_{j=0}^{k}
\int\limits_{m_c^2}^{\infty}   {d s \over s - (P-q)^2} \int\limits_{t_1(s)}^{t_2(s)}
{d s^{\prime} \over s^{\prime}- q^2 }(-1)^{k+j/2}
[1+(-1)^j ] &&  \nonumber \\
\times {1\over (2 m_N^2)^k  } C_k^j     (s - s^{\prime} - m_N^2)^{k-j}
[(s^{\prime}-t_1)(t_2-s^{\prime})]^{ j-1 \over 2 }\,,
\end{eqnarray}
where  $C_k^j$ are the binomial coefficients and the integration limits are
\begin{eqnarray}
t_{1,2} (s)= (s+m_N^2)  \mp 2 m_N \sqrt{s-m_c^2} \,.
\label{eq:intlim}
\end{eqnarray}
The double spectral representations for the master integrals with
$n=2$ and 3,
being more lengthy, are collected in App.~\ref{double
spectral density}.
Using these integrals it is easy to find the
double dispersion representation for all integrals  in  (\ref{eq:QCDcorr}),
where the numerators depend
polynomially on $x$, and linearly on $(P-q)^2$ and $q^2$,
so that the latter variables
can simply be replaced by $s$ and $s'$ respectively.

Equating the OPE results to the double hadronic dispersion relations,
adopting the quark-hadron duality  approximation for the hadronic spectral densities
and performing the double Borel transformation, $(P-q)^2\to M^2$, $q^2\to \widetilde{M}^2$,
we  derive  LCSR for the three strong couplings of our interest:
\ba
 g _{\Lambda_c ND}
=\frac{ e^{m_{\Lc}^2/M^2}e^{m_{D}^2/\widetilde{M}^2}}{m_{\Lambda_c} ( m_{\Lambda_c} + m_{\Lambda_c^{\ast}} )m_{D}^2 f_{D}\lambda_{\Lc}^{(i)}}\,
\frac{1}{\pi^2}\int\limits_{m_c^2}^{s_0} ds\, e^{-s/M^2} \nonumber\\
\times
\int\limits_{t_1(s)}^{t_2(s)}  ds^{\prime}\,
e^{-s^{\prime}/\widetilde{M}^2}
%\nonumber \\
{\rm Im}_{s} \, {\rm Im}_{s^{\prime}} [\Pi_1^{(i)} (s, s^{\prime}) -
(m_{\Lambda_c^{\ast}}  + m_N) \Pi_2^{(i)} (s, s^{\prime})]\,,
\label{eq:DLCSR}
\ea
%%%%%
\ba
g^{V}_{\Lc ND^*}=
- \frac{e^{m_{\Lc}^2/M^2}e^{m_{D^{\ast}}^2/\widetilde{M}^2}}{2  m_{\Lambda_c} ( m_{\Lambda_c} + m_{\Lambda_c^{\ast}} )
m_{D^{\ast}} f_{D^{\ast}}\lambda_{\Lc}^{(i)}}\,
\frac{1}{\pi^2}\int\limits_{m_c^2}^{s_0} ds\, e^{-s/M^2} \nonumber\\
\times
\int\limits_{t_1(s)}^{t_2(s)}  ds^{\prime}\,
e^{-s^{\prime}/\tilde{M}^2}
{\rm Im}_{s} \, {\rm Im}_{s^{\prime}}  \bigg [ (m_{\Lambda_c}  +
m_N) \bigg ( \bar{\Pi}_1^{(i)} (s, s^{\prime})    - (m_{\Lambda_c^{\ast}}
-m_N) \bar{\Pi}_2^{(i)} (s, s^{\prime})  \bigg ) \nonumber \\
+ 2 \bar{\Pi}_3^{(i)} (s, s^{\prime}) +2 (m_{\Lambda_c^{\ast}}
-m_{\Lambda_c}) \bar{\Pi}_4^{(i)} (s, s^{\prime}) \bigg]\,,
\label{eq:DstVLCSR} \ea
%%%%%%%%%%%%%%%%%%%
\ba
g^{T}_{\Lc ND^*}
=- \frac{(m_{\Lambda_c}+m_N)  e^{m_{\Lc}^2/M^2}e^{m_{D^{\ast}}^2/\widetilde{M}^2}}{2  m_{\Lambda_c} ( m_{\Lambda_c} + m_{\Lambda_c^{\ast}} ) m_{D^{\ast}}
f_{D^{\ast}}\lambda_{\Lc}^{(i)} }
\frac{1}{\pi^2}\int\limits_{m_c^2}^{s_0} ds\, e^{-s/M^2} \nonumber\\
\times
\int\limits_{t_1(s)}^{t_2(s)}  ds^{\prime}\,
e^{-s^{\prime}/\widetilde{M}^2}
{\rm Im}_{s} \, {\rm Im}_{s^{\prime}} \bigg [ \bar{\Pi}_1^{(i)} (s,
s^{\prime}) - (m_{\Lambda_c^{\ast}} - m_N) \bar{\Pi}_2^{(i)} (s,
s^{\prime})  -2 \bar{\Pi}_4^{(i)} (s, s^{\prime}) \bigg] \,.
\label{eq:DstTLCSR}
\ea
The $D$ and $D^*$ decay constants entering these sum rules
are defined in a standard way:
\begin{eqnarray}
 \bra{0} m_c\bar{c}\,i\gamma_5 u \ket{D(p)} =m_D^2\,f_D \,, ~~
 \bra{0}\bar{c}\,\gamma^\mu u \ket{D^*(p,\epsilon)} =
 m_{D^*}\,f_{D^*}\,\epsilon^\mu \,.
\end{eqnarray}
Instead of fixing their numerical values we will use the two-point
QCD sum rules for $f_{D^{(\ast)}}$  taken for consistency in
$O(\alpha_s^0)$ (see e.g., \cite{BBKR}). The region in the $\{
s,s'\}$ plane occupied by the double spectral density of the
correlation function calculated in OPE in the limit $m_N\to 0$, is
reduced  to the diagonal $s=s'$, very similar to LCSR for
$D^*D\pi$ coupling \cite{BBKR}. In that case it was sufficient to
use one effective threshold $s_0$ for the duality approximation.
Here we adopt a similar ansatz reflected in the integration limits
in the above sum rules. Clearly, it is possible to use different
borders of the duality region. The sensitivity of the LCSR results
to this region is lowered by the Borel transformation, still   an
additional uncertainty is introduced. Hence, as far as the
quark-hadron duality is concerned, the sum rules for the strong
couplings  are generally less accurate  than the ones for the form
factors.

\section{Numerical results}

We begin the numerical analysis with specifying the  choice of the
input for nucleon DA's, collected in  App.~A. Their normalization
parameters have been calculated from two-point  QCD sum rules
\cite{BFMS}:
\begin{eqnarray}
 f_N = (5.0\pm 0.5) \times 10^{-3}\, \mathrm{GeV}^2\,,\nonumber \\
\lambda_1 = -(27\pm 9)\times 10^{-3} \, \mathrm{GeV}^2\,,~~
\lambda_2 = (54 \pm 19)\times 10^{-3} \, \mathrm{GeV}^2\,.
\label{normalization DA}
\end{eqnarray}
For the remaining five dimensionless parameters determining the
shapes of the nucleon DA's, we use the model suggested in
\cite{BLW06} in which the QCD 2-point sum rule estimates are
adjusted, via LCSR, to the data on the nucleon electromagnetic
form factors:
\begin{eqnarray}
 A_1^u = 0.13\,,~~  V_1^d =0.30\,,~~
f_1^d=0.33 \,,~~   f_1^u=0.09 \,,~~    f_2^d=0.25\,.
\label{eq:shapepar}
\end{eqnarray}
The masses of baryons are taken from
\cite{PDG}:
%\begin{eqnarray}
$m_{N}= m_p=0.938 \, {\rm GeV}$,
%\nonumber \\
$m_{\Lambda_c} =2.286 \, {\rm GeV}$,
$m_{\Lcst} =2.595 \, {\rm GeV}$,
$m_{\Sigma_c} = 2.454 \, {\rm GeV}$,
$m_{\Sigma_c^*} = 2.801 \, {\rm GeV}$,
%\nonumber \\
$m_{\Lambda_b}= 5.620 \, {\rm GeV}$, and the estimated mass of the
negative parity $b$-baryon: $  m_{\Lambda_b^*}= 5.85 \, {\rm
GeV}$, is taken according to the QCD sum rule estimate
\cite{Wang:2010fq}.
%\end{eqnarray}

For the virtual $c$ and $b$ quarks in the correlation functions,
the $\overline{\rm MS}$ mass is preferable. We use
%\begin{eqnarray}
$\bar{m}_b (\bar{m}_b) = 4.16 \pm 0.03 \, {\rm GeV}$, $\bar{m}_c
(\bar{m}_c) = 1.28 \pm 0.03  \, {\rm GeV}$,
%\label{eq:qmass}
%\end{eqnarray}
taking the central values  from the precise determination
\cite{Chetyrkin} based on the quarkonium  sum rules and twice
inflating the uncertainties. In the absence of gluon corrections
the  only renormalization scale which enters our calculation
 is the factorization scale $\mu$ of the nucleon DA's. We adopt
the same scale for the quark masses, using the intervals
$\mu_c=1.5\pm 0.5 \, {\rm GeV}$ and $\mu_b=4.0 \pm 1.0 \, {\rm
GeV}$ for LCSR with $c$ and $b$ quarks, respectively. The
evolution of the scale dependent parameters in the nucleon DA's is
taken according to \cite{Braun:2008ia}. We also adopt the same
vacuum condensates as in \cite{Duplancic:2008ix,KKMO}, in
particular, the quark condensate density $\langle \bar{q}
q\rangle(1 \mbox{GeV}) = -(246^{+28}_{-19} \rm MeV)^3$.
%the gluon condensate $\langle  {
%\alpha_s \over \pi} GG \rangle  = 0.012^{+0.006}_{-0.012}{\rm
%GeV^4}$ and the ratio between quark-gluon condensate and quark
%condensate $m_0^2 =0.8 \pm 0.2 {\rm GeV^2}$ are taken in the.

The intervals of Borel parameters
used in the sum rules considered here are listed in Table~\ref{tab:par}.
Their choice is based on the usual criteria, that is,
both power corrections and continuum contributions in the sum rules
have to be sufficiently suppressed.
The corresponding effective thresholds (see Table~\ref{tab:par})
are  adjusted so that the differentiated sum rules
reproduce the measured mass of the lowest baryon or meson
with at least a 10 \% accuracy.
%%%%%%%%%%%%%%%%%%%%
\begin{table}[tb]
\begin{center}
\begin{tabular}{|c|c|c|}
  \hline
  \hline
&&\\[-4mm]
QCD sum rule & Borel parameter (${\rm GeV^2}$)  & eff. threshold (${\rm GeV^2}$) \\[1.5mm]
\hline
&&\\[-4mm]
{\em LCSR, form factor} & $M^2$ & $s_0$ \\[1.5mm]
 % \hline
$\Lambda_c (\Sigma_c) \to p$ & $7.5 \pm 2.5$ & $10.0 \pm 0.5$ \\
%&&&\\
 $\Lambda_b  \to p$ & $20.0 \pm 5.0$ & $40.0 \pm 1.0$ \\[1.5mm]
\hline
&&\\[-4mm]
{\em LCSR, strong coupling} &$M^2$ \hspace{1.1cm} $\widetilde{M}^2$&$s_0$\\[1.5mm]
%\hline
$\Lc(\Sigma_c) N D$ &  \hspace{1.2cm} $4.5 \pm 1.5$ &
\\[-2mm]
&\hspace{-2cm}\Bigg\}$7.5 \pm 2.5$&$\Bigg\}10.0 \pm 0.5$\\
%\hline
$\Lc(\Sigma_c)N D^{\ast}$&  \hspace{1.2cm} $5.0 \pm 1.5$ &
\\[1.5mm]
\hline
&&\\[-3mm]
&&\\[-3mm]
{\em 2-point SR, decay constant} &$M^2_{2pt}$ & $s_0^{2pt}$\\[1.5mm]

$\Lambda_c (\Sigma_c)$ & $2.5 \pm 0.5$ & $10.0 \pm 0.5$ \\
$\Lambda_b$  & $5.0 \pm 1.0$ & $40.0 \pm 1.0$ \\[2mm]
  $D$  &$2.0 \pm 0.5$ & $6.5 \pm 0.5$ \\
  $D^{\ast}$ & $2.0 \pm 0.5$& $8.0 \pm 0.5$ \\
&& \\
  \hline
  \hline
\end{tabular}
\end{center}
\caption{Borel parameters and effective thresholds
used in  various sum rules.} \label{SR parameters}
\label{tab:par}
\end{table}
%%%%%%%%%%%%%%

Furthermore, instead of substituting in LCSR  a certain fixed
value for  the $\Lambda_{c(b)}$ or $\Sigma_c$ decay constants, we
use the corresponding two-point sum rules. This somewhat reduces
the overall uncertainties. Still, to give
an idea of the magnitude of the decay constants,
let us quote their numerical values:
%%%%%%%%%%%%%%%5
\begin{eqnarray}
\lambda_{\Lambda_c}^{\mathcal{(A)}}&=&1.51^{+0.37}_{-0.39} \times
10^{-2} \,\, {\rm GeV^2} \,, \qquad
\lambda_{\Lambda_c}^{\mathcal{(P)}}=1.19^{+0.19}_{-0.28}\times
10^{-2} \,\, {\rm GeV^2}\,, \nonumber
\end{eqnarray}
\begin{eqnarray}
\lambda_{\Lambda_b}^{\mathcal{(A)}}&=&1.27^{+0.35}_{-0.34} \times
10^{-2} \,\, {\rm GeV^2} \,, \qquad
\lambda_{\Lambda_b}^{\mathcal{(P)}}=1.09^{+0.31}_{-0.30}\times
10^{-2} \,\, {\rm GeV^2}\,, \nonumber \\
%%%%%%%%%%%%%%%%%%
\lambda_{\Sigma_c}^{\mathcal{(I)}}&=&3.08^{+0.49}_{-0.74}\times
10^{-2}  \,\, {\rm GeV^2} \,, \qquad
\lambda_{\Lambda_c}^{\mathcal{(T)}}= 6.08^{+0.90}_{-1.48}\times
10^{-2}\,\, {\rm GeV^2}\,.
\end{eqnarray}
After specifying all input parameters, we compute the numerical
values of the $\Lambda_c (\Sigma_c )  \to p$ form factors at
$q^2=0$ and the $\Lc(\Sigma_c )  N D^{(\ast)}$ strong couplings.
The results are collected in Table \ref{table_res1}, where the
total uncertainties are estimated by varying separate input
parameters within their ranges and adding the resulting separate
uncertainties of the form factors and strong couplings in
quadrature. Correlations between different form factors and strong
couplings with respect to the input variation make the sum rule
predictions for the ratios of these hadronic matrix elements even
more accurate.

%%%%%%%%%%%%%%%%%%%%%
\begin{table}[t]
\begin{center}
\begin{tabular}{|c|c c|c c|}
  \hline
  \hline
   &&& &\\
 Current &$\eta_{\Lambda_c}^{(\mathcal{A})}$ & $\eta_{\Lambda_c}^{(\mathcal{P})}$ &
$\eta_{\Sigma_c}^{(\mathcal{I})}$
 & $\eta_{\Sigma_c}^{(\mathcal{T})}$ \\[2mm]
Form factor & \multicolumn{2}{c|} {$\Lambda_c\to p$ }  & \multicolumn{2}{c|} {$\Sigma_c\to p$}  \\
   &&&& \\
   \hline
   \hline
   &&&&  \\
  $G(0)$ & $0.39^{+0.11}_{-0.09}$ & $0.48^{+0.13}_{-0.13}$ &  $0.066^{+0.035}_{-0.032}$ & $0.061^{+0.011}_{-0.011}$  \\
  &&&&  \\
  \hline
    &&&&  \\
   $f_1(0)$ & $0.46^{+0.15}_{-0.11}$ & $0.59^{+0.15}_{-0.16}$ & $-0.22^{+0.07}_{-0.07}$ & $-0.23^{+0.04}_{-0.05}$ \\
    &&&&  \\
    $f_2(0)$  & $-0.32^{+0.08}_{-0.07}$ & $-0.43^{+0.13}_{-0.12}$ & $-0.24^{+0.05}_{-0.05}$ & $-0.25^{+0.06}_{-0.06}$  \\
   &&&&  \\
    $g_1(0)$ &  $0.49^{+0.14}_{-0.11}$ &  $0.55^{+0.14}_{-0.15}$ & $0.11^{+0.05}_{-0.05}$    & $0.060^{+0.007}_{-0.008}$ \\
   &&&&  \\
    $g_2(0)$  & $-0.20^{+0.09}_{-0.06}$ &  $-0.16^{+0.08}_{-0.05}$ &  $-0.002^{+0.054}_{-0.044}$ &
$-0.030^{+0.039}_{-0.039}$ \\
  &&&&  \\
   \hline
   \hline
  &&&&  \\
Strong coupling & \multicolumn{2}{c|}{$\Lc D^{(*)}N$} & \multicolumn{2}{c|} {$\Sigma_c D^{(*)}N$} \\
 &&&&  \\
   $g_{\Lc (\Sigma_c) N D} $  & $13.8^{+5.2}_{-4.1}$  & $10.7^{+5.3}_{-4.3}$  & $1.3^{+1.0}_{-0.9}$  &   $1.3^{+1.2}_{-0.8}$  \\
   &&&&  \\
   $g^V_{\Lc (\Sigma_c) N D^{\ast}} $   & $-7.9^{+2.7}_{-3.3}$ & $-5.8^{+2.1}_{-2.5}$ &  $1.0^{+1.3}_{-0.6}$ & $0.74^{+1.08}_{-0.45}$
\\
   &&&&  \\
   $g^T_{\Lc (\Sigma_c) N D^{\ast}} $  & $4.7^{+2.7}_{-2.0}$ &  $3.6^{+2.9}_{-1.8}$& $2.1^{+1.9}_{-1.0}$ &  $1.8^{+1.6}_{-0.8}$ \\
   &&&&  \\
  \hline
  \hline
\end{tabular}
\end{center}
\caption{Numerical results for the transition form factors
and strong couplings of charmed baryons obtained from LCSR
with different interpolating currents.}
\label{table_res1}
\end{table}
%%%%%%%%%%%%%%

Replacing $c$-quark with the $b$-quark in LCSR we calculate the
phenomenologically important $\Lambda_b \to p$ form factors.
They are collected in Table \ref{tab_resLambdab}. In this case,
not only the zero momentum transfer but also  small and
intermediate $q^2\ll m_b^2$ are available from LCSR. We estimate
the maximal value of $q^2$ accessible with LCSR  to lie in the
interval $q^2= 11-15 \, \mbox{GeV}^2$ and adopt, conservatively,
the lowest value $q^2_{max}=11 \, \mbox{GeV}^2$. At larger $q^2$
light-cone OPE is not reliable, in particular the contribution of
the highest twist-6 nucleon DA's starts to grow with respect to
the lower twists. Note that the contributions of all twist 3, 4, 5
components of nucleon DA's are numerically important in LCSR.

%%%%%%%%%%%%%%
\begin{table}[tb]
\begin{center}
\begin{tabular}{|c|c|c|}
  \hline
  \hline
   form factors &  $\eta_{\Lambda_b}^{(\mathcal{A})}$ & $\eta_{\Lambda_b}^{(\mathcal{P})}$  \\
   \hline
   &&\\
   $f_1(0)$ & $0.14^{+0.03}_{-0.03}$ &   $0.12^{+0.03}_{-0.04}$   \\
   && \\
$b_1$ &$-1.49^{+1.68}_{-1.88}$& $-9.13^{+0.88}_{-1.12}$\\
&&\\
   $f_2(0)$  & $-0.054^{+0.016}_{-0.013}$ &   $-0.047^{+0.015}_{-0.013}$ \\
   && \\
  $b_2$ & $-14.0^{+1.2}_{-1.8}$  & $-18.5^{+1.7}_{-2.0}$ \\
&&\\
   \hline
   &&\\
   $g_1(0)$ &  $0.14^{+0.03}_{-0.03}$ &  $0.12^{+0.03}_{-0.03}$  \\
   &&\\
  $\tilde{b}_1$ & $-4.05^{+1.38}_{-1.81}$ & $-9.18^{+0.75}_{-1.06}$\\
&&\\
   $g_2(0)$  & $-0.028^{+0.012}_{-0.009}$ & $-0.016^{+0.007}_{-0.005}$ \\
   &&\\
   $\tilde{b}_2$ & $-20.2^{+1.0}_{-2.1}$ & $-22.5^{+1.3}_{-1.7}$ \\
&&\\
  \hline
  \hline
\end{tabular}
\end{center}
\caption{Numerical results of $\Lambda_b \to p$  transition form
factors at zero momentum transfer and their slope parameters
obtained from LCSR with different interpolating currents.}
\label{tab_resLambdab}
\end{table}

%%%%%%%%%%%%%%%%%%%%%
Several additional comments on the numerical results obtained above
are  in order.

\begin{itemize}

\item We found that form factors and strong couplings are (within
uncertainties) insensitive to the interpolating current of the
heavy baryon, once the contribution of the negative-parity baryon
is included in the hadronic dispersion relation. We have checked
that if the negative-parity baryon is simply absorbed in the
duality-approximated continuum, the sum rules yield numerical
predictions that are considerably more sensitive to the choice of
the interpolating current.

\item The achieved accuracy of
LCSR for the form factors
of heavy baryons is well illustrated by the
equation-of-motion relation (\ref{pseudoscalar FF relation})
which yields $G(0)=g_1(0)$. Comparing the numerical results
for both form factors calculated from two different
LCSR, we see that this relation is violated numerically
at the level of 20 \% which is also in the ballpark
of the estimated uncertainty  of the LCSR.

\item
In the infinitely heavy quark limit, the
relation $f_1(q^2)=g_1(q^2)$
is valid  for $\Lambda_{Q} \to N$ form factors
and it is  well reproduced by our numerical results for both $Q=c,b$.
Note that this relation holds for any $q^2$, since only the heavy-quark spin
symmetry is employed in its derivation.

\item  The heavy-mass relations for the three strong couplings
of $\Lambda_c$ baryon,  shown in (\ref{HQET coupling 1}), are
only qualitatively supported  by the LCSR predictions
obtained for the finite $c$ quark mass. In particular,
the magnitude of $g^T_{\Lambda_c N D^{\ast}}$
characterizes the size of $1/m_c$ correction.
Interestingly, the results for $\Sigma_cND^{(*)}$  couplings
are in a better agreement with the heavy mass relation (\ref{HQET coupling 2}).

\end{itemize}

%%%%%%DISCUSSION ON THE LITERATURE
Concluding this section we make a few comments on
the earlier sum rule calculations of the heavy-baryon form factors
and couplings in the literature.

The $\Lambda_{b} \to p$ form factors were calculated
\cite{Huang:1998rq} in a different approach, using three-point QCD
sum rules (see also \cite{Marques de Carvalho:1999ia}), where
$\Lambda_{b}$ was interpolated by the pseudoscalar current and the
tensor interpolating current for the nucleon  was adopted. The
form factors predicted in \cite{Huang:1998rq} are in
agreement with our results. However, the other interpolating
currents, as well as the role of negative-parity partners of both
$\Lambda_{b}$ and proton remain obscure, bringing unaccounted
uncertainties in the numerical predictions.

The strong coupling $g_{\Lambda_c N D}$ was calculated from
three-point QCD sum rules in Ref. \cite{Navarra:1998vi}. In fact
this approach radically differs from the one we use here, first of
all, in the definition of the coupling itself. The starting point
is the correlation function with two baryon interpolating currents
and one pseudoscalar transition current. The latter is then simply
replaced by the $D$ meson. The relation of this definition of the
strong coupling to the one used here is difficult to assess.
Again, the problem of negative-parity baryons in both nucleon and
$\Lc$ channels was practically ignored absorbing these states into
the hadronic continuum.

Light-cone sum rules in HQET with $\Lambda_b$-distribution
amplitudes worked out in \cite{Ball:2008fw} was employed in
\cite{Wang:2009hra} to calculate   the $\Lambda_{b} \to p$ form
factors. The nucleon was interpolated by the CZ current $(u \, C
\! \not \! z \, u) \, \gamma_5 \not \! z \, d $ suggested in
\cite{Chernyak:1984bm}. The form factor $f_1(0)$ obtained in
\cite{Wang:2009hra} is about an order of magnitude smaller than
the one obtained here from LCSR with nucleon distribution
amplitudes. Note that the CZ current can also couple to $I=3/2$
and $J=3/2$ states so  that the sum rules for the $\Lambda_{b} \to
p$ transition form factors are probably influenced by large and
unaccounted $\Delta$- resonance  contribution.  It is known
\cite{BLW06} that, e.g., the isospin relations between nucleon
form factors are  violated when one uses LCSR  with CZ current.
Note that a similar current was employed also in
\cite{Huang:2004vf,Wang:2008sm}.

The $\Lambda_{b,c}\to N $ transitions were investigated in
\cite{Azizi:2009wn,Aliev:2010yx}  using LCSR with the nucleon DA's.
As opposed to our choice, the most general
interpolating current for the $\Lambda_{b,c}$ baryon was employed
introducing an arbitrary parameter $\beta$ for the mixing  of
different components in this current. The stability of the
calculated form factors with respect to the variation of $\beta$
was used as a criterion for choosing a working interval of
$\beta$. In our opinion, such a procedure introduces a sort of a
new systematic error related with the choice of the mixing
parameter. Most importantly, the problem of separating the
negative parity baryon contributions in LCSR remains unsolved,
because the latter are again attributed to the continuum estimated
with the usual quark-hadron duality ansatz. This may explain the
substantial difference of our predictions from the ones presented
in \cite{Azizi:2009wn}.

\section{ Applications to exclusive $\Lambda_b$ decays}

\begin{figure}[t]
\centering
\includegraphics[scale=0.6]{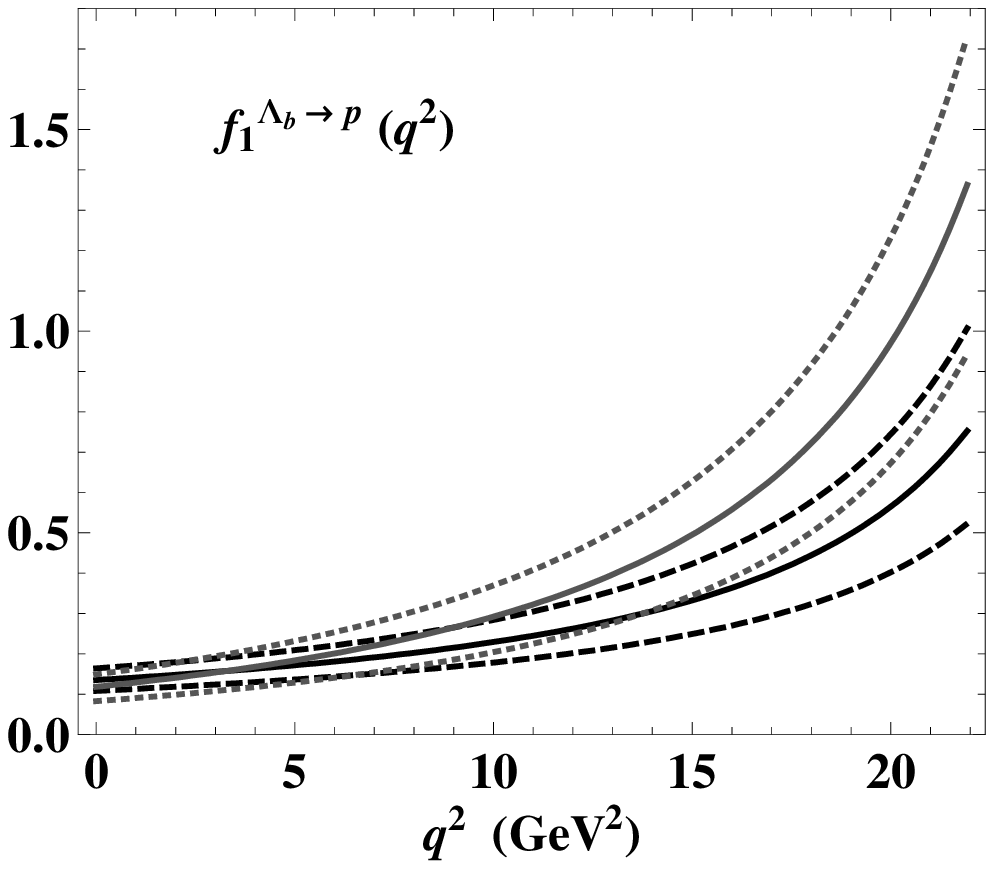}~~
\includegraphics[scale=0.6]{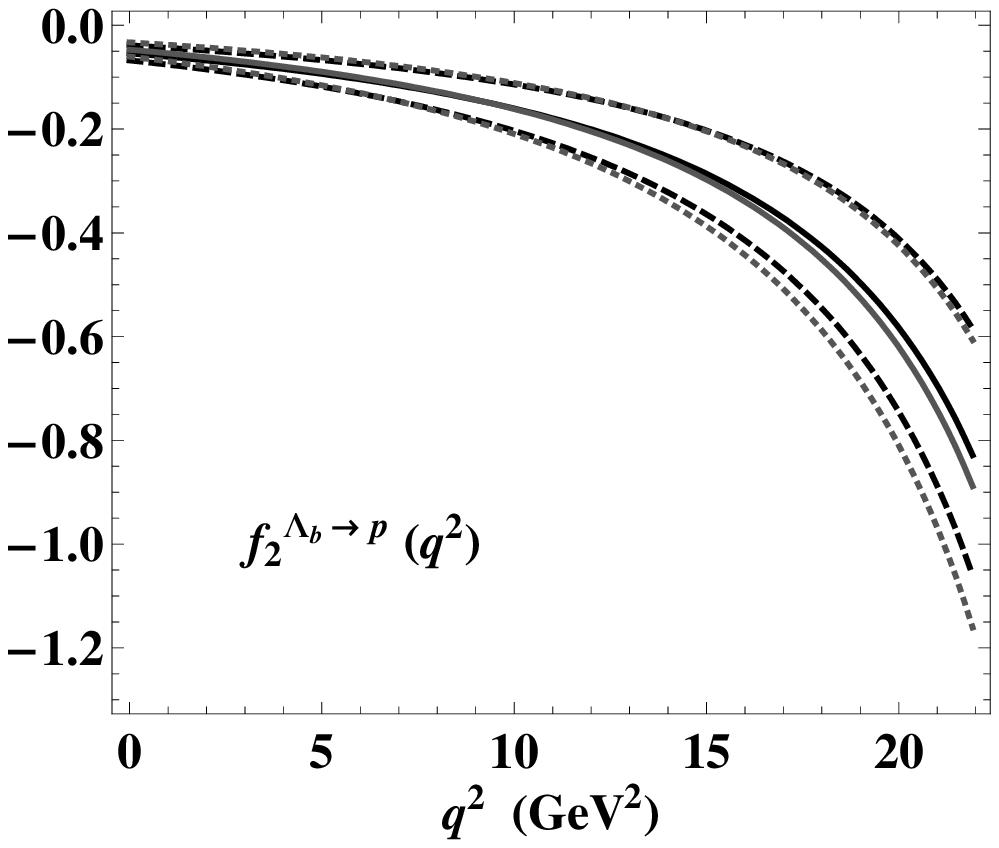}\\
\includegraphics[scale=0.6]{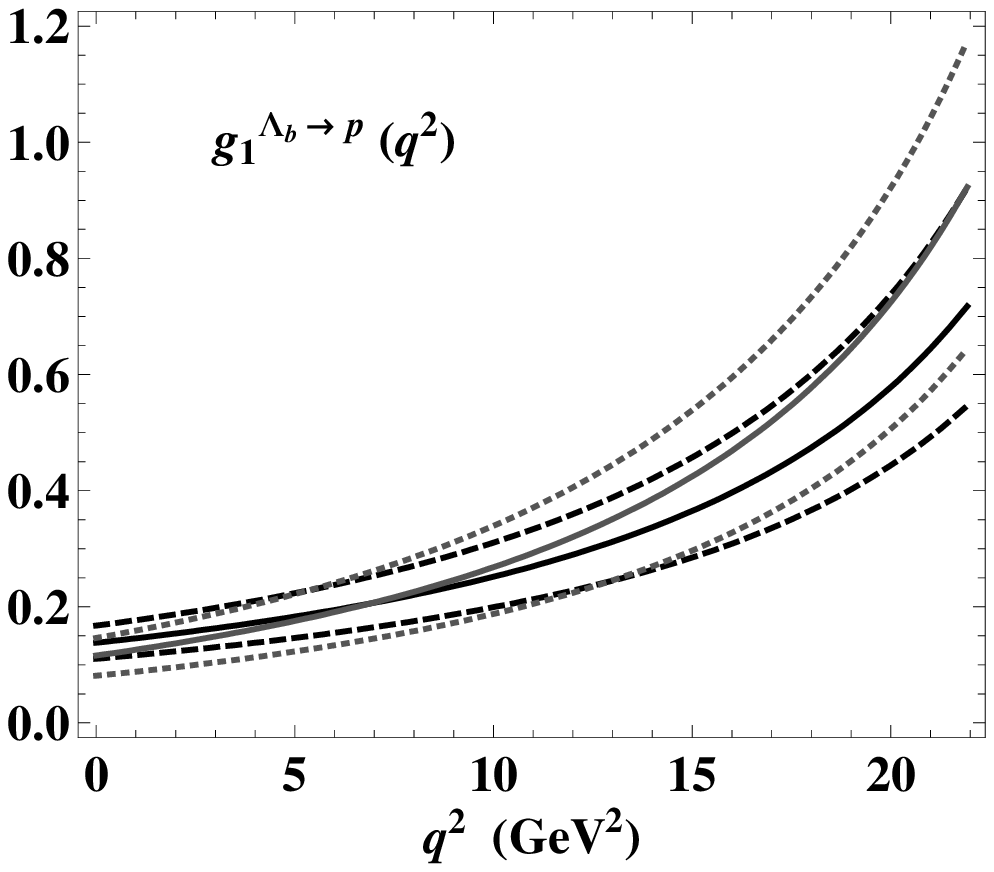}~~
\includegraphics[scale=0.6]{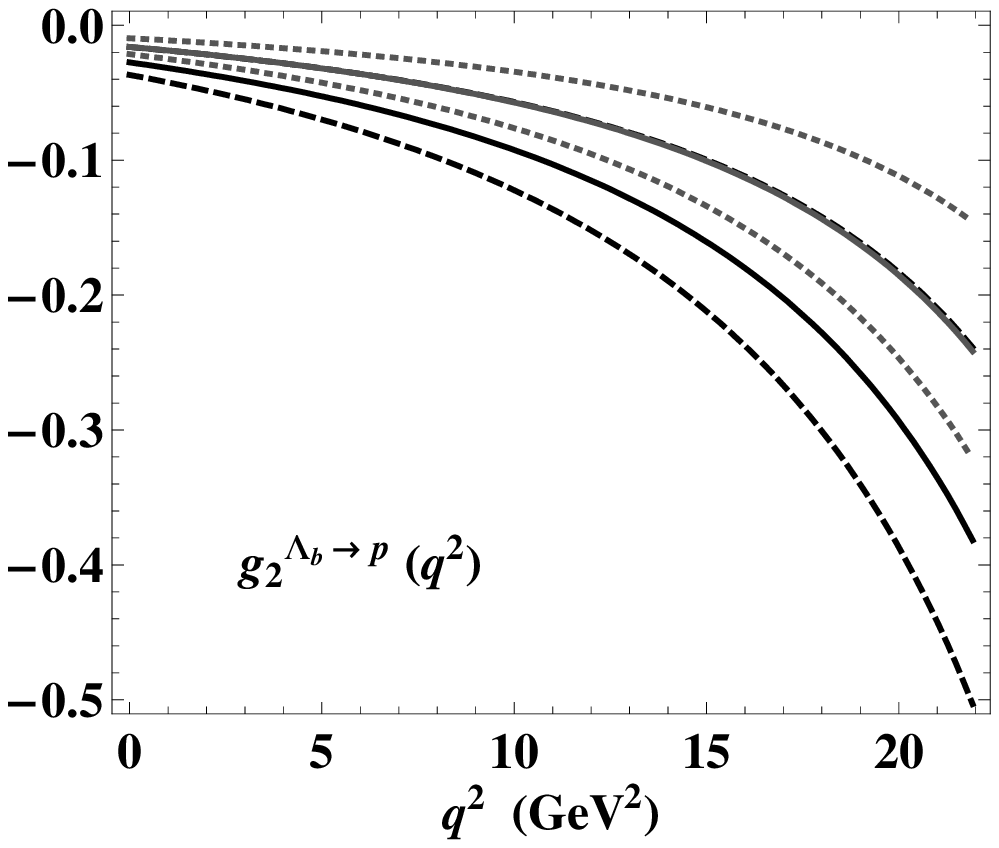}
\vspace{-0.3cm} \caption{\it  $\Lambda_b\to p$ transition form
factors obtained from LCSR at $q^2\leq 11 $ \mbox{GeV}\,$^2$ and
extrapolated to larger $q^2$  using
the series-parametrization: black (grey) solid lines correspond to the
axial-vector (pseudoscalar) interpolating current for $\Lambda_b$,
black long-dashed (grey short-dashed) lines indicate
uncertainties. } \label{fig:LbFF}
\end{figure}

With the results for the $\Lambda_b\to p$ form factors obtained
from LCSR we are now in a position to predict the differential
decay distribution for the exclusive semileptonic $\Lambda_b\to
p\ell\nu_{\ell}$ decay, which is a  $b\to u$ transition with
the CKM parameter $V_{ub}$. In the massless lepton approximation,
the form factors $f_3(q^2)$ and $g_3(q^2)$ do not contribute to
the decay width, hence it is sufficient to use the results for the
four form factors $f_{1,2}$ and $g_{1,2}$ given in the previous
section. Since the form factors are only available at $q^2\leq
q^2_{max}$, we apply the conformal mapping $q^2\to z$ and
$z$-series parametrization to extrapolate the form factors to the
whole semileptonic region $q^2\leq (m_{\Lambda_b}-m_{N})^2$.  More
specifically, we use the $z$-series parametrization in the
BCL-version suggested in \cite{BCL}. The mapping transformation
reads:
\begin{eqnarray}
z(q^2,t_0)=\frac{\sqrt{t_{+}-q^2}-\sqrt{t_{+}-t_0}}{\sqrt{t_{+}-q^2}+\sqrt{t_{+}-t_0}},
 \label{eq:z}
\end{eqnarray}
where $t_{\pm}=(m_{\Lambda_b}  \pm m_N)^2$, and
$t_0=t_{+}-\sqrt{t_{+}-t_{-}}\sqrt{t_{+}-t_{min}}$ is
chosen to maximally reduce the interval of
$z$ obtained after the mapping of the interval $t_{min}<q^2<t_-$,
where
$t_{min}=q_{min}^2<q^2<q^2_{max}$ is the LCSR validity  region. In the numerical
analysis, $t_{min}=-6 \, {\rm GeV}^2$ is adopted. Furthermore, we
employ the following parametrization
\begin{eqnarray}
f_{i}(q^2) &=& \frac{f_i(0)}{1-q^2/m_{B^*(1^-)}^2} \Bigg\{1+
b_i\,\Bigg(z(q^2,t_0)- z(0,t_0) \Bigg)\Bigg\} \,, \nonumber
\\
g_{i}(q^2) &=& \frac{g_i(0)}{1-q^2/m_{B^*(1^+)}^2} \Bigg\{1+
\tilde{b}_i\,\Bigg(z(q^2,t_0)- z(0,t_0) \Bigg)\Bigg\}  \,.
\label{eq:BCLff}
\end{eqnarray}
where $i=1,\, 2$, $m_{B^*(1^-)}=5.325 \, {\rm GeV}$ and
$m_{B^*(1^+)}=5.723 \, {\rm GeV}$ \cite{PDG}. Fitting the shape
parameters $b_i$ and $\tilde{b}_i$ to  LCSR predictions at
$q^2\leq q^2_{max}$, we obtained the $\Lambda_b \to p$ form
factors shown in Fig. \ref{fig:LbFF}. The shape parameters (see
Table \ref{tab_resLambdab}) turn out to be more sensitive to the
choice of the interpolating current, also the uncertainties of
these parameters (correlated with the uncertainties of the form
factor normalization) are larger. This however  plays role only at
large $q^2$, beyond the region of validity of LCSR.

%%FIG 2%%%%%%%%%%%%%%%%%%%%%%
%%%%%%%%%%%%%%%%%%%%%%%%%

\begin{figure}[ht]
\centering
\includegraphics[scale=0.8]{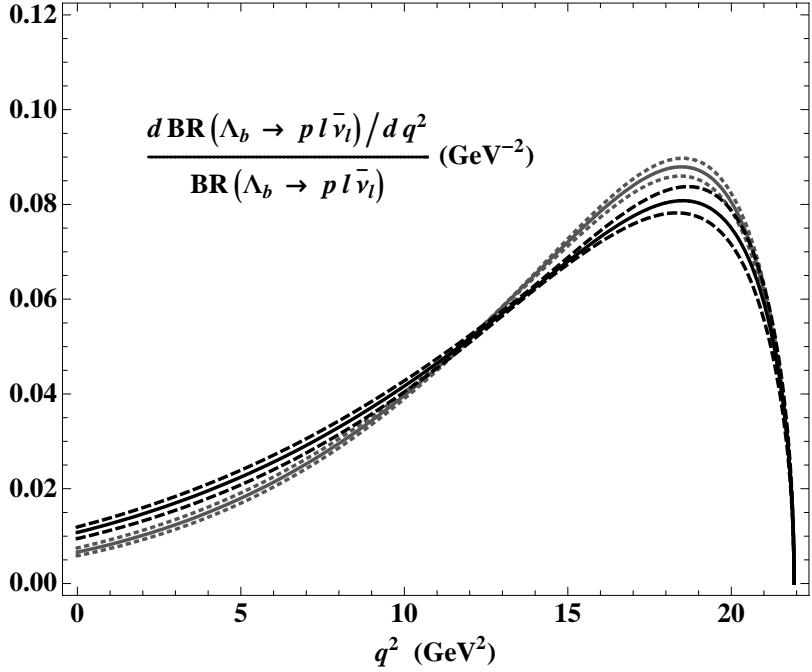}~~
\vspace{-0.3cm} \caption{\it Normalized differential width of
$\Lambda_b\to p \ell \nu_\ell$ calculated with the form factors
obtained from LCSR and extrapolated to larger $q^2$ using the
series-parametrization. The notation is the same as in Fig.
\ref{fig:LbFF}.} \label{fig:Lbdiff}
\end{figure}

Finally, we calculate the differential width of $\Lambda_b\to p \ell \nu$ decay
using the following expression.
\begin{eqnarray}
 \frac{d\Gamma}{dq^2}(\Lambda_b\to p l \nu_l) = {G_F^2
 m_{\Lambda_b}^3 \over 192 \pi^3 } |V_{ub}|^2 \lambda^{1/2}(1, r^2,
t)  \bigg \{ [(1-r)^2-t][(1+r)^2+2 t] |f_1(q^2)|^2 && \nonumber \\
  + [(1+r)^2-t][(1-r)^2+2 t] |g_1(q^2)|^2
-6 t [(1-r)^2 - t]  (1+r) f_1(q^2) f_2(q^2) && \nonumber \\
 -6 t [(1+r)^2 - t]  (1-r) g_1(q^2) g_2(q^2) + t [(1-r)^2-t] [2 (1+r)^2+t] |f_2(q^2)|^2  && \nonumber \\
  + t [(1+r)^2-t] [2 (1-r)^2+t] |g_2(q^2)|^2\bigg \}  \,, &&
\end{eqnarray}
where $r=m_N/m_{\Lambda_b}$, $t=q^2/m_{\Lambda_b}^2$ and
$\lambda(a,b,c)=a^2+b^2+c^2-2 ab -2 ac -2 bc$. Substituting the
form factors (\ref{eq:BCLff}) and integrating over
$q^2$ we obtain the total branching fraction
\begin{eqnarray}
BR (\Lambda_b\to p l \nu_l) =  \left\{
\begin{array}{l}
\left(3.3^{+1.5}_{-1.2} \big |_{th.} \pm 0.1\big |_{exp.}\right)\\
\\ \left(4.0^{+2.3}_{-2.0} \big |_{th.} \pm 0.1 \big
|_{exp.}\right) \,\,\,\,
\end{array}\right\}\left(\frac{|V_{ub}|}{3.5\cdot 10^{-3}}\right)^2 \times
10^{-4}\,,
\end{eqnarray}
where the upper (lower) interval corresponds to the form factors
obtained from LCSR with the axial-vector (pseudoscalar)
$\Lambda_b$-interpolating current, and the lifetime
$\tau_{\Lambda_b}=(1.391^{+0.038}_{-0.037}) ~\mbox{ps} $ from
\cite{PDG} is used. The normalized $q^2$ distribution is  plotted
in Fig.~\ref{fig:Lbdiff}. The enhancement in the region of large $q^2$
due to the growth of the form factors is in this case quite
pronounced because the width of this  decay  contains only the $S$-wave phase-space factor
$\lambda^{1/2}$, as opposed to
the $B \to \pi l \nu$ decay width, where there is a
$P$-wave factor $\lambda^{3/2}$.

Following our recent analysis of semileptonic $B \to \pi l \nu$ decay
\cite{Khodjamirian:2011ub}, we also calculate the specific integral
\begin{eqnarray}
\Delta \zeta (0, q^2_{max}) = {1 \over |V_{ub}|^2 }
\int_0^{q^2_{max}} \, d q^2 \,
\frac{d \Gamma}{dq^2}(\Lambda_b\to p l \nu_l) \,,
\end{eqnarray}
where the form factors directly calculated from LCSR are used,
independent of their parametrization and/or extrapolation. Our
prediction for the above integral
from LCSR with the axial-vector $\Lambda_b$- interpolating current
is :
\begin{eqnarray}
\Delta \zeta (0, 11 {\rm GeV^2})
&=&5.5^{+1.1}_{-0.9}\big|_{f_N}\,\,{}^{+0.5}_{-0.4}\big|_{\lambda_1}
\,\,{}^{+0.1}_{-0.0}\big|_{m_b}
\,\,{}^{+0.0}_{-0.1}\big|_{\mu}\,\,{}^{+0.5}_{-0.2}\big|_{M^2}
\,\,{}^{+0.5}_{-0.4}\big|_{s_0}
\,\,{}^{+0.4}_{-0.4}\big|_{M^2_{2pt}} \nonumber \\
&&  \,\,{}^{+0.1}_{-0.0}\big|_{s_0^{2pt}}
\,\,{}^{+2.1}_{-1.7}\big|_{\langle q \bar{q} \rangle}\,\, {\rm
ps^{-1}}
%\nonumber \\
=5.5^{+2.5}_{-2.0}\,\, {\rm ps^{-1}}    \,,
\end{eqnarray}
where also the uncertainties due to the variations
of separate input parameters  are shown (only those which are larger than
$O(1\%)$). The total error quoted above is obtained by adding all
separate uncertainties in quadrature.
A very close interval is obtained in the case of
the pseudoscalar $\Lambda_b$-interpolating current:
\begin{eqnarray}
\Delta \zeta (0, 11 {\rm GeV^2}) &=&
5.6^{+2.0}_{-1.7}\big|_{\lambda_1}
\,\,{}^{+2.3}_{-1.9}\big|_{\lambda_2}\,\,{}^{+0.0}_{-0.1}\big|_{m_b}
\,\,{}^{+0.2}_{-0.4}\big|_{\mu}\,\,{}^{+0.6}_{-0.2}\big|_{M^2}
\,\,{}^{+0.3}_{-0.3}\big|_{s_0}
\,\,{}^{+0.4}_{-1.2}\big|_{M^2_{2pt}} \nonumber \\
&&  \,\,{}^{+0.3}_{-0.3}\big|_{s_0^{2pt}}
\,\,{}^{+0.7}_{-1.2}\big|_{\langle q \bar{q} \rangle}\,\, {\rm
ps^{-1}}
%\nonumber \\
= 5.6^{+3.2}_{-2.9}\,\, {\rm ps^{-1}}    \,.
\end{eqnarray}
Note that the accuracy of this prediction is not yet competitive
with the one for $B\to \pi \ell \nu_l$ \cite{Khodjamirian:2011ub}.
Still the exclusive semileptonic decay of $\Lambda_b$  offers  a
possibility of $|V_{ub}|$ determination independent of the
$B$-meson semileptonic decays.

%%FIG 3%%%%%%%%%%%%%%%%%%%%%%
%%%%%%%%%%%%%%%%%%%%%%%%%

Another interesting possibility to use  the $\Lambda_b\to p$ form factors,
is to estimate the rate of the nonleptonic  $\Lambda_b \to p
\pi $ decay in the factorization approximation and to compare the
result with the recently
measured branching fraction. We write down the factorizable amplitude
in the same form as in \cite{Lu:2009cm}:
\begin{eqnarray}
\label{eq:fCNFpion} \mathcal{M}_{f}(\Lambda_b \to p \pi) ={G_F
\over \sqrt{2}} f_{\pi} \bar{u}_N \bigg  \{ \bigg
[V_{ub}V_{ud}^{\ast} a_1 -
V_{tb}V_{td}^{\ast}(a_4+a_{10}+R_1^{\pi} (a_6+a_8)) \bigg] &&
\nonumber \\ \bigg [f_1(m_{\pi}^2)(m_{\Lambda_b}-m_N)-
f_3(m_{\pi}^2)m_{\pi}^2\bigg] + \bigg [V_{ub}V_{ud}^{\ast} a_1 -
V_{tb}V_{td}^{\ast}(a_4+a_{10}-R_2^{\pi} (a_6+a_8)) \bigg] &&
\nonumber \\
 \bigg
[g_1(m_{\pi}^2)(m_{\Lambda_b}+m_N)- g_3(m_{\pi}^2)m_{\pi}^2\bigg]
\gamma_5 \bigg\} u_{\Lambda_b},  &&
 \label{fC-NF-pion}
\end{eqnarray}
where $a_i$ are the combinations  of Wilson coefficients of
the effective weak Hamiltonian and
\begin{eqnarray}
R_{1}^{\pi} = {2 m_{\pi}^2 \over (m_b - m_u) (m_u +m_d)}, \qquad
R_{2}^{\pi} = {2 m_{\pi}^2 \over (m_b + m_u) (m_u +m_d)}\,.
\end{eqnarray}
The pion  mass is neglected, hence we only use
$f_1(0)$ and $g_1(0)$ in the numerical analysis.
Our prediction for the branching fraction:
\begin{eqnarray}
{\rm BR}(\Lambda_b \to p \pi)\times 10^6 =  3.8^{+1.3}_{-1.0}
\left (2.8^{+1.1}_{-0.9}\right )\,,
\end{eqnarray}
obtained with the axial-vector (pseudoscalar) interpolating current is
consistent with the experimental measurement
\cite{Aaltonen:2008hg} ${\rm BR}(\Lambda_b \to p \pi)\times 10^6=
3.5 \pm 0.6 \pm 0.9$, indicating that nonfactorizable
contributions in this decay are at least moderate. This is in
agreement with the expectations based on QCD factorization for a
nonleptonic decay of a heavy hadron in two light hadrons. On the other
hand, our prediction is in a sharp contrast to the analysis of
$\Lambda_b\to p\pi$ done  in \cite{Lu:2009cm} where the
$k_{T}$-factorization approach was applied to the $\Lambda_b\to p$
form factors in the ``conventional PQCD" scenario, hence the soft
form factors were not taken into account. As a result the form
factors predicted in that approach are almost two orders of
magnitude smaller than the ones obtained from LCSR. On the
contrary, the nonfactorizable contributions in $\Lambda_b\to p\pi$
calculated in \cite{Lu:2009cm} applying the same PQCD approach,
are considerably enhanced, so that
their predicted branching fraction is also in the same ballpark as the
experimental one.

Further applications to semileptonic, radiative  and nonleptonic decays
of heavy baryons  demand also the $\Lambda_{c,b}\to \Lambda$ form factors.
In this respect our results can only be used in the $SU(3)_{flavour}$
approximation.

\section{Discussion}

In this paper, we considered the form factors and strong
couplings of heavy baryons. These non-perturbative quantities are
important inputs for a rich variety of phenomenological
applications, from  the $\Lambda_{b,c}$ (electro)weak decays
relevant for the flavour physics to the estimates of charmed hadron
production. The calculation we have done is based on the method of
QCD LCSR, where we use the light-cone DA's of the nucleon and interpolate
the heavy baryons by appropriate currents. These sum rules are
more complicated than the ones for the meson form factors and
strong couplings. First, the number of independent DA's to
be included in the OPE is much larger. These DA's are nevertheless
well under control due to twist and conformal-spin expansion, and the
resulting number of relevant input parameters is in fact not that
large. The other complication is more serious because little is
known about the background contributions of negative-parity
baryons in the hadronic dispersion relations. We proposed a novel
method to  eliminate these ``contamination''  by combining sum
rules obtained from different kinematical structures. As a
by-product, our results are less sensitive to the particular
choice of the interpolating currents.
Furthermore, as a step forward with respect to the procedure used
in \cite{BBKR} we worked out the double
spectral density for the power suppressed contributions in the
correlation functions where also the finite mass of the nucleon
is taken into account.
Our numerical results
for $\Lambda_c(\Sigma_c)\to p$ and $\Lambda_b\to p$ form factors
and $\Lambda_c(\Sigma_c)ND^{(*)}$ strong couplings
provide a nontrivial test of self-consistency of different sum
rules, and also of the reliability of our procedure of eliminating
the negative parity states.

We also obtained the predictions for semileptonic
and nonleptonic $\Lambda_b$ decays using the LCSR form factors.
In particular, the partially integrated width of $\Lambda_b\to p\ell \nu_l$
can be used for the $|V_{ub}|$ determination
from the future data on this decay.
Furthermore, in  a forthcoming publication \cite{pbarp2charm} we will
employ the LCSR results for the strong couplings of charmed baryons
to estimate their production in
proton-antiproton collisions in
the PANDA energy region.

%Future
Future improvements of the nucleon DA's, and the calculation of
gluon radiative corrections to the correlation functions will make
the sum rule results obtained here more accurate. A considerable
step forward will be to work out and incorporate in LCSR the
$\Lambda$-baryon  DA's. This
will make possible an accurate calculation of the form factors
needed in the FCNC decays such as $\Lambda_b \to \Lambda \gamma$
and $\Lambda_b \to \Lambda \ell^+ \ell^-$. Finally, a very
important task is to confront the LCSR predictions with the
relations for the heavy baryon form factors and couplings derived
from heavy-quark symmetries. In this paper we only made one step
in this direction, by comparing the LCSR predictions with the
relations for the form factors and strong couplings which follow
from the heavy-quark spin symmetry. A more elaborated study
including the quantitative analysis of HQET
relations for the baryonic form factors  based on LCSR
calculations will be presented elsewhere.

\section*{Acknowledgements}
This  work is supported  by the German research foundation DFG
under contract  MA1187/10-1 and by the German Ministry of Research
(BMBF), contract O6SI9192.

\appendix

%%%%APPA
\section{Nucleon Distribution Amplitudes}

The nucleon $\to$ vacuum matrix element of a three-quark
operator with light-like separations ($z^2 \to 0$)
has the following decomposition \cite{BFMS,NuclDAs09}:
\begin{eqnarray}
\label{eq:DecomNuclDAs}
\lefteqn{ 4 \bra{0} \epsilon^{ijk}
u_\alpha^i(a_1 z) u_\beta^j(a_2 z) d_\gamma^k(a_3 z) \ket{N(P)} }
\nonumber \\
&=& {\cal S}_1 m_N C_{\alpha \beta} \left(\gamma_5 u_N
\right)_\gamma + {\cal S}_2 m_N^2 C_{\alpha \beta}
\left(\!\not\!{z} \gamma_5 u_N \right)_\gamma + {\cal P}_1 m_N
\left(\gamma_5 C\right)_{\alpha \beta} {(u_N)}_\gamma \nonumber \\
&& + {\cal P}_2 m_N^2 \left(\gamma_5 C \right)_{\alpha \beta}
\left(\!\not\!{z} u_N \right)_\gamma
 + \left(\mathcal{V}_1+\frac{z^2m_N^2}{4}\mathcal{V}_1^M \right)
\left(\!\not\!{P}C \right)_{\alpha \beta} \left(\gamma_5 u_N
\right)_\gamma \nonumber \\
&& + {\cal V}_2 m_N \left(\!\not\!{P} C \right)_{\alpha \beta}
\left(\!\not\!{z} \gamma_5 u_N \right)_\gamma + {\cal V}_3 m_N
\left(\gamma_\mu C \right)_{\alpha \beta}\left(\gamma^{\mu}
\gamma_5 u_N \right)_\gamma
\nonumber \\
&& + {\cal V}_4 m_N^2 \left(\!\not\!{z}C \right)_{\alpha \beta}
\left(\gamma_5 u_N \right)_\gamma + {\cal V}_5 m_N^2
\left(\gamma_\mu C \right)_{\alpha \beta} \left(i \sigma^{\mu\nu}
z_\nu \gamma_5 u_N \right)_\gamma \nonumber \\
&& + {\cal V}_6 m_N^3 \left(\!\not\!{z} C \right)_{\alpha \beta}
\left(\!\not\!{z} \gamma_5 u_N \right)_\gamma  +
\left(\mathcal{A}_1+\frac{z^2m_N^2}{4}\mathcal{A}_1^M\right)
\left(\!\not\!{P}\gamma_5 C \right)_{\alpha \beta} {(u_N)}_\gamma
\nonumber \\
&& + {\cal A}_2 m_N \left(\!\not\!{P}\gamma_5 C \right)_{\alpha
\beta} \left(\!\not\!{z} u_N \right)_\gamma  + {\cal A}_3 m_N
\left(\gamma_\mu \gamma_5 C \right)_{\alpha \beta}\left(
\gamma^{\mu} u_N \right)_\gamma
\nonumber \\
&& + {\cal A}_4 m_N^2 \left(\!\not\!{z} \gamma_5 C \right)_{\alpha
\beta} {(u_N)}_\gamma + {\cal A}_5 m_N^2 \left(\gamma_\mu \gamma_5
C \right)_{\alpha \beta} \left(i \sigma^{\mu\nu} z_\nu u_N
\right)_\gamma \nonumber \\
&& + {\cal A}_6 m_N^3 \left(\!\not\!{z} \gamma_5 C \right)_{\alpha
\beta} \left(\!\not\!{z} u_N \right)_\gamma  +
\left(\mathcal{T}_1+\frac{z^2m_N^2}{4}\mathcal{T}_1^M\right)
\left(P^\nu i \sigma_{\mu\nu} C\right)_{\alpha \beta}
\left(\gamma^\mu\gamma_5 u_N \right)_\gamma \nonumber \\
&& + {\cal T}_2 m_N \left(z^\mu P^\nu i \sigma_{\mu\nu}
C\right)_{\alpha \beta} \left(\gamma_5 u_N \right)_\gamma + {\cal
T}_3 m_N \left(\sigma_{\mu\nu} C\right)_{\alpha \beta}
\left(\sigma^{\mu\nu}\gamma_5 u_N \right)_\gamma \nonumber \\
&& + {\cal T}_4 m_N \left(P^\nu \sigma_{\mu\nu} C\right)_{\alpha
\beta} \left(\sigma^{\mu\rho} z_\rho \gamma_5 u_N \right)_\gamma +
{\cal T}_5 m_N^2 \left(z^\nu i \sigma_{\mu\nu} C\right)_{\alpha
\beta} \left(\gamma^\mu\gamma_5 u_N \right)_\gamma \nonumber \\
&& + {\cal T}_6 m_N^2 \left(z^\mu P^\nu i \sigma_{\mu\nu}
C\right)_{\alpha \beta} \left(\!\not\!{z} \gamma_5 u_N
\right)_\gamma + {\cal T}_{7} m_N^2 \left(\sigma_{\mu\nu}
C\right)_{\alpha \beta} \left(\sigma^{\mu\nu} \!\not\!{z} \gamma_5
u_N \right)_\gamma \nonumber \\
&& + {\cal T}_{8} m_N^3 \left(z^\nu \sigma_{\mu\nu}
C\right)_{\alpha \beta} \left(\sigma^{\mu\rho} z_\rho \gamma_5 u_N
\right)_\gamma \,,
\end{eqnarray}
including the $O(z^2)$ corrections to the lowest twist-3 part of
this decomposition. In the above, the gauge links maintaining
gauge invariance are not shown for brevity.  The calligraphic
coefficients $\mathcal{S}_{1,2}$, $\mathcal{P}_{1,2}$,
$\mathcal{V}_{1,...,6}$, $\mathcal{A}_{1,...,6}$,
$\mathcal{T}_{1,...,8}$, $\mathcal{V}_1^M, \mathcal{A}_1^M,
\mathcal{T}_1^M $ at different Dirac structures, are related to
the integrals containing  the nucleon DA's depending on the
longitudinal momentum fractions $x_i$.  The  general relation
reads
\begin{equation}
\mathcal{F}(a_1,a_2,a_3,(P\cdot z))= \int \! d x_1 d x_2 d x_3
 \delta(1\!-\!x_1\!-\!x_2\!-\!x_3)e^{-i
(P  \cdot  z) \sum_i x_i a_i} F(x_i)\,,
\label{eq:F1}
\end{equation}
where our choice of the three-quark configuration on the
light-cone corresponds to $a_{1,3}=0,~a_2=1$. The function
$\mathcal{F}$  and the integrand on r.h.s. for all coefficients
in (\ref{eq:DecomNuclDAs}) are  collected in Table \ref{Details of
functions F}.
%%%%%%%%%%%%%%%%%%%%%%%%%%%%%%%%
\begin{table}[h]
\begin{center}
\begin{tabular}{|c|c||c|c|}
\hline
\hline
 $\mathcal{F}$  & integrand on r.h.s. of (\ref{eq:F1})&  $\mathcal{F}$   & integrand on r.h.s. of (\ref{eq:F1}) \\
\hline ${\cal S}_1$ & $S_1$    & $2 (P \! \cdot \! z) \, {\cal
S}_2$ &
$S_1-S_2 $\\
\hline
 ${\cal P}_1$ & $P_1$ & $2 (P \! \cdot \! z) \, {\cal P}_2$ &  $P_2-P_1 $ \\
 \hline
 ${\cal V}_1$ & $ V_1$ & $2 (P \! \cdot
\! z) \, {\cal V}_2 $ &  $V_1 - V_2 - V_3$\\
 \hline
 $2 {\cal V}_3$ & $V_3$ & $4 (P \! \cdot \! z)
\, {\cal V}_4$ & $- 2 V_1 + V_3 + V_4  + 2 V_5$ \\
 \hline
 $4 (P \! \cdot \! z) {\cal V}_5$ & $V_4 - V_3$ & $4 \left(P \! \cdot \! z\right)^2  {\cal V}_6$ &  $- V_1 + V_2 +
V_3 +  V_4 + V_5 - V_6$ \\
 \hline
 $ {\cal A}_1$ & $A_1$ & $2 (P \! \cdot \! z) {\cal A}_2 $ &  $- A_1 + A_2 -  A_3$ \\
 \hline
 $ 2 {\cal A}_3$ & $A_3$ & $4 (P \! \cdot \! z){\cal A}_4 $&  $- 2 A_1 - A_3 - A_4  + 2 A_5$ \\
 \hline
 $4 (P \! \cdot \! z) {\cal A}_5$ & $A_3 - A_4$ & $4 \left(P \! \cdot \! z\right)^2  {\cal A}_6$ & $A_1 - A_2 +  A_3 +  A_4 - A_5 + A_6$ \\
 \hline
 ${\cal T}_1 $ & $T_1$ &  $ 2 (P \! \cdot \! z) {\cal T}_2$ &  $T_1 + T_2 - 2 T_3$\\
 \hline
  $2 {\cal T}_3$& $T_7$ & $2 (P \! \cdot \! z){\cal T}_4 $ &  $T_1 - T_2 - 2  T_7$ \\
 \hline
 $ 2 (P \! \cdot \! z) {\cal T}_5$ & $- T_1 + T_5 + 2  T_8$ & $4 \left(P \! \cdot \! z\right)^2 {\cal T}_6$ & $ 2 T_2 - 2 T_3 - 2 T_4 + 2 T_5 + 2 T_7 + 2 T_8$ \\
 \hline
 $ 4 (P \! \cdot \! z) {\cal T}_7 $ & $ T_7 - T_8$ & $4 \left(P \! \cdot \! z\right)^2 {\cal T}_8$ &  $-T_1 + T_2 + T_5 - T_6 + 2 T_7 + 2 T_8 $ \\
 \hline
  $\mathcal{V}_1^M$ & $V_1^M$ & $\mathcal{A}_1^M$ & $A_1^M$ \\
 \hline
   $\mathcal{T}_1^M$ & $T_1^M$ & &  \\
 \hline
 \hline
\end{tabular}
\end{center}
\caption{ Terms of the decomposition (\ref{eq:DecomNuclDAs}) and
their relation to the nucleon DA's via Eq.(\ref{eq:F1}). }
\label{Details of functions F}
\end{table}
%%%%%%%%%%%%%%%%
The resulting decomposition contains altogether 27 DA's. We use
their expressions obtained in \cite{BLMS,NuclDAs09} to the
next-to-leading order in the conformal spin expansion. The three
twist-3 DA's $V_1,A_1,T_1$  are presented in eq. (\ref{eq:tw3DA})
and the others are:
\begin{itemize}

\item twist-4 DA's:
\ba
 V_2(x_i)= 24 x_1 x_2 [\phi_4^0 + \phi_4^{+} (1- 5 x_3)]\,,~~
  A_2(x_i) = 24 x_1 x_2 (x_2 -x_1) \phi_4^{-}\,, \nonumber\\
  T_2(x_i) =24 x_1 x_2 [ \xi_4^{0} + \xi_4^{+} (1- 5 x_3) ]\,, \nonumber\\
%%%%%%%%%%%%%%%%%%%
V_3(x_i)=12 x_3 [\psi_4^0 (1-x_3) + \psi_4^{+}(1-x_3 -10 x_1
x_2)+ \psi_4^{-}(x_1^2+x_2^2-x_3 (1-x_3)) ] \,,\nonumber\\
%%%%%%%%%%%%%%%
 A_3(x_i)=12 x_3 (x_2-x_1) [ (\psi_4^0+\psi_4^+) +  \psi_4^{-} (1-2 x_3)]\,,\nonumber
\ea
\ba
%%%%%%%%%%%%%%%%%
T_3(x_i)=6 x_3  [(\phi_4^0 + \psi_4^0 + \xi_4^0 )(1-x_3)
+(\phi_4^+ + \psi_4^+ + \xi_4^+ ) (1-x_3 -10 x_1 x_2) \nonumber\\
 + (\phi_4^- - \psi_4^-+ \xi_4^- ) (x_1^2 + x_2^2 - x_3 (1-x_3))  ] \,,\nonumber\\
 %%%%%%%%%%%%%%%%%
 T_7(x_i)= 6 x_3 [ (\phi_4^0 + \psi_4^0 - \xi_4^0 )(1-x_3)
 +(\phi_4^+ + \psi_4^+ - \xi_4^+ ) (1-x_3 -10 x_1 x_2) ] \nonumber\\
 + (\phi_4^- - \psi_4^- - \xi_4^- ) (x_1^2 + x_2^2 - x_3 (1-x_3))  ]\,, \nonumber\\
%%%%%%%%%%%%%%
 S_1(x_i)=6 x_3 (x_2 -x_1) [( \phi_4^0 + \psi_4^0 + \xi_4^0 +\phi_4^+ + \psi_4^+ + \xi_4^+ )
 + (\phi_4^- - \psi_4^- + \xi_4^- ) (1- 2 x_3)] \,,\nonumber\\
%%%%%%%%%%%%%
 P_1(x_i)= 6 x_3 (x_1 -x_2) [( \phi_4^0 + \psi_4^0 - \xi_4^0 +\phi_4^+ + \psi_4^+ - \xi_4^+ )
 + (\phi_4^- - \psi_4^- - \xi_4^- ) (1- 2 x_3)]\,,\nonumber\\
\label{eq:tw3DAs}
\ea

\item twist-5 DA's:
\ba
 V_4(x_i)= 3 [\psi_5^0 (1-x_3) + \psi_5^{+} (1-  x_3 - 2 (x_1^2 +x_2^2)) + \psi_5^{-} (2 x_1 x_2 -x_3 (1-x_3)) ] \,,\nonumber\\
%%%%%%%%%%%%%%
A_4(x_i)=3  (x_2 -x_1) [ -\psi_5^{0} + \psi_5^{+}(1- 2 x_3) + \psi_5^{-} x_3] \,,\nonumber
\ea
%%%%%%%%%%%%
\ba
 T_4(x_i)=\frac{3}{2} [ (\phi_5^{0} + \psi_5^{0} + \xi_5^0 ) (1 - x_3) + (\phi_5^{+} + \psi_5^{+} + \xi_5^{+} )  (1-  x_3 - 2 (x_1^2 + x_2^2))]\nonumber\\
 + (\phi_5^{-} - \psi_5^{-} + \xi_5^{-} )   (2 x_1 x_2 - x_3 (1- x_3)) \,, \nonumber\\
%%%%%%%%%%%%%%%%%%%%%
T_8(x_i)=\frac{3}{2}  [ ( \phi_5^0 + \psi_5^{0} - \xi_5^0 )
(1-x_3) + (\phi_5^{+} +\psi_5^{+}-\xi_5^+ )(1-x_3 -2 (x_1^2 + x_2^2) ) \nonumber\\
+ (\phi_5^{-} - \psi_5^{-} + \xi_5^{-} )  (2 x_1 x_2 - x_3 (1-
x_3)) ]  \,, \nonumber \ea
 %%%%%%%%%%%%%%%
\ba
V_5(x_i)= 6 x_3 [\phi_5^0 + \phi_5^+  (1-2 x_3)  ]\,, ~~
  %%%%%%%%%%%%%%
A_5(x_i)= 6 x_3 (x_2 - x_1) \phi_5^- \,,\nonumber \\
 %%%%%%%%%%%%
 T_5(x_i)=6 x_3 [  \xi_5^0 + \xi_5^+  (1- 2 x_3) ] \,, \nonumber\\
 %%%%%%%%%%%%
 S_2(x_i)= \frac{3}{2} (x_2 -x_1) [-( \phi_5^0 + \psi_5^0 + \xi_5^0 )
 + (\phi_5^{+}+ \psi_5^{+} + \xi_5^{+} ) (1- 2 x_3) \nonumber \\
 + (\phi_5^{-} - \psi_5^{-} + \xi_5^{-} ) x_3] \,, \nonumber \\
 %%%%%%%%%%%%
 P_2(x_i)= \frac{3}{2} (x_1 -x_2) [-( \phi_5^0 + \psi_5^0 - \xi_5^0 )
 + (\phi_5^{+} + \psi_5^{+} - \xi_5^{+} ) (1- 2 x_3)   \nonumber \\
 + (\phi_5^{-} - \psi_5^{-} - \xi_5^{-} ) x_3] \,, \nonumber
 \ea

\item twist-6 DA's:

\ba
 V_6(x_i)=2 [  \phi_6^0  + \phi_6^+ (1- 3 x_3) ] \,,~~
 %%%%%%%%%%%
 A_6(x_i)=2 (x_2-x_1)\phi_6^- \,,\nonumber \\
%%%%%
 T_6(x_i)=2 [ \phi_6^0  - {1 \over 2}(\phi_6^+-\phi_6^-) (1- 3 x_3)\,.
\ea
\end{itemize}
The expressions for the remaining  three DA's
($V_1^M,~A_1^M,~T_1^M$) determining  the $z^2$-corrections are
shown below.

Furthermore, following Ref. \cite{BLW06}, shorthand notations for  the
combinations of nucleon DA's are used:

\begin{eqnarray}
   S_{12} = S_1 - S_2\,, & \qquad  &   P_{21} = P_2 - P_1 \,,
\nonumber\\[7pt]
 V_{1345} = -2V_1+V_3+V_4+2 V_5\,, & \qquad  &  V_{43} = V_4-V_3\,, \nonumber \\
 V_{123456} =
-V_1+V_2+V_3+V_4+V_5-V_6\,, & \qquad & V_{123} = V_1-V_2-V_3
\nonumber\\[7pt]
  A_{1345} = -2A_1-A_3-A_4+2 A_5\,, & \qquad  &  A_{34} = A_3-A_4 \nonumber \\
A_{123456} = A_1-A_2+A_3+A_4-A_5+A_6\,, & \qquad & A_{123} =
-A_1+A_2-A_3
\nonumber\\[7pt]
 T_{78}   = T_7 - T_8\,, & \qquad  & T_{123}   = T_1 + T_2 - 2T_3
\,,
\nonumber\\
 T_{234578}   = 2 T_2 - 2 T_3 - 2 T_4 + 2 T_5 + 2 T_7 + 2 T_8 \,,
& \qquad  & T_{127}   = T_1 - T_2 - 2 T_7\,, \nonumber \\
 T_{125678}   = -T_1 + T_2 + T_5 - T_6 + 2
T_7 + 2 T_8 \,, & \qquad & T_{158}   = -T_1 + T_5 + 2 T_8\,.
\end{eqnarray}
In addition,  the following notations are introduced for the integrals:
\begin{eqnarray}
\tilde{F}(x_2) &=& \int_0^{1-x_2} d x_1 F(x_1, x_2, 1-x_1-x_2) \,, \nonumber \\
\tilde{\tilde{F}}(x_2) &=& \int_1^{x_2} d x_2^{\prime}
\int_0^{1-x_2^{\prime}} d x_1 F(x_1, x_2^{\prime},
1-x_1-x_2^{\prime}) \,, \nonumber \\
\tilde{\tilde{\tilde{F}}}(x_2) &=& \int_1^{x_2} d x_2^{\prime}
\int_1^{x_2^{\prime}} d x_2^{\prime \prime} \int_0^{1-x_2^{\prime
\prime}} d x_1 F(x_1, x_2^{\prime \prime}, 1-x_1-x_2^{\prime
\prime})
\end{eqnarray}
where $F$ is one of the DA's.

%%%%%%%%%%%%%%%
Finally, the DA's originating from $z^2$ corrections
enter in the integrated form
denoted as $\tilde{V}_1^M(x_2)$, $\tilde{A}_1^M(x_2)$ and
$\tilde{T}_1^M(x_2)$, where

\begin{eqnarray}
\tilde{V}_1^M(x_2) &=& \frac{x_2^2}{24} [f_N C_f^u(x_2) +
\lambda_1 C_{\lambda}^u(x_2) ] \,, \nonumber \\
\tilde{A}_1^M(x_2) &=& \frac{x_2^2}{24} (1-x_2)^3 [f_N
D_f^u(x_2) + \lambda_1 D_{\lambda}^u(x_2) ] \,, \nonumber \\
\tilde{T}_1^M(x_2) &=& \frac{x_2^2}{48}  [f_N E_f^u(x_2) +
\lambda_1 E_{\lambda}^u(x_2) ] \,,
\end{eqnarray}
with
\begin{eqnarray}
C_f^u(x_2)&=& (1-x_2)^3 [113 + 495 x_2 - 552 x_2^2 -10 A_1^u (1- 3
x_2) \nonumber \\
&& + 2 V_1^d  (113 -951 x_2 +828 x_2^2) ] \,, \nonumber \\
C_{\lambda}^u(x_2)&=& -(1-x_2)^3 [13 -20 f_1^d + 3 x_2 + 10 f_1^u
(1- 3 x_2)) ] \,, \nonumber \\
D_f^u(x_2)&=& 11 + 45 x_2 - 2 A_1^u (113 -951 x_2 + 828 x_2^2) + 10 V_1^d (1-30 x_2)  \,, \nonumber \\
D_{\lambda}^u(x_2)&=& 29- 45 x_2 -10 f_1^u (7-9 x_2) -20 f_1^d (5 - 6 x_2) \,, \nonumber
\end{eqnarray}
\begin{eqnarray}
E_f^u(x_2)&=& -[(1-x_2)(3 (439+ 71 x_2 -621 x_2^2  + 587 x_2^3 - 184 x_2^4 ) \nonumber \\
&& + 4 A_1^u (1-x_2)^2 (59 - 483 x_2 +414 x_2^2)\nonumber \\
&&  - 4 V_1^d (1301-619 x_2 -769 x_2^2 +1161 x_2^3 -414 x_2^4) )  ] \nonumber \\
&&-12 (73 -220 V_1^d) \ln x_2 \,, \nonumber \\
E_{\lambda}^u(x_2)&=& -[(1-x_2) (5 - 211 x_2 + 281 x_2^2 -111
x_2^3 \nonumber \\
&& +10 (1+61 x_2 -83 x_2^2 + 33 x_2^3) f_1^d \nonumber \\
&& -40 (1-x_2)^2 (2-3 x_2) f_1^u) ] -12 (3 -10 f_1^d) \ln x_2 \,.
\label{log dependence}
\end{eqnarray}
%%%%%%%%%%%%%

The terms proportional to $\ln x_2$ in the above DA
$\tilde{T}_1^M(x_2)$  are the only non-polynomial ones in the
whole OPE expressions. In the case of LCSR for the form factors
their transformation to a dispersion form is straightforward.
However, for the strong coupling sum rules a double dispersion
form of such nonpolynomial terms demands a separate derivation. In
fact, these terms turn out to have negligible coefficients with
our choice of the DA parameters $V_1^d$ and $f_1^d$ and hence are
simply neglected.

The coefficients $\phi_i^{(\pm,0)}$, $\psi_i^{(\pm,0)}$ and
$\xi_i^{(\pm,0)}$ ($i=3,4,5,6$)  determining the normalization
and shape of DA's  can be expressed through  the eight independent
parameters listed in (\ref{normalization DA}) and
(\ref{eq:shapepar}). The corresponding relations
for the leading conformal spin in DA's are
\begin{eqnarray}
\phi_3^0=\phi_6^0=f_N \,, & \qquad & \phi_4^0=\phi_4^0=\frac{1}{2}
(f_N + \lambda_1)\,, \nonumber \\
\xi_4^0 =\xi_5^0 = \frac{1}{6} \lambda_2\,, & \qquad &
\psi_4^0=\psi_5^0=\frac{1}{2} (f_N - \lambda_1)\,.
\end{eqnarray}
For the next-to-leading conformal spin,

\begin{itemize}

\item in twist-3 DA's:
\begin{eqnarray}
\phi_3^-=\frac{21}{2}f_N A_1^u \,, & \qquad &
\phi_3^+=\frac{7}{2}f_N (1- 3 V_1^d) \,,
\end{eqnarray}

\item in twist-4 DA's:

\begin{eqnarray}
 \phi_4^+ &=& \frac{1}{4} [f_N  (3- 10 V_1^d) + \lambda_1 (3- 10 f_1^d)] \,,
 \nonumber  \\
  \phi_4^- &=& - \frac{5}{4} [f_N  (1- 2 A_1^u) - \lambda_1 (1- 2 f_1^d - 4 f_1^u)] \,,
\nonumber \\
 \psi_4^+ &=& - \frac{1}{4} [f_N  (2+ 5 A_1^u - 5 V_1^d) - \lambda_1 (2- 5 f_1^d- 5 f_1^u)]
\nonumber \\
 \psi_4^- &=& \frac{5}{4} [f_N  (2 -  A_1^u - 3 V_1^d) - \lambda_1 (2- 7 f_1^d +
 f_1^u)] \nonumber \\
 \xi_4^+ &=& \frac{1}{16} \lambda_2 (4 -15 f_2^d) \,, \qquad   \xi_4^- = \frac{5}{16}
\lambda_2 (4 -15 f_2^d)
 \,,
\end{eqnarray}
\item in twist-5 DA's:

\begin{eqnarray}
\phi_5^+ &=& - \frac{5}{6} [f_N  (3+4  V_1^d) - \lambda_1 (1- 4
f_1^d)]\,,  \nonumber \\
\phi_5^- &=& - \frac{5}{3} [f_N  (1- 2 A_1^u) - \lambda_1 ( f_1^d - f_1^u)] \,,\nonumber \\
\psi_5^+ &=& - \frac{5}{6} [f_N  (5+ 2 A_1^u - 2 V_1^d) -
\lambda_1(1-2 f_1^d- 2 f_1^u)] \,,  \nonumber \\
\psi_5^- &=&  \frac{5}{3} [f_N  (2-  A_1^u - 3 V_1^d) + \lambda_1 ( f_1^d- f_1^u)]\,,
\nonumber \\
\xi_5^+ &=& \frac{5}{36} \lambda_2 (2 - 9 f_2^d) \,,   \qquad
\xi_5^- = - \frac{5}{4} \lambda_2 f_2^d \,,
\end{eqnarray}

\item in twist-6 DA's:

\begin{eqnarray}
 \phi_6^+ &=& \frac{1}{2} [f_N  (1-4  V_1^d) - \lambda_1 (1- 2 f_1^d)]\,,
 \nonumber \\
  \phi_6^- &=& \frac{1}{2} [f_N  (1+4  A_1^u) + \lambda_1 (1- 4 f_1^d - 2
  f_1^u)]\,.
\end{eqnarray}

\end{itemize}

%%%%%%%%%%%%%%%%END APP A

\section{ Correlation function in the LCSR for $\Lambda_c \to N $
form factors } \label{lambda_c N QCD representation}

\subsection{ pseudoscalar transition current}
\label{lambda_c N QCD representation for pseudoscalar transition}

The invariant amplitudes  $\Pi^{(i)}_j ((P-q)^2,q^2)$ of the
correlation function with the pseudoscalar transition current
$j_5$ are given in (\ref{eq:QCDcorr}),  where the coefficient
functions $\tilde{\omega}^{(i)}_{jn}$ with $i={\cal P},{\cal A}$,
$j=1,2$ and $n=1,2,3$, (after replacement of $(P-q)^2$ described
in Sect.~4) are listed below for:

\begin{itemize}
\item the pseudoscalar interpolating current
\begin{eqnarray}
 \omega^{(\cal P)}_{11} &=& \frac{m_N}{2} \Big[ (m_c-x\,m_N) \Phi_1^{(\cal P)} - m_N \Phi_2^{(\cal P)} \Big] \,,\nonumber\\
 \omega^{(\cal P)}_{12}  &=& -\frac{m_N^2}{2} \Big[m_c\Big(m_c-x \,m_N\Big) \Phi_2^{(\cal P)} + 2x m_N^2 \Phi_3^{(\cal P)} \Big]\,, \nonumber\\
\omega^{(\cal P)}_{13}   &=& 2 m_N^3 m_c^2 \big(m_c-x \,m_N\big)
\Phi_3^{(\cal P)}, \nonumber
\end{eqnarray}

\begin{eqnarray}
\omega^{(\cal P)}_{21} &=& -\frac{m_N}{2} \Phi_1^{(\cal P)} \,,
\qquad
 \omega^{(\cal P)}_{22} = \frac{m_N^2}{2}\Big( m_c \Phi_2^{(\cal P)} -
2 m_N \Phi_3^{(\cal P)} \Big)\,,
\nonumber\\
 \omega^{(\cal P)}_{23} &=& - 2 m_N^3 m_c^2 \Phi_3^{(\cal P)} \,,
\end{eqnarray}

where the functions $\Phi_i^{(\cal P)}$ in the above equations are
\begin{eqnarray}
 \Phi_1^{(\cal P)} &=& 2 \tilde{A}_1+4 \tilde{A}_3+2 \tilde{A}_{123}+2 \tilde{P}_1+2 \tilde{S}_1
 +6 \tilde{T}_1-12 \tilde{T}_7-\tilde{T}_{123}-5 \tilde{T}_{127}\nonumber \\
   &&-2 \tilde{V}_1+4 \tilde{V}_3+2
   \tilde{V}_{123} \,, \nonumber\\
 \Phi_2^{(\cal P)} &=& 3 \tilde{\tilde{A}}_{34}+2 \tilde{\tilde{A}}_{123}-\tilde{\tilde{A}}_{1345}
 -2 \tilde{\tilde{P}}_{21}+2 \tilde{\tilde{S}}_{12}-12 \tilde{\tilde{T}}_{78}-2
   \tilde{\tilde{T}}_{123}-4 \tilde{\tilde{T}}_{127}\nonumber \\
   &&-6 \tilde{\tilde{T}}_{158}
   +\tilde{\tilde{T}}_{234578} -3 \tilde{\tilde{V}}_{43}+2
   \tilde{\tilde{V}}_{123}+\tilde{\tilde{V}}_{1345} \,, \nonumber\\
 \Phi_3^{(\cal P)} &=& -\tilde{A}_1^M-3 \tilde{T}_1^M+\tilde{V}_1^M+\tilde{\tilde{\tilde{A}}}_{123456}-3
   \tilde{\tilde{\tilde{T}}}_{125678}+\tilde{\tilde{\tilde{T}}}_{234578}+\tilde{\tilde{\tilde{V}}}_{123456}
   \,; \nonumber
\end{eqnarray}

\item the axial-vector interpolating current

\begin{eqnarray}
\omega^{(\cal A)}_{11}&=& 2{ m_c^2 -q^2 \over x} \Phi_1^{(\cal A)}
+ x \,m_N^2 \Big[2\Phi_1^{(\cal A)}+\Phi_2^{(\cal
A)}\Big] + m_N m_c \Phi_3^{(\cal A)}  + m_N^2\Big[\Phi_4^{(\cal A)}+\frac{2\Phi_5^{(\cal A)}}{x}\Big] \,, \nonumber\\
 \omega^{(\cal A)}_{12}&=& -m_N^2 \bigg[2(x^2\,m_N^2-q^2) \Phi_6^{(\cal A)} +
x\,m_N m_c \Phi_7^{(\cal A)} + m_c^2 \Phi_8^{(\cal A)} -
2\frac{q^2+m_c^2}{x} \Phi_5^{(\cal A)}\nonumber \\
&& + 2 x \,m_N^2 \Phi_9^{(\cal A)} \bigg] \,,  \nonumber\\
\omega^{(\cal A)}_{13} &=& 4 \frac{m_N^2}{x} \bigg[ m_c^2
(q^2-m_c^2) \Phi_5^{(\cal A)} - x^2 m_N^2 m_c^2 \Phi_9 ^{(\cal
A)}+ x m_N m_c^3 \Phi_{10}^{(\cal A)} \bigg]\,, \nonumber
\end{eqnarray}

\begin{eqnarray}
 \omega^{(\cal A)}_{21} &=& \frac{m_N}{x} \bigg[2\Phi_6^{(\cal A)} + x \,\Phi_2^{(\cal A)} \bigg] \,, \qquad
 \omega^{(\cal A)}_{23} =  4 m_N^3 m_c^2 \Phi_{11}^{(\cal A)}\,,  \nonumber\\
 \omega^{(\cal A)}_{22}&=& \frac{m_N}{x} \bigg[2(q^2-m_c^2-x^2\,m_N^2) \Phi_6^{(\cal A)}
 - x \,m_N m_c \Phi_7^{(\cal A)} + 2 x\,m_N^2 \Phi_{11}^{(\cal A)} \bigg]
 \,, \,\,\,\,\,\,
\end{eqnarray}

where the functions $\Phi_i^{(\cal A)}$ are
\begin{eqnarray}
 \Phi_1^{(\cal A)} &=& \tilde{A}_1+2 \tilde{T}_1+\tilde{V}_1 \,, \nonumber\\
 \Phi_2^{(\cal A)} &=& 2 \tilde{A}_3-2 \tilde{P}_1+2 \tilde{S}_1-2 \tilde{T}_1+\tilde{T}_{123}+\tilde{T}_{127}-2 \tilde{V}_3  \,, \nonumber\\
 \Phi_3^{(\cal A)} &=& 2 \tilde{A}_1+4 \tilde{A}_3+2 \tilde{A}_{123}-4 \tilde{P}_1+4 \tilde{S}_1+2 \tilde{V}_1-4 \tilde{V}_3-2
   \tilde{V}_{123}  \,, \nonumber\\
 \Phi_4^{(\cal A)} &=& -2 \tilde{\tilde{A}}_{1345}+2 \tilde{\tilde{P}}_{21}+2 \tilde{\tilde{S}}_{12}-2 \tilde{\tilde{T}}_{123}
 +4 \tilde{\tilde{T}}_{127}-6 \tilde{\tilde{T}}_{158}+3 \tilde{\tilde{T}}_{234578}-2 \tilde{\tilde{V}}_{1345}  \,, \nonumber\\
 \Phi_5^{(\cal A)} &=& \tilde{A}_1^M+2 \tilde{T}_1^M+\tilde{V}_1^M-\tilde{\tilde{\tilde{T}}}_{234578}  \,, \nonumber\\
 \Phi_6^{(\cal A)} &=& \tilde{\tilde{A}}_{123}-\tilde{\tilde{T}}_{123}-\tilde{\tilde{T}}_{127}-\tilde{\tilde{V}}_{123}  \,, \nonumber\\
 \Phi_7^{(\cal A)} &=& -3 \tilde{\tilde{A}}_{34}-2 \tilde{\tilde{A}}_{123}+\tilde{\tilde{A}}_{1345}-4 \tilde{\tilde{P}}_{21}
 -4 \tilde{\tilde{S}}_{12}-3 \tilde{\tilde{V}}_{43}+2 \tilde{\tilde{V}}_{123}+\tilde{\tilde{V}}_{1345}  \,, \nonumber\\
 \Phi_8^{(\cal A)} &=& 2 \tilde{\tilde{A}}_{123}+2 \tilde{\tilde{A}}_{1345}-2 \tilde{\tilde{P}}_{21}-2 \tilde{\tilde{S}}_{12}
 -6 \tilde{\tilde{T}}_{127}+6 \tilde{\tilde{T}}_{158}-3
 \tilde{\tilde{T}}_{234578}-2 \tilde{\tilde{V}}_{123}+2 \tilde{\tilde{V}}_{1345}  \,, \nonumber \\
 \Phi_9^{(\cal A)} &=& \tilde{A}_1^M+ \tilde{T}_1^M+ \tilde{V}_1^M-2 \tilde{\tilde{\tilde{A}}}_{123456}
 +3 \tilde{\tilde{\tilde{T}}}_{125678}-2 \tilde{\tilde{\tilde{T}}}_{234578}+2 \tilde{\tilde{\tilde{V}}}_{123456}  \,, \nonumber\\
 \Phi_{10}^{(\cal A)} &=& -\tilde{A}_1^M-\tilde{V}_1^M+\tilde{\tilde{\tilde{A}}}_{123456}-\tilde{\tilde{\tilde{V}}}_{123456}  \,, \nonumber\\
 \Phi_{11}^{(\cal A)} &=& \tilde{T}_1^M+2 \tilde{\tilde{\tilde{A}}}_{123456}-3 \tilde{\tilde{\tilde{T}}}_{125678}
 + \tilde{\tilde{\tilde{T}}}_{234578}-2\tilde{\tilde{\tilde{V}}}_{123456}
 \,. \nonumber
\end{eqnarray}

\end{itemize}

\subsection{vector transition current}
\label{lambda_c N QCD representation for vector transition}

The invariant amplitudes $\tilde{\Pi}^{(i)}_j ((P-q)^2,q^2)$ for
the correlation function with the vector transition current
$j_{\mu}$ are given by  Eq. (\ref{eq:QCDcorr}) with the
replacement of the coefficient $m_c/4 \to 1/4$. The coefficient functions
$\tilde{\omega}^{(i)}_{jn}$ with  $i={\cal P},{\cal A}$,
$j=1,2,...6$ and $n=1,2,3$  are listed below for:

\begin{itemize}

\item{the pseudoscalar interpolating current}

\begin{eqnarray}
\tilde{ \omega}^{(\cal P)}_{11} &=& x \,m_N \tilde{\Phi}_1^{(\cal
P)}  \, , \qquad \tilde{ \omega}^{(\cal P)}_{12} = x \,m_N^3
\bigg[x\,\tilde{\Phi}_2^{(\cal P)}  + 2 \tilde{\Phi}_3^{(\cal
P)}  \bigg] \,, \nonumber\\
\tilde{ \omega}^{(\cal P)}_{13} &=& 4 x \,m_N^3 m_c^2
\tilde{\Phi}_3^{(\cal P)}  \,, \nonumber
\end{eqnarray}
\begin{eqnarray}
\tilde{ \omega}^{(\cal P)}_{21} &=&  \tilde{ \omega}^{(\cal
P)}_{23}= 0 \,, \qquad   \tilde{ \omega}^{(\cal P)}_{22} = - x
\,m_N^2 \tilde{\Phi}_2^{(\cal P)}  \,, \nonumber
\end{eqnarray}

\begin{eqnarray}
\tilde{ \omega}^{(\cal P)}_{31}&=& \frac{m_N}{2}(m_c-x \,m_N)
\tilde{\Phi}_1^{(\cal
P)}  \,, \nonumber\\
\tilde{ \omega}^{(\cal P)}_{32} &=&
-\frac{m_N^2}{2}\bigg[m_c(m_c-x \,m_N) \tilde{\Phi}_2^{(\cal P)} +
2 x \,m_N^2 \tilde{\Phi}_3^{(\cal
P)} \bigg] \,, \nonumber\\
\tilde{ \omega}^{(\cal P)}_{33}&=& 2 m_N^3 m_c^2 (m_c-x \,m_N)\,
\tilde{\Phi}_3^{(\cal P)} \,, \nonumber
\end{eqnarray}

\begin{eqnarray}
\tilde{ \omega}^{(\cal P)}_{41}&=& \frac{m_N}{2}
\tilde{\Phi}_1^{(\cal P)} \,, \qquad  \tilde{ \omega}^{(\cal
P)}_{42} =  \frac{m_N^2}{2} \Big[- m_c \tilde{\Phi}_2^{(\cal P)} +
2 m_N \tilde{\Phi}_3^{(\cal
P)} \Big] \,, \nonumber\\
\tilde{ \omega}^{(\cal P)}_{43} &=& 2 m_N^3 m_c^2
\tilde{\Phi}_3^{(\cal P)} \,, \nonumber
\end{eqnarray}

\begin{eqnarray}
\tilde{ \omega}^{(\cal P)}_{51}&=& - m_N \tilde{\Phi}_1^{(\cal P)}
\,, \qquad   \tilde{ \omega}^{(\cal P)}_{52}= -m_N^3
\bigg[x\,\tilde{\Phi}_2^{(\cal P)} + 2 \tilde{\Phi}_3^{(\cal
P)} \bigg] \,, \nonumber\\
\tilde{ \omega}^{(\cal P)}_{53}&=& - 4 m_N^3 m_c^2
\tilde{\Phi}_3^{(\cal P)} \,, \nonumber
\end{eqnarray}

\begin{eqnarray}
\tilde{ \omega}^{(\cal P)}_{61}=\tilde{ \omega}^{(\cal P)}_{63}= 0
\,, \qquad \tilde{ \omega}^{(\cal P)}_{62}= m_N^2
\tilde{\Phi}_2^{(\cal P)} \,,
\end{eqnarray}

where the functions $\tilde{\Phi}_i^{(\cal P)}$ are given by
\begin{eqnarray}
\tilde{\Phi}_1^{(\cal P)} &=& 2 \tilde{A}_1+4 \tilde{A}_3+2
\tilde{A}_{123}+2 \tilde{P}_1+2 \tilde{S}_1+6 \tilde{T}_1-12
\tilde{T}_7-\tilde{T}_{123}-5 \tilde{T}_{127}\nonumber
\\ && -2 \tilde{V}_1+4 \tilde{V}_3+2 \tilde{V}_{123} \,,\nonumber\\
\tilde{\Phi}_2^{(\cal P)}  &=& 3 \tilde{\tilde{A}}_{34}+2
\tilde{\tilde{A}}_{123}-\tilde{\tilde{A}}_{1345}-2
\tilde{\tilde{P}}_{21}+2 \tilde{\tilde{S}}_{12}-12
\tilde{\tilde{T}}_{78}-2 \tilde{\tilde{T}}_{123}-4
\tilde{\tilde{T}}_{127}\nonumber
\\ &&-6 \tilde{\tilde{T}}_{158}+\tilde{\tilde{T}}_{234578} -3 \tilde{\tilde{V}}_{43}+2
   \tilde{\tilde{V}}_{123}+\tilde{\tilde{V}}_{1345} \,,\nonumber\\
\tilde{\Phi}_3^{(\cal P)} &=& -\tilde{A}_1^M-3
\tilde{T}_1^M+\tilde{V}_1^M+\tilde{\tilde{\tilde{A}}}_{123456}-3
   \tilde{\tilde{\tilde{T}}}_{125678}+\tilde{\tilde{\tilde{T}}}_{234578}+\tilde{\tilde{\tilde{V}}}_{123456}\,; \nonumber
\end{eqnarray}

\item{the axial-vector interpolating current}

\begin{eqnarray}
\tilde{ \omega}^{(\cal A)}_{11}  &=& 2\Big[2 m_c
\tilde{\Phi}_1^{(\cal A)} - x \,m_N (2\tilde{\Phi}_1^{(\cal A)}
+ \tilde{\Phi}_2^{(\cal A)}) + 2 m_N \tilde{\Phi}_3^{(\cal A)} \Big] \,, \nonumber\\
\tilde{ \omega}^{(\cal A)}_{12}  &=& 2 m_N \Big[ x^2\,m_N^2
\tilde{\Phi}_4^{(\cal A)}
 + x\,m_N m_c \tilde{\Phi}_5^{(\cal A)} + 2 m_c^2 \tilde{\Phi}_3^{(\cal A)}
 + 2 x\,m_N^2 \tilde{\Phi}_6^{(\cal A)} \Big] \,, \nonumber\\
\tilde{ \omega}^{(\cal A)}_{13}  &=& 8 m_N^2 m_c \Big[ m_c^2
\tilde{\Phi}_7^{(\cal A)} + x\,m_N m_c \tilde{\Phi}_6^{(\cal A)} +
x^2\,m_N^2 \tilde{\Phi}_8^{(\cal A)} \Big] \,, \nonumber
\end{eqnarray}

\begin{eqnarray}
\tilde{ \omega}^{(\cal A)}_{21} &=& 4 \tilde{\Phi}_1^{(\cal A)}
\,,  \qquad \tilde{ \omega}^{(\cal A)}_{23}  =  8 m_N^2 m_c \Big[
m_c \tilde{\Phi}_7^ {(\cal A)} - x \,m_N \tilde{\Phi}_8^{(\cal
A)} \Big] \,, \nonumber\nonumber\\
\tilde{ \omega}^{(\cal A)}_{22}  &=& 2 m_N \Big[ 2 m_c
\tilde{\Phi}_3^{(\cal A)} - x \,m_N \tilde{\Phi}_4^{(\cal A)} + 2
m_N \tilde{\Phi}_7^{(\cal A)} \Big] \,, \nonumber
\end{eqnarray}

\begin{eqnarray}
\tilde{ \omega}^{(\cal A)}_{31}&=& 2 { m_c^2 -q^2 \over x}
\tilde{\Phi}_1^{(\cal A)}
 + m_N m_c \tilde{\Phi}_2^{(\cal A)} + m_N^2 \tilde{\Phi}_{9}^{(\cal A)} -2 x \,m_N^2 \tilde{\Phi}_{10}^{(\cal A)}- 2 \frac{m_N^2}{x} \tilde{\Phi}_7^{(\cal A)}  \,, \nonumber\\
\tilde{ \omega}^{(\cal A)}_{32} &=& m_N^2 \Big[
2(q^2-x^2\,m_N^2)\tilde{\Phi}_3^{(\cal A)} - 2
\frac{q^2+m_c^2}{x}\tilde{\Phi}_7^{(\cal A)}
  + x m_N m_c \tilde{\Phi}_{11}^{(\cal A)} \nonumber \, \\
  && + m_c^2 (\tilde{\Phi}_5^{(\cal A)}-2 \tilde{\Phi}_3^{(\cal A)}) + 2 m_N(m_c + x \,m_N) \tilde{\Phi}_8^{(\cal A)} \Big] \nonumber \,, \\
\tilde{ \omega}^{(\cal A)}_{33}&=&  \frac{4 m_c^2 m_N^2}{x} \Big[
(m_c^2-q^2)\tilde{\Phi}_7^{(\cal A)} + x m_N m_c
\tilde{\Phi}_{12}^{(\cal A)} + x^2\,m_N^2 \tilde{\Phi}_8^{(\cal
A)} \Big] \,, \nonumber
\end{eqnarray}

\begin{eqnarray}
\tilde{ \omega}^{(\cal A)}_{41} &=& 2 m_N\Big[
(\tilde{\Phi}_1^{(\cal A)}  + \tilde{\Phi}_{10}^{(\cal A)} ) -
\frac{\tilde{\Phi}_3^{(\cal A)} }{x} \Big] \,, \qquad \tilde{
\omega}^{(\cal A)}_{43} = 4 m_N^3 m_c^2
\tilde{\Phi}_{13}^{(\cal A)} \,,  \nonumber\\
\tilde{ \omega}^{(\cal A)}_{42} &=& \frac{m_N}{x} \Big[ 2
(m_c^2+x^2\,m_N^2-q^2) \tilde{\Phi}_3^{(\cal A)}  - x\,m_N m_c
\tilde{\Phi}_{11}^{(\cal A)}  +2 x\,m_N^2 \tilde{\Phi}_{13}^{(\cal
A)}  \Big]  \,, \nonumber
\end{eqnarray}

\begin{eqnarray}
\tilde{ \omega}^{(\cal A)}_{51}  &=& 2 m_N \tilde{\Phi}_2^{(\cal
A)} \,, \qquad  \tilde{ \omega}^{(\cal A)}_{53} = 8 m_N^3 m_c
\Big[ m_c \tilde{\Phi}_{14}^{(\cal A)} - x \,m_N
\tilde{\Phi}_8^{(\cal A)}
\Big] \,, \nonumber \\
\tilde{ \omega}^{(\cal A)}_{52}  &=& 2 m_N^2 \Big[ -x \,m_N
\tilde{\Phi}_4^{(\cal A)} - m_c (\tilde{\Phi}_5^{(\cal
A)}+2\tilde{\Phi}_3^{(\cal A)}) + 2 m_N \tilde{\Phi}_{14}^{(\cal
A)}  \Big] \,, \nonumber
\end{eqnarray}

\begin{eqnarray}
\tilde{ \omega}^{(\cal A)}_{61} &=& 0 \,, ~~    \tilde{
\omega}^{(\cal A)}_{62}  = 2 m_N^2 \tilde{\Phi}_4^{(\cal A)} \,,
~~ \tilde{ \omega}^{(\cal A)}_{63}  = 8 m_N^3 m_c
\tilde{\Phi}_8^{(\cal A)} \,.
\end{eqnarray}

The functions $\tilde{\Phi}_i^{(\cal A)}$ are given by
\begin{eqnarray}
\tilde{\Phi}_1^{(\cal A)} &=& \tilde{A}_1+2 \tilde{T}_1+\tilde{V}_1 \,, \nonumber\\
\tilde{\Phi}_2^{(\cal A)} &=& 2 \tilde{A}_3-2 \tilde{P}_1+2 \tilde{S}_1-2 \tilde{T}_1
+\tilde{T}_{123}+\tilde{T}_{127}-2 \tilde{V}_3 \,, \nonumber\\
\tilde{\Phi}_3^{(\cal A)} &=& \tilde{\tilde{A}}_{123}-\tilde{\tilde{T}}_{123}
-\tilde{\tilde{T}}_{127}-\tilde{\tilde{V}}_{123} \,, \nonumber\\
\tilde{\Phi}_4^{(\cal A)} &=&
-\tilde{\tilde{A}}_{34}+\tilde{\tilde{A}}_{1345}-2
\tilde{\tilde{P}}_{21}-2 \tilde{\tilde{S}}_{12}-2
\tilde{\tilde{T}}_{127}+2\tilde{\tilde{T}}_{158}-\tilde{\tilde{T}}_{234578}
-\tilde{\tilde{V}}_{43}+\tilde{\tilde{V}}_{1345} \,, \nonumber\\
\tilde{\Phi}_5^{(\cal A)} &=& -\tilde{\tilde{A}}_{34}-2
\tilde{\tilde{A}}_{123}-\tilde{\tilde{A}}_{1345}+4
\tilde{\tilde{T}}_{127}-4 \tilde{\tilde{T}}_{158}+2
 \tilde{\tilde{T}}_{234578}-\tilde{\tilde{V}}_{43}+2 \tilde{\tilde{V}}_{123}-\tilde{\tilde{V}}_{1345} \,, \nonumber\\
\tilde{\Phi}_6^{(\cal A)} &=& \tilde{A}_1^M+ \tilde{T}_1^M+
\tilde{V}_1^M- \tilde{\tilde{\tilde{A}}}_{123456}+
\tilde{\tilde{\tilde{T}}}_{125678}
 - \tilde{\tilde{\tilde{T}}}_{234578}+ \tilde{\tilde{\tilde{V}}}_{123456} \,, \nonumber\\
\tilde{\Phi}_7^{(\cal A)} &=& -\tilde{A}_1^M-2
\tilde{T}_1^M-\tilde{V}_1^M+\tilde{\tilde{\tilde{T}}}_{234578} \,,
\nonumber \\
\tilde{\Phi}_8^{(\cal A)} &=& \tilde{\tilde{\tilde{A}}}_{123456}-2
\tilde{\tilde{\tilde{T}}}_{125678}+\tilde{\tilde{\tilde{T}}}_{234578}-\tilde{\tilde{\tilde{V}}}_{123456}
\,, \nonumber \\ \tilde{\Phi}_{9}^{(\cal A)} &=& -2
\tilde{\tilde{A}}_{34}-2 \tilde{\tilde{A}}_{123}-2
\tilde{\tilde{P}}_{21}-2 \tilde{\tilde{S}}_{12}+2
\tilde{\tilde{T}}_{127}-2
\tilde{\tilde{T}}_{158}+\tilde{\tilde{T}}_{234578}-2
\tilde{\tilde{V}}_{43}+2 \tilde{\tilde{V}}_{123} \,, \nonumber \\
\tilde{\Phi}_{10}^{(\cal A)} &=& \tilde{A}_{123}-\tilde{T}_{123}-\tilde{T}_{127}-\tilde{V}_{123} \,, \nonumber
\end{eqnarray}
\begin{eqnarray}
\tilde{\Phi}_{11}^{(\cal A)} &=& 2 \tilde{\tilde{A}}_{34}+2
\tilde{\tilde{P}}_{21}+2 \tilde{\tilde{S}}_{12}+2
\tilde{\tilde{T}}_{123}+2
\tilde{\tilde{T}}_{158}-\tilde{\tilde{T}}_{234578}+2 \tilde{\tilde{V}}_{43} \,, \nonumber\\
\tilde{\Phi}_{12}^{(\cal A)} &=& \tilde{T}_1^M+\tilde{\tilde{\tilde{T}}}_{125678}-\tilde{\tilde{\tilde{T}}}_{234578} \,, \nonumber\\
\tilde{\Phi}_{13}^{(\cal A)} &=& -\tilde{A}_1^M-2
\tilde{T}_1^M-\tilde{V}_1^M- \tilde{\tilde{\tilde{A}}}_{123456}+2
\tilde{\tilde{\tilde{T}}}_{125678}+
\tilde{\tilde{\tilde{V}}}_{123456}\,, \nonumber \\
\tilde{\Phi}_{14}^{(\cal A)} &=& \tilde{T}_1^M+
\tilde{\tilde{\tilde{A}}}_{123456}-
\tilde{\tilde{\tilde{T}}}_{125678}-
\tilde{\tilde{\tilde{V}}}_{123456} \,. \nonumber
\end{eqnarray}

\end{itemize}

\subsection{axial-vector transition current}

The invariant amplitudes $\bar{\Pi}^{(i)}_j ((P-q)^2,q^2)$ for the
correlation function with the axial-vector transition current
$j_{\mu 5}$ are given by  (\ref{eq:QCDcorr}) where $m_c/4 \to
1/4$, with $i={\cal P},{\cal A}$ and $j=1,2,...6$. The coefficient
functions $\bar{\omega}^{(i)}_{jn}$ can be obtained from
$\tilde{\omega}^{(i)}_{jn}$ in the above subsection by  changing
the sign at $m_c$ and at  $\tilde{\omega}^{({\cal P})}_{2n}$,
$\tilde{\omega}^{({\cal P})}_{3n}$, $\tilde{\omega}^{({\cal
P})}_{6n}$, $\tilde{\omega}^{({\cal A})}_{1n}$,
$\tilde{\omega}^{({\cal A})}_{4n}$ and $\tilde{\omega}^{({\cal
A})}_{5n}$.

\section{Correlation function in the LCSR for  $\Sigma_c \to N $
form factors} \label{sigma_c N QCD representation}

\subsection{ pseudoscalar transition current}

The invariant amplitudes $\Pi^{(i)}_j ((P-q)^2,q^2)$ for the
correlation function with the pseudoscalar transition current
$j_5$ are  given by  Eq. (\ref{eq:QCDcorr}) with $i={\cal I},
{\cal T}$ and $j=1,2$.
Here we specify the corresponding coefficient funcitons for:

\begin{itemize}
\item{ Ioffe
 current:}\label{eq:sum_rules_D_Ioffe_Sigma}

The functions   $\omega^{({\cal I})}_{jn}$ can be obtained from
$\omega^{({\cal A})}_{jn}$ given in previous App.B by changing the
sign of the terms involving scalar $S_{1,2}(x_i)$, pseudoscalar
$P_{1,2}(x_i)$ and tensor $T_{1,...,8}(x_i)$ DA's and also $m_c
\to -m_c$;

\item {tensor interpolating current:}

\begin{eqnarray}
\omega^{(\cal T)}_{11}&=& 2\Big[4  {m_c^2 -q^2 \over x}
\Phi_1^{(\cal T)}
 + m_N m_c \Phi_2^{(\cal T)} + 2\,x \,m_N^2 ( \Phi_1^{(\cal T)} + \Phi_3^{(\cal T)}) \, \nonumber \\
 &&  + m_N^2 \Phi_4^{(\cal T)} + \frac{4 m_N^2}{x} \Phi_5^{(\cal T)} \Big]  \,, \nonumber\\
\omega^{(\cal T)}_{12}&=& 2 m_N^2 \Big[ 4 (x^2 m_N^2 - q^2)
\Phi_6^{(\cal T)} + x m_N m_c \Phi_7^{(\cal T)} + m_c^2
\Phi_4^{(\cal T)}-\frac{2}{x} (2 (q^2 + m_c^2) \, \nonumber \\
 && + x^2 m_N^2 )\Phi_5^{(\cal T)} - 6 x m_N^2 \Phi_8^{(\cal T)} \Big] \,, \nonumber\\
\omega^{(\cal T)}_{13} &=& \frac{8 m_N^2 m_c^2}{x}\Big[ 2
(q^2-m_c^2) \Phi_5^{(\cal T)} + x m_N m_c \Phi_9^{(\cal T)} - x^2
m_N^2 (3 \Phi_8^{(\cal T)} + \Phi_5^{(\cal T)} ) \Big]\,,
\nonumber
\end{eqnarray}

\begin{eqnarray}
 \omega^{(\cal T)}_{21} &=& 4 m_N\Big[ -\Phi_1^{(\cal T)} +
\Phi_3^{(\cal T)} -\frac{2}{x}\Phi_6^{(\cal T)} \Big]  \,, \qquad
\omega^{(\cal T)}_{23} = 8 m_N^3 m_c^2 \Big[ \Phi_5^{(\cal
T)} -3 \Phi_8^{(\cal T)} \Big] \,, \nonumber\\
\omega^{(\cal T)}_{22} &=& \frac{2 m_N}{x} \Big[ 4 (m_c^2 + x^2
m_N^2 - q^2) \Phi_6^{(\cal T)} + x m_N m_c \Phi_7^{(\cal
T)} \nonumber \\
&& \hspace{1.2 cm} - 2 x m_N^2 (3 \Phi_8^{(\cal T)} +
\Phi_5^{(\cal T)}) \Big] \,,
\end{eqnarray}

where the functions $\Phi_i^{(\cal T)}$ are
\begin{eqnarray}
 \Phi_1^{(\cal T)} &=& \tilde{V}_1 - \tilde{A}_1 \,, \qquad  \Phi_3^{(\cal T)} = \tilde{V}_{123} + \tilde{A}_{123} \,, \nonumber\\
 \Phi_2^{(\cal T)} &=& 6 \tilde{P}_1+6 \tilde{S}_1-6 \tilde{T}_1+12 \tilde{T}_7+\tilde{T}_{123}+5 \tilde{T}_{127} \,, \nonumber\\
\Phi_4^{(\cal T)} &=& 3 \tilde{\tilde{A}}_{34}+2
\tilde{\tilde{A}}_{123}+3 \tilde{\tilde{A}}_{1345}
 -3 \tilde{\tilde{V}}_{43}+2 \tilde{\tilde{V}}_{123}-3
   \tilde{\tilde{V}}_{1345}  \,, \nonumber\\
 \Phi_5^{(\cal T)} &=& \tilde{V}_1^M - \tilde{A}_1^M \,, \qquad
 \Phi_6^{(\cal T)} = \tilde{\tilde{V}}_{123} + \tilde{\tilde{A}}_{123} \,, \nonumber\\
 \Phi_7^{(\cal T)} &=& -6 \tilde{\tilde{P}}_{21}+6 \tilde{\tilde{S}}_{12}+12 \tilde{\tilde{T}}_{78}+2 \tilde{\tilde{T}}_{123}
 +4 \tilde{\tilde{T}}_{127}+6 \tilde{\tilde{T}}_{158}-\tilde{\tilde{T}}_{234578} \,, \nonumber\\
 \Phi_8^{(\cal T)} &=&  \tilde{\tilde{\tilde{V}}}_{123456} +  \tilde{\tilde{\tilde{A}}}_{123456} \,,
 \qquad  \Phi_9^{(\cal T)} = 3 \tilde{T}_1^M+3
\tilde{\tilde{\tilde{T}}}_{125678}-\tilde{\tilde{\tilde{T}}}_{234578}
 \,. \nonumber
\end{eqnarray}

\end{itemize}

\subsection{vector  transition current }

The invariant amplitudes $\tilde{\Pi}^{(i)}_j ((P-q)^2,q^2)$, with
$i={\cal I},{\cal T}$ and $j=1,2,...6$, for the correlation
function with the vector transition current $j_{\mu}$ are given by
Eq. (\ref{eq:QCDcorr})  with the replacement $m_c/4 \to 1/4$,
where the coefficient $\tilde{\omega}^{(i)}_{jn}$ functions are
given below for:

\begin{itemize}

\item { Ioffe current:}

The functions   $\tilde{\omega}^{({\cal I})}_{jn}$ can be obtained
from $\tilde{\omega}^{({\cal A})}_{jn}$ presented in App.~B by
changing the sign of the terms involving vector $V_{1,...,6}(x_i)$
and axial-vector $A_{1,...,6}(x_i)$ DA's as well as $m_c \to
-m_c$;

\item{tensor  interpolating current}

\begin{eqnarray}
\tilde{ \omega}^{(\cal T)}_{11}  &=& 4\Big[- 4 m_c \tilde{\Phi}_1^{(\cal T)}
+ x \,m_N  \tilde{\Phi}_2^{(\cal T)} + 4 m_N \tilde{\Phi}_3^{(\cal T)} \Big] \,, \nonumber\\
\tilde{ \omega}^{(\cal T)}_{12} &=& 4 m_N \Big[ 4 q^2
\tilde{\Phi}_3^{(\cal T)} - x^2\,m_N^2 (4 \tilde{\Phi}_3^{(\cal
T)} + 2 \tilde{\Phi}_4^{(\cal T)} + 3 \tilde{\Phi}_5^{(\cal T)})
+ 2 x \,m_N m_c (\tilde{\Phi}_6^{(\cal T)}-2 \tilde{\Phi}_7^{(\cal T)})  \nonumber \\
 &&  + 2 x \,m_N^2 (\tilde{\Phi}_8^{(\cal T)}+\tilde{\Phi}_9^{(\cal T)}) \Big]  \,,  \nonumber\\
\tilde{ \omega}^{(\cal T)}_{13} &=& 16 m_N^2 \Big[ x \,m_N m_c^2
\tilde{\Phi}_8^{(\cal T)} + 2 x^2\,m_N^2 m_c
\tilde{\Phi}_{10}^{(\cal T)} +
x \,m_N\left(2(q^2-x^2\,m_N^2)-m_c^2\right) \tilde{\Phi}_9^{(\cal T)}  \, \nonumber\\
&& + 2 m_c^3 \Phi_{11}^{(\cal T)}  \Big] \,, \nonumber
\end{eqnarray}

\begin{eqnarray}
\tilde{ \omega}^{(\cal T)}_{21}  &=& 0 \,, \qquad \tilde{
\omega}^{(\cal T)}_{23} =  32 m_N^2 \Big[ \left(x^2\,m_N^2 -
q^2\right) \tilde{\Phi}_9^{(\cal T)} - x\,m_N m_c \tilde{\Phi}_{10}^{(\cal T)} \Big] \,,  \nonumber\\
\tilde{ \omega}^{(\cal T)}_{22}  &=& 4 m_N \Big[ 4 m_c
\tilde{\Phi}_{12}^{(\cal T)} + x\,m_N (2 \tilde{\Phi}_4^{(\cal T)}
+ 3 \tilde{\Phi}_5^{(\cal T)}) - 8 m_N \tilde{\Phi}_9^{(\cal T)}
\Big] \,, \nonumber
\end{eqnarray}

\begin{eqnarray}
\tilde{ \omega}^{(\cal T)}_{31} &=& 2 m_N \Big[ 2 m_c
(\tilde{\Phi}_1^{(\cal T)}-\tilde{\Phi}_{13}^{(\cal T)})
 + x\,m_N (\tilde{\Phi}_2^{(\cal T)} + 4 \tilde{\Phi}_{14}^{(\cal T)}) \nonumber \\
 && + 2 m_N (-2(\tilde{\Phi}_3^{(\cal T)}
 +\tilde{\Phi}_4^{(\cal T)})+\tilde{\Phi}_5^{(\cal T)} ) \Big] \,, \nonumber\\
\tilde{ \omega}^{(\cal T)}_{32}  &=& 2 m_N^2 \Big[ -x\,m_N m_c
(\tilde{\Phi}_6^{(\cal T)}+2\tilde{\Phi}_{12}^{(\cal T)})
+ 4 m_N m_c \tilde{\Phi}_{10}^{(\cal T)} + 2 x\,m_N^2 (\tilde{\Phi}_8^{(\cal T)} + \tilde{\Phi}_9^{(\cal T)})  \nonumber\\
&& + m_c^2 \left(-2\tilde{\Phi}_4^{(\cal T)} + 5
\tilde{\Phi}_5^{(\cal T)} - \frac{8}{x}\tilde{\Phi}_9^{(\cal T)}
\right) + 4 (x^2\,m_N^2-q^2) \tilde{\Phi}_3^{(\cal T)} \Big] \,, \nonumber\\
\tilde{ \omega}^{(\cal T)}_{33} &=& \frac{8 m_N^2 m_c^2}{x} \Big[
2\left(x^2\,m_N^2 - q^2 + m_c^2\right) \tilde{\Phi}_9^{(\cal T)} +
x^2\,m_N^2 (\tilde{\Phi}_8^{(\cal T)}+\tilde{\Phi}_9^{(\cal T)})
\nonumber\\ &&- x\,m_N m_c (\tilde{\Phi}_{10}^{(\cal
T)}+\tilde{\Phi}_{11}^{(\cal T)}) \Big] \,, \nonumber
\end{eqnarray}

\begin{eqnarray}
\tilde{ \omega}^{(\cal T)}_{41}&=& \frac{2 m_N}{x} \Big[ 4
\tilde{\Phi}_3^{(\cal T)}
- x (\tilde{\Phi}_2^{(\cal T)}+4\tilde{\Phi}_{14}^{(\cal T)}) \Big] \,, \nonumber\\
\tilde{ \omega}^{(\cal T)}_{42} &=& \frac{2 m_N}{x} \Big[ 4
\left(q^2 -x^2\,m_N^2 -m_c^2\right) \tilde{\Phi}_3^{(\cal T)}
+ x\,m_N m_c (\tilde{\Phi}_6^{(\cal T)}+ 2\tilde{\Phi}_7^{(\cal T)})  \nonumber \\
&& - 2 x\,m_N^2 (\tilde{\Phi}_8^{(\cal T)} +\tilde{\Phi}_9^{(\cal T)} )  \Big] \,, \nonumber\\
\tilde{ \omega}^{(\cal T)}_{43} &=& - 8 m_N^3 m_c^2
(\tilde{\Phi}_8^{(\cal T)} + \tilde{\Phi}_9^{(\cal T)} ) \,,
\nonumber
\end{eqnarray}

\begin{eqnarray}
\tilde{ \omega}^{(\cal T)}_{51} &=& - 4 m_N \Big[\tilde{\Phi}_2^{(\cal T)} + \frac{4 \tilde{\Phi}_3^{(\cal T)}}{x} \Big] \,, \nonumber\\
\tilde{ \omega}^{(\cal T)}_{52} &=& \frac{4 m_N}{x} \Big[ 4
\left(m_c^2-q^2\right) \tilde{\Phi}_3^{(\cal T)}
  +  x^2\,m_N^2 (4\tilde{\Phi}_3^{(\cal T)} + 2\tilde{\Phi}_4^{(\cal T)} + 3\tilde{\Phi}_5^{(\cal T)} )  \nonumber \\
 && - 2 x \,m_N m_c \tilde{\Phi}_6^{(\cal T)} - 2 x\,m_N^2 (\tilde{\Phi}_8^{(\cal T)} + \tilde{\Phi}_9^{(\cal T)} ) \Big] \,, \nonumber\\
\tilde{ \omega}^{(\cal T)}_{53} &=& 16 m_N^3 \Big[
2\left(x^2\,m_N^2 - q^2\right) \tilde{\Phi}_9^{(\cal T)} + m_c^2
(3\tilde{\Phi}_9^{(\cal T)} -\tilde{\Phi}_8^{(\cal T)} ) - 2 x
\,m_N m_c \tilde{\Phi}_{10}^{(\cal T)} \Big] \,, \nonumber
\end{eqnarray}

\begin{eqnarray}
\tilde{ \omega}^{(\cal T)}_{61}&=& 0 \,, \qquad  \tilde{
\omega}^{(\cal T)}_{62} = 4 m_N^2 \Big[ - 2 \tilde{\Phi}_4^{(\cal
T)} - 3 \tilde{\Phi}_5^{(\cal T)}
 + \frac{8}{x} \tilde{\Phi}_9^{(\cal T)} \Big] \,, \nonumber\\
\tilde{ \omega}^{(\cal T)}_{63} &=& \frac{32 m_N^2}{x} \Big[
\left(q^2 -x^2\,m_N^2-m_c^2\right) \tilde{\Phi}_9^{(\cal T)} + x
\,m_N m_c \tilde{\Phi}_{10}^{(\cal T)} \Big] \,,
\end{eqnarray}

where the functions $\tilde{\Phi}_i^{(\cal T)}$ are
\begin{eqnarray}
 \tilde{\Phi}_1^{(\cal T)} &=& \tilde{V}_1 - \tilde{A}_1 \,,
 \qquad
 \tilde{\Phi}_2^{(\cal T)} = 2 \tilde{P}_1+2 \tilde{S}_1-2 \tilde{T}_1+4 \tilde{T}_7-\tilde{T}_{123}+3 \tilde{T}_{127} \,, \nonumber\\
 \tilde{\Phi}_3^{(\cal T)} &=& \tilde{\tilde{T}}_{123}- \tilde{\tilde{T}}_{127} \,,
 \qquad
 \tilde{\Phi}_4^{(\cal T)} = \tilde{\tilde{P}}_{21}- \tilde{\tilde{S}}_{12}-2 \tilde{\tilde{T}}_{78}- \tilde{\tilde{T}}_{123}
 - \tilde{\tilde{T}}_{158} \,, \nonumber
 \end{eqnarray}
 \begin{eqnarray}
\tilde{\Phi}_5^{(\cal T)} &=& \tilde{\tilde{T}}_{234578} \,,
\qquad
 \tilde{\Phi}_6^{(\cal T)} = \tilde{\tilde{V}}_{43}+ \tilde{\tilde{V}}_{1345}
 - \tilde{\tilde{A}}_{34}- \tilde{\tilde{A}}_{1345} \,, \nonumber\\
 \tilde{\Phi}_7^{(\cal T)} &=& \tilde{\tilde{V}}_{123} + \tilde{\tilde{A}}_{123} \,,
 \qquad
 \tilde{\Phi}_8^{(\cal T)} = \tilde{T}_1^M+ \tilde{\tilde{\tilde{T}}}_{125678}  \,, \nonumber\\
 \tilde{\Phi}_9^{(\cal T)} &=& \tilde{\tilde{\tilde{T}}}_{234578} \,,
 \qquad
 \tilde{\Phi}_{10}^{(\cal T)} = \tilde{\tilde{\tilde{V}}}_{123456}+ \tilde{\tilde{\tilde{A}}}_{123456} \,, \nonumber\\
 \tilde{\Phi}_{11}^{(\cal T)} &=& \tilde{V}_1^M-\tilde{A}_1^M  \,,
 \qquad
 \tilde{\Phi}_{12}^{(\cal T)} =\tilde{\tilde{V}}_{123}+ \tilde{\tilde{A}}_{123} \,, \nonumber\\
 \tilde{\Phi}_{13}^{(\cal T)} &=& \tilde{V}_{123}+ \tilde{A}_{123} \,,
 \qquad
\tilde{ \Phi}_{14}^{(\cal T)} = \tilde{T}_{123}-\tilde{T}_{127}
\,. \nonumber
\end{eqnarray}

\end{itemize}

\subsection{axial-vector transition current}
The invariant amplitudes $\bar{\Pi}^{(i)}_j ((P-q)^2,q^2)$, with
$i={\cal I},{\cal T}$ and $j=1,2,...6$,  for the correlation
function with the axial-vector transition current $j_{\mu 5}$ are
given by Eq. (\ref{eq:QCDcorr})  with the replacement $m_c/4 \to
1/4$. The coefficient functions $\bar{\omega}^{(i)}_{jn}$ can be
obtained from $\tilde{\omega}^{(i)}_{jn}$ in the above subsection
by  changing the sign  for $\tilde{\omega}^{({\cal T})}_{2n}$,
$\tilde{\omega}^{({\cal T})}_{3n}$, $\tilde{\omega}^{({\cal
T})}_{6n}$, $\tilde{\omega}^{({\cal I})}_{1n}$,
$\tilde{\omega}^{({\cal I})}_{4n}$ and $\tilde{\omega}^{({\cal
I})}_{5n}$ together with  $m_c\to - m_c$ \,.

%%%%%%%%%%%%%%%%%%%%%%%%

\section {Two-point sum rules }
\label{two-point sum rules}

Here we present the expressions for the spectral densities
in the two-point sum rules for the decay constants of charmed baryons:

\subsection{$\Lambda_c$ baryon}

\begin{eqnarray}
 {\rm Im} F_1^{({\cal P})}(s) &=&{m_c^4  \over 512 \pi^3}
 \bigg [ (1-\tau^2) (1- {8 \over \tau} +{1 \over \tau^2}) -12 \ln \tau \bigg ]
\nonumber \\
&& + {1 \over 768 \pi^2} \langle  \alpha_s G^2 \rangle (1-\tau)
(1+5 \tau) + {\pi } { \langle  \bar{q} q \rangle^2 \over 6}
\delta(s -m_c^2)
\,, \nonumber \\
{\rm Im} F_2^{({\cal P})}(s) &=&{m_c^5  \over 128 \pi^3}
 \bigg [ (1-\tau) (1 + {10 \over \tau} +{1 \over \tau^2}) + 6 (1+{1 \over \tau})\ln \tau \bigg ]
\nonumber \\
&& + {m_c \over 384 \pi^2} \langle  \alpha_s G^2 \rangle \bigg[
(1-\tau) (7+{2 \over \tau}) + 6 \ln \tau \bigg] \nonumber \\
&& + {\pi } m_c { \langle \bar{q} q \rangle^2 \over 6} \delta(s
-m_c^2) \,,
\end{eqnarray}
%The invariant functions $\tilde{F}_1(s)$ and $\tilde{F}_2(s)$ are
%given by \cite{Bagan:1992tp}
\begin{eqnarray}
{\rm Im} \tilde{F}_1(s) &=&{(5+2 b + 5b^2)m_c^4  \over 2048 \pi^3}
 \bigg [ (1-\tau^2) (1- {8 \over \tau} +{1 \over \tau^2}) -12 \ln \tau \bigg ]
\nonumber \\
&& -{ (5 - 4 b -b^2) m_c \over 96 \pi} \langle  \bar{q} q \rangle
(1-\tau)^2 \nonumber \\
&&  + {\langle \alpha_s G^2 \rangle  \over 3072 \pi^2} (1-\tau)
 \bigg [ (5+ 2 b+5 b^2)  + 3 ( 7+ 6b + 7b^2) \tau \bigg ] \nonumber
\end{eqnarray}
\begin{eqnarray}
&& + { m_0^2 \langle \bar{q} q \rangle (1-b)  \over 384 \pi }
{\tau \over m_c} (11 \tau-6+b(7\tau-6)) \nonumber \\
&& + {\pi }{ \langle \bar{q} q \rangle^2 \over 72} \delta(s
-m_c^2)(11+ 2 b+ 3 b^2) \,, \nonumber \\
{\rm Im} \tilde{F}_2(s) &=&{(11+ 2 b-13 b^2)m_c^5  \over 1536
\pi^3}
 \bigg [ (1-\tau) (1+{10 \over \tau} +{1 \over \tau^2})  + 6 (1+{1 \over \tau})\ln \tau \bigg ]
\nonumber \\
&&  -{ 5- 4 b -b^2 \over 96 \pi  }  s  \langle \bar{q} q \rangle
(1-\tau)^2 - { m_0^2 \langle \bar{q} q \rangle (1-b)  \over 384
\pi }
\bigg [ \tau-  6  + b (5 \tau- 6)\bigg ]   \nonumber \\
&& + {(1-b ) m_c \langle  \alpha_s G^2  \rangle \over 4608
\pi^2}\bigg[ (1-\tau) \bigg(2(11+13 b) {1 \over \tau}+ (89+79 b)
\bigg)
+ 72 (1+b) \ln \tau \bigg] \nonumber  \\
&& + {\pi } { \langle \bar{q} q \rangle^2 \over 24} \delta(s
-m_c^2) (5 + 2 b +5 b^2) m_c \,,
\end{eqnarray}
where for brevity we denote $m_c^2/s=\tau$ and use the standard
notations for the vacuum condensate densities. The above relations
for ${\rm Im} \tilde{F}_{1,2}(s)$ are used at $b=-1/5$. Hereafter
the integration convention $\int_{m_c^2}^{\infty} \, d s \,
\delta(s-m_c^2)=1$ is implied.

\subsection{$\Sigma_c$ baryon}

\begin{eqnarray}
{\rm Im} \bar{F}_1(s) &=&{(5+2 b + 5b^2)m_c^4  \over 512 \pi^3}
\bigg [ -{\tau^2 \over 4}+ 2 \tau -{2 \over \tau} + {1 \over 4
\tau^2}-3 \ln \tau \bigg ]
\nonumber \\
&& -{ 3 m_c \over 32 \pi} \langle  \bar{q} q \rangle
(1-b^2)(1-\tau)^2 + {\pi } { \langle \bar{q}
q \rangle^2 \over 24} \delta(s -m_c^2)(1-b)^2 \nonumber \\
&&  - {\langle \alpha_s G^2 \rangle  \over 3072 \pi^2} \tau
(1-\tau)
 \bigg [ (1+ b^2)(11 -{5 \over \tau})  + 2 b ( 7- {1 \over \tau}) \bigg ] \nonumber \\
&& + { m_0^2 \langle \bar{q} q \rangle (1-b^2)  \over 128 \pi } {1
\over m_c} 13 \tau^2 (1- {6 \over 13 \tau})
\,, \nonumber \\
{\rm Im} \bar{F}_2(s) &=&{(1- b)^2m_c^5  \over 512 \pi^3 \tau^2}
 \bigg [ (1-\tau) (1+10  \tau +\tau^2)  + 6 \tau (1+\tau)\ln \tau \bigg ]
\nonumber \\
&& -{ 3m_c^2  \over 32 \pi  }   \langle \bar{q} q \rangle
(1-b^2)\tau (1-{1 \over \tau})^2  + { m_0^2 \langle \bar{q} q
\rangle(1-b^2)  \over 128 \pi } (6+\tau)  \nonumber \\
&& - {(1-b )^2 m_c \langle  \alpha_s G^2  \rangle \over 1536
\pi^2} (1-\tau) (5 -{2 \over \tau}) \nonumber  \\
&&+ {\pi } { \langle \bar{q} q \rangle^2 \over 24} \delta(s
-m_c^2) (5 + 2 b +5 b^2) m_c \,,
\end{eqnarray}
used for $b=\pm 1 $.
%%%%%%%%%%%%%%%%%%%%%%%

\section{Double spectral representations }
\label{double spectral density}

Here  we collect the double dispersion relations for the master
integrals with the powers  $n=2, 3$ in denominators.

\begin{eqnarray}
 \int\limits_0^1 d x   { x^k \over D^2 } & =  &  - {1 \over  \pi}\int\limits_{m_c^2}^{\infty}   {d s \over s -
(P-q)^2}   \int\limits_{t_1(s)}^{t_2(s)}  {d s^{\prime} \over s^{\prime}-
q^2} \bigg \{  \sum \limits_{j=2}^{k} \, (-1)^{k+1+j/2} \,\,  {1
+(-1)^j \over 2}
\nonumber \\
&&  \times \frac{j-1}{ (2 m_N^2)^{k-1}} \, C_k^j \,
[\bar{s}(s^{\prime})]^{k-j} \, [\kappa(s^{\prime},t_1, t_2)]^{ j-3
\over 2 }  \theta(k-2)  \nonumber \\
&&   + {(-1)^k \over (2 m_N^2)^{k-1} } {[\bar{s}(s^{\prime})]^k
\over [\kappa(s^{\prime},t_1,
t_2)]^{3/2} }   \nonumber \\
&& -  { (-1)^k \over (2 m_N^2)^{k-1} }  \bigg [   \bigg ({[
\bar{s}(t_1)]^k \over t_2 -t_1 } \delta(s^{\prime}-t_1) X_1(t_1,
t_2) \bigg )  -  \bigg ( t_1 \leftrightarrow t_2  \bigg )  \bigg ] \nonumber \\
 &&-\pi \delta(m_c^2 - s) \delta(m_c^2+m_N^2 - s^{\prime}) \bigg \}\,+
 ...\,,
\end{eqnarray}

\begin{eqnarray}
 \int\limits _0^1 d x  { x^k \over D^3 } & = & {1 \over 2 \pi}\int\limits_{m_c^2}^{\infty}   {d s \over s
- (P-q)^2} \int\limits_{t_1(s)}^{t_2(s)}  {d s^{\prime} \over
s^{\prime}- q^2} \bigg \{  \sum \limits_{j=4}^{k} (-1)^{k+j/2} {1
+(-1)^j \over 2} \, C_k^j
\nonumber \\
&& \cdot \frac{(j-1)(j-3)}{( 2 m_N^2)^{k-2}  } \,
[\bar{s}(s^{\prime})]^{k-j} \, [\kappa(s^{\prime},t_1, t_2)]^{ j-5
\over 2 } \theta(k-4)
\nonumber \\
%%%%%%%%%%%%%%%%%%%%%%%%
&& +   \frac{(-1)^k C_k^2 }{(2 m_N^2)^{k-2} }\theta(k-2) \bigg [
{[\bar{s}(s^{\prime})]^{k-2} \over
[\kappa(s^{\prime},t_1, t_2)]^{3/2} } \nonumber \\
&& \hspace{0.3 cm} -\bigg ({ [\bar{s}(t_1)]^{k-2} \over t_2 -t_1 }
\delta(s^{\prime} -t_1) X_1(t_1, t_2) \bigg )
+ \bigg ( t_1 \leftrightarrow t_2 \bigg ) \bigg ]\nonumber \\
%%%%%%%%%%%%%%%%%%%%%%%%
&& + 3 \frac{(-1)^k }{(2 m_N^2)^{k-2} } \,\, \bigg \{
{[\bar{s}(s^{\prime})]^{k} \over
[\kappa(s^{\prime},t_1, t_2)]^{5/2} }  \nonumber \\
&& \hspace{0.3 cm}  - \bigg ({[\bar{s}(t_1)]^{k} \over
(t_2-t_1)^3} \delta(s^{\prime} -t_1) X_2(t_1, t_2) \bigg )
 +\bigg (t_1 \leftrightarrow t_2 \bigg ) \nonumber \\
&& \hspace{0.3 cm}  + \bigg [{[\bar{s}(t_1)]^{k} \over
(t_2-t_1)^2}  \xi(s, s^{\prime}, t_1) X_1(t_1, t_2) \bigg]
- \bigg [ t_1 \leftrightarrow t_2  \bigg ]  \bigg \} \nonumber \\
%%%%%%%%%%%%%%%%%
&& + {\pi \over 2} \frac{(-1)^k}{ (2 m_N^2)^k} \bigg ( - 2
\delta^{\prime}(m_c^2-s)\delta(m_c^2 +m_N^2-s^{\prime})(-2
m_N^2)^k   \nonumber \\
&& \hspace{0.3 cm} + k (k-3) \delta(m_c^2-s) \delta(m_c^2
+m_N^2-s^{\prime}) (-2m_N^2)^{k-1} \theta(k-1) \nonumber \\
&&  \hspace {0.2 cm}+  \delta(m_c^2-s) \,\,
\delta^{(2)}(\bar{s}(s^{\prime})+2 m_N^2) \,\,
[\bar{s}(s^{\prime}) +2 m_N^2]^k \,\, [\bar{s}(s^{\prime}) + 4
m_N^2] \bigg ) \bigg \}  \,\,\nonumber \\
 &&+ ...\,.
\end{eqnarray}
In the above,
the ellipses denote the terms that vanish after double Borel
transformation and are therefore inessential;
$t_{1,2}$ are the functions of $s$ defined in
(\ref{eq:intlim}), $\theta(k-a)= 1(0)$ at $k\geq a$($k<a$),
$\bar{s}(y)=s-y-m_N^2$ and  $\kappa(a, b, c)=(a- b)(c-a)$. The
auxiliary functions entering the above expressions are defined as
\begin{eqnarray}
X_1(a, b) &=& \int_{a}^{b} {d \sigma  \over
[\kappa(\sigma,a, b)]^{3/2} } (b -\sigma) \,, \nonumber \\
X_2(a, b) &=& \int_{a}^{b} {d \sigma  \over [\kappa(\sigma,a,
b)]^{5/2} } (b -\sigma)^2 (2 \sigma-3 a
+b) \,, \nonumber \\
\xi(s, s^{\prime}, a)&=& \delta^{\prime}(s^{\prime} -a) + {k \,
\theta(k-1) \over \bar{s}(a)} \delta(s^{\prime} -a) \,. \nonumber
\end{eqnarray}

%%%%%%%%%%%%%%%%%%%


\begin{thebibliography}{100}

\bibitem{exp}
%\cite{Aaltonen:2008eu}
%\bibitem{Aaltonen:2008eu}
  T.~Aaltonen {\it et al.} [CDF Collaboration],
  %``First Measurement of the Ratio of Branching Fractions $B(\Lambda^0_b \to \Lambda^+_{c} \mu^{-} \bar{\nu}_\mu / B(Lambda^0_b \to \Lambda^+_{c} \pi^{-})$,''
  Phys.\ Rev.\  {\bf D79 } (2009)  032001;
  %[arXiv:0810.3213 [hep-ex]].
%\cite{Aaltonen:2011qs}
%\bibitem{Aaltonen:2011qs}
%  T.~Aaltonen {\it et al.} [ CDF Collaboration ],
  %``Observation of the Baryonic Flavor-Changing Neutral Current Decay $\Lambda_b^0 \to \Lambda \mu^+ \mu^-$,''
  arXiv:1107.3753 [hep-ex];
 %\cite{:2009ny}
%\bibitem{:2009ny}
 B.~Adeva {\it et al.} [The LHCb Collaboration],
  %``Roadmap for selected key measurements of LHCb,''
  arXiv:0912.4179 [hep-ex];
 %\cite{Graziani:2011ft}
%\bibitem{Graziani:2011ft}
  G.~Graziani {\it et al.} [on behalf of the LHCb Collaboration],
  %``Recent LHCb Results,''
  arXiv:1107.2328 [hep-ex].


\bibitem{PANDA}
%\cite{Wiedner:2011mf}
%\bibitem{Wiedner:2011mf}
  U.~Wiedner,
  %``Future Prospects for Hadron Physics at PANDA,''
  Prog.\ Part.\ Nucl.\ Phys.\  {\bf 66 } (2011)  477.
  %[arXiv:1104.3961 [hep-ex]].


\bibitem{lcsr}
  I.~I.~Balitsky, V.~M.~Braun and A.~V.~Kolesnichenko,
  %``Radiative Decay Sigma+ ---> p gamma in Quantum Chromodynamics,''
  Nucl.\ Phys.\  B {\bf 312} (1989) 509;
  %%CITATION = NUPHA,B312,509;%%
   V.~M.~Braun and I.~E.~Filyanov,
  %``QCD Sum Rules in Exclusive Kinematics and Pion Wave Function,''
  Z.\ Phys.\  C {\bf 44} (1989) 157;
  %%CITATION = YAFIA,50,818;%%
  V.~L.~Chernyak and I.~R.~Zhitnitsky,
  %``B meson exclusive decays into baryons,''
  Nucl.\ Phys.\  B {\bf 345 } (1990)  137.



\bibitem{Bpi}
 V.~M.~Belyaev, A.~Khodjamirian and R.~R\"{u}ckl,
  %``QCD calculation of the B ---> pi, K form-factors,''
  Z.\ Phys.\  {\bf C60 } (1993)  349;
  A.~Khodjamirian, R.~R\"uckl, S.~Weinzierl and O.~I.~Yakovlev,
  %``Perturbative QCD correction to the B --> pi transition form factor,''
  Phys.\ Lett.\  B {\bf 410} (1997) 275;
  %[arXiv:hep-ph/9706303];
  %%CITATION = PHLTA,B410,275;%%
  E.~Bagan, P.~Ball and V.~M.~Braun,
  %``Radiative corrections to the decay B --> pi e nu and the heavy quark
  %limit,''
  Phys.\ Lett.\  B {\bf 417} (1998) 154;
%  [arXiv:hep-ph/9709243];
  %%CITATION = PHLTA,B417,154;%%
    P.~Ball,
  %``B --> pi and B --> K transitions from {QCD} sum rules on the light-cone,''
  JHEP {\bf 9809} (1998) 005.
  %[arXiv:hep-ph/9802394];
  %%CITATION = JHEPA,9809,005;%%

%%%%%%%%%%%%%%%%%%%%%nucleon  DAS
\bibitem{BFMS}
  V.~Braun, R.~J.~Fries, N.~Mahnke and E.~Stein,
  %``Higher twist distribution amplitudes of the nucleon in QCD,''
  Nucl.\ Phys.\  B {\bf 589} (2000) 381,
  [Erratum-ibid.\  B {\bf 607} (2001) 433].
  %[arXiv:hep-ph/0007279].
  %%CITATION = NUPHA,B589,381;%%


\bibitem{BLMS}
  V.~M.~Braun, A.~Lenz, N.~Mahnke and E.~Stein,
  %``Light cone sum rules for the nucleon form-factors,''
  Phys.\ Rev.\  D {\bf 65} (2002) 074011.
  %%CITATION = PHRVA,D65,074011;%%



\bibitem{LWS}
  A.~Lenz, M.~Wittmann and E.~Stein,
  %``Improved light-cone sum rules for the electromagnetic form factors of  the
  %nucleon,''
  Phys.\ Lett.\  B {\bf 581} (2004) 199.
  %%CITATION = PHLTA,B581,199;%%


\bibitem{BLW06}
V.~M.~Braun, A.~Lenz and M.~Wittmann,
% ``Nucleon Form Factors in QCD''
Phys.\ Rev.\ D {\bf 73} (2006) 094019 .



\bibitem{NuclDAs09}
A. Lenz, M. Gockeler, Th. Kaltenbrunner and N. Warkentin,
% ``The Nucleon Distribution Amplitudes and their application to nucleon form factors and the N ---> Delta transition at intermediate values of Q**2$''
Phys.\ Rev.\ D {\bf 79} (2009) 093007






\bibitem{Ioffe}
  B.~L.~Ioffe,
  %``Calculation Of Baryon Masses In Quantum Chromodynamics,''
  Nucl.\ Phys.\  B {\bf 188} (1981) 317
  [Erratum-ibid.\  B {\bf 191} (1981) 591].
  %%CITATION = NUPHA,B188,317;%%




\bibitem{BBKR}
  V.~M.~Belyaev, V.~M.~Braun, A.~Khodjamirian and R.~Ruckl,
  %``D* D pi and B* B pi couplings in QCD,''
  Phys.\ Rev.\  D {\bf 51} (1995) 6177.
  %%CITATION = PHRVA,D51,6177;%%

%%%LCSR for baryon ff

\bibitem{Huang:2004vf}
  M.~Q.~Huang and D.~W.~Wang,
  %``Light cone QCD sum rules for the semileptonic decay Lambda(b) -> p l
  %anti-nu,''
  Phys.\ Rev.\  D {\bf 69} (2004) 094003.
  %%CITATION = PHRVA,D69,094003;%%


\bibitem{Wang:2008sm}
  Y.~M.~Wang, Y.~Li and C.~D.~L\"{u},
  %``Rare Decays of Lambda(b) ---> Lambda + gamma and Lambda(b) ---> Lambda + l+
  %l- in the Light-cone Sum Rules,''
  Eur.\ Phys.\ J.\  C {\bf 59} (2009) 861.
  %%CITATION = EPHJA,C59,861;%%


\bibitem{Wang:2009hra}
  Y.~M.~Wang, Y.~L.~Shen and C.~D.~L\"{u},
  %``Lambda(b) ---> p, Lambda transition form factors from QCD light-cone sum
  %rules,''
  Phys.\ Rev.\  D {\bf 80} (2009) 074012.
  %%CITATION = PHRVA,D80,074012;%%




\bibitem{Azizi:2009wn}
  K.~Azizi, M.~Bayar, Y.~Sarac and H.~Sundu,
  %``Semileptonic Lambda(b,c) to Nucleon Transitions in Full QCD at Light
  %Cone,''
  Phys.\ Rev.\  D {\bf 80} (2009) 096007.
  %%CITATION = PHRVA,D80,096007;%%


\bibitem{Aliev:2010yx}
  T.~M.~Aliev, K.~Azizi and M.~Savci,
  %``Strong coupling constants of light pseudoscalar mesons with heavy baryons
  %in QCD,''
  Phys.\ Lett.\  B {\bf 696} (2011) 220.
  %%CITATION = PHLTA,B696,220;%%




%%%3 -point


\bibitem{Dai:1996xv}
  Y.~B.~Dai, C.~S.~Huang, M.~Q.~Huang and C.~Liu,
  %``QCD sum rule analysis for the Lambda(b) ---> Lambda(c) semileptonic
  %decay,''
  Phys.\ Lett.\  B {\bf 387} (1996) 379.
  %%CITATION = PHLTA,B387,379;%%


%\cite{Huang:1998rq}
\bibitem{Huang:1998rq}
  C.~S.~Huang, C.~F.~Qiao and H.~G.~Yan,
  %``Decay Lambda(b) ---> p lepton anti-neutrino in QCD sum rules,''
  Phys.\ Lett.\  B {\bf 437} (1998) 403.
  %%CITATION = PHLTA,B437,403;%%



\bibitem{Navarra:1998vi}
  F.~S.~Navarra and M.~Nielsen,
  %``gND-lambda(c) from QCD sum rules,''
  Phys.\ Lett.\  B {\bf 443} (1998) 285.
  %%CITATION = PHLTA,B443,285;%%


\bibitem{Marques de Carvalho:1999ia}
  R.~S.~Marques de Carvalho, F.~S.~Navarra, M.~Nielsen, E.~Ferreira and H.~G.~Dosch,
  %``Form-factors and decay rates for heavy Lambda semileptonic decays from QCD sum rules,''
  Phys.\ Rev.\  D {\bf 60 } (1999)  034009.












\bibitem{PDG}
  K.~Nakamura {\it et al.}  [Particle Data Group],
  %``Review of particle physics,''
  J.\ Phys.\ G {\bf 37} (2010) 075021.
  %%CITATION = JPHGB,G37,075021;%%



\bibitem{Bagan:1993ii}
  E.~Bagan, M.~Chabab, H.~G.~Dosch and S.~Narison,
  %``Baryon sum rules in the heavy quark effective theory,''
  Phys.\ Lett.\  B {\bf 301} (1993) 243.
  %%CITATION = PHLTA,B301,243;%%



\bibitem{Jido:1996ia}
  D.~Jido, N.~Kodama and M.~Oka,
  %``Negative-parity Nucleon Resonance in the QCD Sum Rule,''
  Phys.\ Rev.\  D {\bf 54} (1996) 4532.
  %%CITATION = PHRVA,Dacryagogue






%\cite{Shuryak:1981fza}
\bibitem{Shuryak:1981fza}
  E.~V.~Shuryak,
  %``Hadrons Containing A Heavy Quark And QCD Sum Rules,''
  Nucl.\ Phys.\  B {\bf 198} (1982) 83.
  %%CITATION = NUPHA,B198,83;%%





\bibitem{Chung:1981cc}
  Y.~Chung, H.~G.~Dosch, M.~Kremer and D.~Schall,
  %``Baryon Sum Rules And Chiral Symmetry Breaking,''
  Nucl.\ Phys.\  B {\bf 197} (1982) 55.
  %%CITATION = NUPHA,B197,55;%%





\bibitem{Braun:2008ia}
  V.~M.~Braun, A.~N.~Manashov and J.~Rohrwild,
  %``Baryon Operators of Higher Twist in QCD and Nucleon Distribution Amplitudes,''
  Nucl.\ Phys.\  B {\bf 807 } (2009)  89.



\bibitem{Braun:2011aw}
  V.~M.~Braun, T.~Lautenschlager, A.~N.~Manashov and B.~Pirnay,
  %``Higher twist parton distributions from light-cone wave functions,''
  Phys.\ Rev.\  {\bf D83 } (2011)  094023.




\bibitem{Bagan:1992tp}
  E.~Bagan, M.~Chabab, H.~G.~Dosch and S.~Narison,
  %``Spectra of heavy baryons from QCD spectral sum rules,''
  Phys.\ Lett.\  B {\bf 287} (1992) 176.
  %%CITATION = PHLTA,B287,176;%%



\bibitem{Bagan:1991sc}
  E.~Bagan, M.~Chabab, H.~G.~Dosch and S.~Narison,
  %``The Heavy baryons Sigma(c) Sigma(b) from QCD spectral sum rules,''
  Phys.\ Lett.\  B {\bf 278} (1992) 367.
  %%CITATION = PHLTA,B278,367;%%



\bibitem{pbarp2charm}A.~Khodjamirian, Ch.~Klein, Th.~Mannel and
Y.-M.~Wang, {\em paper in preparation.}




\bibitem{Wang:2010fq}
   Z.-G.~Wang,
  %``Analysis of the ${1\over 2}^{\pm}$ antitriplet heavy baryon states with QCD sum rules,''
  Eur.\ Phys.\ J.\   C {\bf 68 } (2010)  479.




%\cite{Chetyrkin:2009fv}
\bibitem{Chetyrkin}
  K.~G.~Chetyrkin, J.~H.~K\"uhn, A.~Maier, P.~Maierhofer, P.~Marquard, M.~Steinhauser and C.~Sturm,
  %``Charm and Bottom Quark Masses: An Update,''
  Phys.\ Rev.\  D {\bf 80 } (2009)  074010.
  %[arXiv:0907.2110 [hep-ph]].



\bibitem{Duplancic:2008ix}
  G.~Duplancic, A.~Khodjamirian, Th.~Mannel, B.~Melic and N.~Offen,
  %``Light-cone sum rules for B ---> pi form factors revisited,''
  JHEP {\bf 0804 } (2008)  014.





%\cite{Khodjamirian:2009ys}
\bibitem{KKMO}
  A.~Khodjamirian, Ch.~Klein, Th.~Mannel and N.~Offen,
  %``Semileptonic charm decays D ---> pi l nu(l) and D ---> K l nu(l) from QCD Light-Cone Sum Rules,''
  Phys.\ Rev.\   D {\bf 80 } (2009)  114005.





\bibitem{Ball:2008fw}
  P.~Ball, V.~M.~Braun and  E.~Gardi,
  %``Distribution Amplitudes of the Lambda(b) Baryon in QCD,''
  Phys.\ Lett.\  B {\bf 665 } (2008)  197.



\bibitem{Chernyak:1984bm}
  V.~L.~Chernyak and I.~R.~Zhitnitsky,
  %``Nucleon Wave Function and Nucleon Form-Factors in QCD,''
  Nucl.\ Phys.\  B {\bf 246 } (1984)  52.



\bibitem{BCL}
%\cite{Bourrely:2008za}
%\bibitem{Bourrely:2008za}
  C.~Bourrely, I.~Caprini and L.~Lellouch,
  %``Model-independent description of $B\to \pi l\nu$ decays and a determination
  %of $|V_{ub}|$,''
  Phys.\ Rev.\  D {\bf 79} (2009) 013008.
 % [arXiv:0807.2722 [hep-ph]].
  %%CITATION = PHRVA,D79,013008;%%





\bibitem{Khodjamirian:2011ub}
  A.~Khodjamirian, Th.~Mannel, N.~Offen and Y.-M.~Wang,
  %``$B \to \pi \ell \nu_l$ Width and $|V_{ub}|$ from QCD Light-Cone Sum Rules,''
  Phys.\ Rev.\  D {\bf 83 } (2011)  094031.




\bibitem{Lu:2009cm}
  C.~D.~L\"{u}, Y.~M.~Wang, H.~Zou, A.~Ali and G.~Kramer,
  %``Anatomy of the pQCD Approach to the Baryonic Decays Lambda(b) ---> p pi, p
  %K,''
  Phys.\ Rev.\  D {\bf 80} (2009) 034011.
  %%CITATION = PHRVA,D80,034011;%%





\bibitem{Aaltonen:2008hg}T.~Aaltonen et al. [CDF Collaboration],
Phys.\ Rev.\ Lett.\  {\bf 103} (2009) 031801.














\end{thebibliography}
\end{document}